\PassOptionsToPackage{hyphens}{url}

\documentclass[pre,aps,reprint,amsmath,amssymb,longbibliography, superscriptaddress]{revtex4-2}

\PassOptionsToPackage{bbgreekl}{mathbbol}

\usepackage[sidebarcomments]{ajtex-revtex}

\usepackage{multirow}
\usepackage{tikz}

\makeatletter 
    
\renewcommand\onecolumngrid{
\do@columngrid{one}{\@ne}%
\def\set@footnotewidth{\onecolumngrid}
\def\footnoterule{\kern-6pt\hrule width 1.5in\kern6pt}%
}

\renewcommand\twocolumngrid{
\def\footnoterule{
\dimen@\skip\footins\divide\dimen@\thr@@
\kern-\dimen@\hrule width.5in\kern\dimen@}
\do@columngrid{mlt}{\tw@}
}%

\makeatother

\makeatletter
\def\l@subsubsection#1#2{}
\makeatother

\def\tot{\mathrm{tot}}
\newcommand\ren{\text{ren}}
\def\KMSto{\xrightarrow{\mathrm{\mathsmaller{KMS}}}}

\renewcommand{\e}{{\rm e}}
\renewcommand{\i}{{\text i}}
\newcommand{\f}{{\text f}}
\newcommand{\cq}{{\mathpzc q}}
\newcommand{\calp}{{\mathpzc p}}

\newcommand\env{{\mathsmaller{\mathsf E}}}
\newcommand\fuel{{\mathsmaller{\mathsf F}}}

\def\Ng{\nabla^{\text g}}

\newcommand\blue[1]{{\color{RoyalBlue}#1}}

\renewcommand{\ul}{\underline}

\renewcommand{\paragraph}[1]{\vspace{1em}\noindent\textbf{#1}}

 \definecolor{ogreen}{RGB}{71,191,145}

\newcommand{\aj}[1]{\textcolor{Magenta}{#1}}

\begin{document} 

\title{Hydrodynamics of thermal active matter}

\author{Jay Armas}\email{j.armas@uva.nl}
\affiliation{Institute for Theoretical Physics, University of Amsterdam, 1090 GL Amsterdam, The Netherlands}
\affiliation{Dutch Institute for Emergent Phenomena, 1090 GL Amsterdam, The Netherlands}
\affiliation{Institute for Advanced Study, University of Amsterdam, Oude Turfmarkt 147, 1012 GC Amsterdam, The Netherlands}
\affiliation{Niels Bohr International Academy, The Niels Bohr Institute, University of Copenhagen,
Blegdamsvej 17, DK-2100 Copenhagen \O{}, Denmark}
\author{Akash Jain}\email{ajain@uva.nl}
\author{Ruben Lier}\email{rlier@uva.nl}
\affiliation{Institute for Theoretical Physics, University of Amsterdam, 1090 GL Amsterdam, The Netherlands}
\affiliation{Dutch Institute for Emergent Phenomena, 1090 GL Amsterdam, The Netherlands}
\affiliation{Institute for Advanced Study, University of Amsterdam, Oude Turfmarkt 147, 1012 GC Amsterdam, The Netherlands}

\date{\today}

\begin{abstract}
Active matter concerns many-body systems comprised of living or self-driven agents that collectively exhibit macroscopic phenomena distinct from conventional passive matter. Using Schwinger-Keldysh effective field theory, we develop a novel hydrodynamic framework for thermal active matter that accounts for energy balance, local temperature variations, and the ensuing stochastic effects. By modelling active matter as a driven open system, we show that the source of active contributions to hydrodynamics, violations of fluctuation-dissipation theorems, and detailed balance is rooted in the breaking of time-translation symmetry due to the presence of fuel consumption and an external environmental bath. In addition, our framework allows for non-equilibrium steady states that produce entropy, with a well-defined notion of steady-state temperature. We use our framework of active hydrodynamics to develop effective field theory actions for active superfluids and active nematics that offer a first-principle derivation of various active transport coefficients and feature activity-induced phase transitions. We also show how to incorporate temperature, energy and noise in fluctuating hydrodynamics for active matter. Our work suggests a broader perspective on active matter that can leave an imprint across scales.

\end{abstract} 

\pacs{Valid PACS appear here}

\maketitle

{
\parskip=0.1\baselineskip \advance\parskip by 0pt plus 0.1pt
\tableofcontents
}




\section{Introduction}

The second law of thermodynamics posits that matter should move towards disorder, finally reaching a state of thermal equilibrium that maximises entropy. 
To postpone this ultimate fate of demise, living matter produces an excessive amount of entropy that is released to its surroundings as heat, thus allowing it to maintain its ordered state~\cite{schrodinger1946life}. Living organisms achieve this by burning fuel at the cellular level, primarily adenosine-triphosphate (ATP), to maintain themselves at an operating temperature sufficiently higher than the ambient temperature, ensuring an uninterrupted outflux of entropy to the environment. These metabolic processes of creating and releasing entropy allow living organisms to undertake otherwise statistically unfavourable activities, like self-replication, adaptation, self-organisation, and spontaneous motion, and may be viewed as a \emph{local} violation of the second law of thermodynamics. At macroscopic scales, this dynamics results in entirely new phases of matter, called \emph{active matter} \cite{RevModPhys.85.1143}, of which living systems are the most prominent example, exhibiting collective behaviour distinct from their non-living passive counterparts.

The inherent non-equilibrium nature of active matter maintained via the entropy exchange with the environment and the self-driven nature of the microscopic constituents, leads to novel collective behaviour in the form of pattern formation, non-equilibrium phase transitions, breaking of fluctuation-dissipation theorems, and new forms of mechanical/elastic responses. These properties are manifested in a plethora of systems including active liquid crystals, active solids, and active gels. In particular, active liquid crystals with polar order exhibit a disordered/ordered flocking transition even in two spatial dimensions \cite{1998PhRvE..58.4828T,Toner2005,JULICHER20073,doi:10.2976/1.3054712,2012EPJST.202....1R,Amoretti:2024obt}, while with nematic order activity can give rise to nematic turbulence with spectacular spatially modulated patterns at mesoscopic scales \cite{Doostmohammadi2018,Alert_2020,turbulencealert,morozovalexander,PhysRevLett.89.058101,Ramaswamy_2007}. In turn, active solids may display odd-elastic responses that violate mechanical reciprocity \cite{2019arXiv190207760S} and may become susceptible to instabilities due to the presence of noise \cite{2022PhRvE.106e4607J}. Besides being realised in biological systems, active matter may also be engineered using motile particles with inbuilt batteries, light-activated beads, mechanically/electrically driven systems \cite{PhysRevE.67.040301}, or designed using metamaterials \cite{Shankar_2022, brandenbourger2022limitcyclesturnactive} and colloidal particles \cite{doi:10.1021/ja047697z, doi:10.1126/science.1230020}.

To arrive at macroscopic descriptions of phases of matter, one can use a top-down reductionist approach, starting from the microscopic constituents and their mutual interactions. This approach is often quite difficult and only really manageable for weakly interacting systems, using techniques such as kinetic theory and agent-based modelling~\cite{pitaevskii2012physical, 1995PhRvL..75.1226V}~\footnote{An alternative bottom-up approach for strongly-interacting systems is holography~\cite{Maldacena:1997re, Policastro:2001yc, Bhattacharyya:2008jc}, though the application of such methods to non-equilibrium open systems remains challenging.}. However, most collective behaviour in nature turn out to be largely agnostic to the underlying microscopic dynamics and are effectively described by a handful of collective variables, such as local temperature and density, and a few transport parameters, such as viscosity and conductivity. This allows one to take a bottom-up emergent approach, commonly known as \emph{hydrodynamics}, to directly obtain effective theories for phases of matter based on the laws of thermodynamics and the underlying symmetries, such as rotations, translations, or number conservation, and the relevant symmetry breaking patterns. In this context, the most important guiding principle for passive phases of matter is the \emph{local second law of thermodynamics}, which requires that the local rate of entropy production must be non-negative everywhere throughout the system. Despite being just an inequality, this requirement is immensely constraining for hydrodynamic models as it needs to be satisfied for every possible configuration admitted by the system. 

The presence of local fuel consumption allows active systems to bypass the usual constraints on transport imposed by the local second law of thermodynamics, thus allowing for a wider range of transport properties. The standard prescription to derive these novel effects is to incorporate a fuel source into the framework of hydrodynamics, e.g. in models of polar/nematic gels and viscoelasticity \cite{2018RPPh...81g6601J, PhysRevLett.92.078101, 2023arXiv230915142L, JULICHER20073, Callan-Jones_2011, 2023arXiv231008640D,Lier:2021wxd}. Fuel consumption leads to novel contributions to the hydrodynamic constitutive relations, such as odd strains arising from interactions between non-reciprocal springs \cite{2019arXiv190207760S, Tan2022, matus2024molecular, Han2021} and active nematic stresses in liquid crystals responsible for nematic turbulence \cite{PhysRevLett.92.118101, PhysRevLett.89.058101, Ramaswamy_2007}. In contrast with passive systems, active hydrodynamic constitutive relations induced by fuel consumption do not obey the usual fluctuation-dissipation theorems (FDTs) nor Onsager's reciprocity relations~\cite{1931PhRv...37..405O, 1931PhRv...38.2265O, landau1980statistical, Wang:1998wg} that are rooted in microscopic time-reversal dynamics. 

The absence of standard fluctuation-dissipation statistics, microscopic time-reversibility, and second law constraints suggests a more phenomenological approach to active hydrodynamics based solely on the relevant symmetries for a given phase of matter. This viewpoint has been pursued in various works, for instance in the widely-studied Toner-Tu hydrodynamic model \cite{1998PhRvE..58.4828T, Toner2005} of active matter with polar order (see also \cite{RevModPhys.85.1143} for a review of many hydrodynamic models of active matter). However, many active systems still operate relatively close to equilibrium, suggesting that instead of abandoning entirely the framework of passive hydrodynamics, one should depart from it slowly as the strength of activity is increased. This can be done by retaining the \textit{local equilibrium hypothesis} and introducing activity perturbatively by means of a control parameter \cite{2018RPPh...81g6601J,de2013non}.



One of the main uses of hydrodynamics is extracting correlation functions. 
To do so, one must first specify the global equilibrium state, or a global steady state for active matter, around which the hydrodynamic variables locally fluctuate. However, one is immediately faced with a problem. Since fuel consumption constantly injects energy into the system, the local temperature rises indefinitely and it is impossible to formulate a steady state at fixed temperature. See~\cite{2018RPPh...81g6601J} for a review of temperature dynamics in the presence of fuel sources.
Indeed, for this reason, temperature is often not taken as a collective variable in theories of active hydrodynamics (see e.g.~\cite{RevModPhys.85.1143}). Consequently, it is not possible to access the correlation functions of energy that are a hallmark of non-equilibrium systems \cite{sengersortiz}.

The solution we propose is to model active systems as driven open systems and to incorporate, in addition to the energy source arising from fuel consumption, an energy sink within the hydrodynamic model to remove the excess energy. To model this, we take inspiration from a large body of literature developing hydrodynamic models with momentum or charge sinks, in systems with approximate translation or U(1) symmetries~\cite{Grossi:2021gqi, Delacretaz:2021qqu, Armas:2021vku, Armas:2022vpf, Armas:2023tyx, Armas:2023ouk}. In our case, the approximate symmetry is time-translations, associated with the approximately conserved energy density. However, energy sinks are conceptually different from momentum or charge sinks because the latter do not extract entropy from the system, while energy sinks cause entropy to be released to the environment. Interactions between the fuel source and the energy sink define the global steady state temperature of the active system, in turn determining the rate at which entropy is released to the environment.

Since active systems operate around non-equilibrium steady states, their dynamics cannot be well-described by deterministic models and one must also account for stochastic thermal noise. To this end, and to systematise the role of fuel source and energy sink in active hydrodynamics, we use the methodology of \emph{Schwinger-Keldysh effective field theory} (SK-EFT)~\cite{Grozdanov:2013dba, Harder:2015nxa, Crossley:2015evo, Haehl:2015uoc, Haehl:2018lcu, Jensen:2017kzi, Glorioso:2018wxw}. SK-EFT is a systematic symmetry-based effective action approach to hydrodynamics that has stochastic noise built into the framework. It features a discrete Kubo-Martin-Schwinger (KMS) symmetry that is responsible for imposing FDTs, Onsager's relations, and the local second law of thermodynamics in passive systems. We propose a suitably generalised \emph{active KMS symmetry} that allows for systematic violations of each of these requirements in terms of the rates of fuel consumption and heat loss, perturbatively controlled by the strength of activity. The proposal stems from the physical requirement that when an active system runs out of fuel and all external forces are turned off, dubbed the \emph{famine state}, it must behave as an ordinary passive system. The active KMS symmetry can be understood as a field theoretic realisation of the principle of microscopic reversibility when subjected to external work~\cite{Crooks1998, Crooks:1999ttd} as we demonstrate in our work.

We apply this new framework of active hydrodynamics to active superfluids and active nematics, while sketching potential extensions to several other active phenomena. We systematically study how active contributions may arise in the hydrodynamic constitutive relations, and cause violations of FDTs, Onsager's relations, and the local second law of thermodynamics. Depending on the choice of parameters, these models admit activity-induced phase transitions between the ordered and disordered states and present the perfect opportunity for approaching active phase transitions~\cite{2018PhRvL.121j8002S} from a Wilsonian renormalisation group perspective. We also provide the first-ever computation of energy correlation functions in active matter and show how to incorporate temperature, energy and noise within the framework of fluctuating hydrodynamics.

A few recent works used EFT techniques for modelling certain aspects of active matter. In~\cite{2023arXiv230915142L}, a SK-EFT model for active nematics was constructed by including a fuel source, as done previously for conventional models of active hydrodynamics \cite{JULICHER20073,2018RPPh...81g6601J}. In~\cite{Huang:2023eyz}, a complementary EFT approach to non-equilibrium systems based on the Fokker-Planck equation and stationary probability distributions was presented and applied to systems without time-reversal symmetry. Furthermore,~\cite{Huang:2023eyz} discussed a version of FDT applicable to states described by non-thermal stationary distributions, generalising previous work of~\cite{Basu2008,PhysRevLett.103.090601}. In this paper, we move beyond, and extend, these works by accounting for both fuel consumption and entropy loss to the environment within the framework of SK-EFTs. This allows us to make several advances in active hydrodynamics: (i) all active contributions can be seen to arise from the explicit breaking of time-translation symmetry; (ii) the framework allows for non-equilibrium steady states that produce entropy, with a well-defined notion of steady-state temperature; (iii) the active hydrodynamic equations include temperature as a dynamical collective variable, (iv) we can systematically access symmetric and retarded correlation functions of energy and heat loss to the environment; and (v) concretely identify the source of violations of FDTs and detailed balance in the presence of activity.

\vspace{1em}
\noindent
\emph{Organisation:} The structure of this work is as follows. 
In \cref{sec:active-formalism}, we review previous formulations of active matter hydrodynamics and then establish a general hydrodynamic framework for active matter starting from the principle of energy balance and the existence of a famine state. In particular, we introduce the notion of active KMS symmetry in SK-EFTs and discuss how it relates to the systematic violations of the local second law of thermodynamics, FDTs, and detailed balance. Then, in \cref{sec:simplediffusion}, we consider a simple toy model of active superfluids without momentum conservation, and discuss the salient features of our model such as active transport coefficients and activity-induced phase transitions. We also show that the obtained correlators do not rely on the SK-EFT and can be obtained in a more conventional way using the principles of nonequilibrium fluctuating hydrodynamics. In \cref{sec:active_nematics}, we extend this construction to active nematics. We conclude in \cref{sec:buzz} with a discussion of future directions. 

The paper is accompanied by several appendices. In \cref{app:Julicher}, we provide a comparison of our formalism with the conventional framework of active hydrodynamics. In \cref{app:SK}, we provide an in-depth review of the SK formalism for passive hydrodynamics and give the details of its extension to active hydrodynamics. In \cref{app:examples}, we provide a detailed construction of SK effective actions for active diffusion, active superfluids, active hydrodynamics, and active nematics. Lastly, in \cref{eq:glossary}, we provide a comprehensive glossary of symbols and notation employed in this work.


\section{A field theory for activity}
\label{sec:active-formalism}

\subsection{Activity engines}
\label{sec:fuelconsumption}

Consider a physical system that burns fuel from a reservoir for energy that it uses to stay active. For instance, this may be ATP used by cells, food consumed by a bacterial population, or internal batteries of motile particles. Burning fuel produces heat proportional to the chemical energy differential $\Delta E$ at rate $r_\fuel$, leading to the \emph{energy balance} equation
\begin{align}\label{eq:energy-balance-raw}
    \dow_t{\epsilon} +  \partial_i\epsilon^i 
    = \ell r_\fuel \Delta E~.
\end{align}
where $\epsilon$ and $\epsilon^i$ are the energy density and flux. The parameter $\ell$ serves as a bookkeeping tool to keep track of any contributions tied to the in/outflow of energy. We will assume $\ell$ to be sufficiently small, so that we remain in a regime where local energy conservation is still approximately applicable. Alternatively, following~\cite{2018RPPh...81g6601J}, we can also work with total conserved energy density $\epsilon_\tot$ that includes the fuel contribution; a detailed comparison is provided in \cref{app:Julicher}. Depending on the underlying symmetry structure, we may also need to account for conservation equations, e.g. momentum, mass, or particle conservation, and Goldstone equations for spontaneously broken symmetries. We shall return to these considerations in detail later.

The first law of thermodynamics, $\df\epsilon = T\df s$, relates the energy density $\epsilon$ to the entropy density $s$ of the fluid and its local temperature $T$. Using this, we can recover
\begin{align}
    \dow_t s + \dow_is^i = 
    - \frac{1}{T^2}\epsilon^i\dow_i T
    + \ell r_\fuel \frac{\Delta E}{T}.
\end{align}
where $Ts^i = \epsilon^i$ denotes the heat flux. The total entropy of the system plus fuel must locally be produced, i.e. 
\begin{align}\label{eq:second-law-passive}
    \dow_t s + \dow_i s^i \geq 0.
\end{align}
Therefore, we are led to the constitutive relations
\begin{align}\label{eq:constitutive-relations-basic}
    \epsilon^i = -\kappa\, \dow^i T + \ldots, \qquad 
    r_\fuel = \ell\gamma_\fuel \Delta E,
\end{align}
where $\kappa\geq 0$ is thermal conductivity and $\gamma_\fuel\geq 0$ is a coefficient that controls the rate of fuel depletion. Ellipsis denote further derivative corrections. For simplicity, we will assume $r_\fuel$ to not admit any derivative corrections throughout this work, which amounts to the physical assumption that the fluid does not backreact on the fuel burning process. 

Plugging the constitutive relations from \cref{eq:constitutive-relations-basic} back into \cref{eq:energy-balance-raw}, we find that fuel consumption results in a steady build-up of energy in the system $\dow_t\epsilon\sim \ell^2\gamma_\fuel\Delta E^2$. Correspondingly, the system continuously heats up in time, with $\dow_tT\sim \ell^2\gamma_\fuel\Delta E^2/c_v$, where $c_v=\dow\epsilon/\dow T$ is the heat capacity.

\begin{figure}
    \centering
    \tikz[thick]{
        \draw[rounded corners=5pt] (-2.3,-.7) rectangle (-3.7,.7);
        \node at (-3,-.9) {\scriptsize{\bf{Fuel}}};
        \node at (-3,0) {$\Delta E$};

        \node at (-1.65,.25) {${r_\fuel\Delta E}$};
        \node at (-1.65,-.24) {$\mathsmaller{r_\fuel\sim\gamma_\fuel\Delta\!E}$};
        
        \draw[rounded corners=10pt] (-1,-1) rectangle (1,1);
        \node at (0,-1.25) {\scriptsize{\bf{System}}};
        \node at (0,0) {$\epsilon(T_0)$};

        \node at (1.65,.25) {${r_\env T_\env}$};
        \node at (1.65,-.24) {$\mathsmaller{r_\env\sim\gamma_\env\Delta\!T}$};
        
        \draw[rounded corners=5pt] (2.3,-.7) rectangle (3.7,.7);
        \node at (3,-.9) {\scriptsize{\bf{Environment}}};
        \node at (3,0) {$T_\env$};

        \draw[->, ultra thick, Red] (-2.6,0) -- (-.6,0);
        \draw[->, ultra thick, Blue] (.6,0) -- (2.7,0);
    }
    \caption{Schematic representation of energy balance between the system, fuel reservoir, and environment. Without the environment component, fuel consumption would lead to an indefinite build-up of energy or heat in the system.}
    \label{fig:energy-balance}
\end{figure}

For the system to reach a homogeneous active steady-state in the presence of fuel consumption, we must also include a sink in the energy balance equation. To wit,
\begin{align}\label{eq:energy-balance}
    \dow_t{\epsilon} +  \partial_i\epsilon^i 
    = \ell r_\fuel \Delta E
    - \ell  r_\env \kB  T_\env~,
\end{align}
where $\kB r_\env$ denotes the rate of entropy loss to the environment of ambient temperature $T_\env$, and $\kB$ is the Boltzmann constant. See \cref{fig:energy-balance}. The energy sink simultaneously removes entropy from the system, so the second law of thermodynamics takes a modified form
\begin{align}\label{eq:second-law}
    \dow_t s + \dow_i s^i + \ell \kB r_\env \geq 0.
\end{align}
This says that the entropy of an active system, plus the entropy lost to the environment, is locally produced. We will derive this relation more systematically in \cref{sec:second-law} using Schwinger-Keldysh effective field theory.
Once again, using the first law $\df\epsilon=T\df s$, we find
\begin{equation}
    \dow_t s + \dow_i s^i + \ell \kB r_\env
    = 
    - \frac{1}{T^2}\epsilon^i\dow_iT
    + \ell r_\fuel \frac{\Delta E}{T}
    + \ell \kB r_\env \frac{\Delta T}{T}
\end{equation}
implying that the rate of heat loss is proportional to the thermal gradient $\Delta T = T - T_\env$ between the system and the environment, i.e.
\begin{align}\label{eq:losscurrent}
    r_\env = \ell  \gamma_\env \kB \Delta T  +  \ldots~,
\end{align}
controlled by the coefficient $\gamma_\env\geq 0$. 

A homogeneous active steady-state is achieved when the energy $r_\fuel\Delta E$ received from the fuel consumption in \cref{eq:energy-balance} is balanced by the heat loss $r_\env \kB T_\env$. This happens at the steady-state temperature
\begin{equation}\label{eq:eqb-temp}
    T_0
    = T_\env
    \lb 1 + \frac{\gamma_\fuel}{\gamma_\env}\frac{\Delta E^2}{\kB^2 T_\env^2} \rb~,
\end{equation}
which is above the ambient temperature as the system is subjected to a chemical gradient $\Delta E\neq 0$. It will be useful to identify some dimensionless measure for quantifying the strength of activity. There are two natural candidates: an external measure $\hat\aleph=\ell\Delta E/(\kB T_\env)$ that measures the amount of energy being injected by the fuel relative to the environment, and a dynamical measure $\aleph=\ell \Delta T/T_\env$ that measures the operating temperature of the active system relative to the environment. In the steady state \eqref{eq:eqb-temp}, the two are related by $\aleph = \ell\gamma_\fuel/\gamma_\env\,\hat\aleph^2$.

\begin{figure}
    \centering
    \tikz[thick]{
        \draw[rounded corners=5pt,Gray] (-2.3,-.7) rectangle (-3.7,.7);
        \node at (-3,-.9) {\scriptsize{\textcolor{Gray}{\bf{Fuel}}}};
        \node at (-3,0) {$\textcolor{Gray}{\Delta E=0}$};
        
        \draw[rounded corners=10pt] (-1,-1) rectangle (1,1);
        \node at (0,-1.25) {\scriptsize{\bf{System}}};
        \node at (0,0) {$\epsilon(T_\env)$};

        \node at (1.65,.25) {${r_\env T_\env}$};
        \node at (1.65,-.24) {$\mathsmaller{r_\env\sim\gamma_\env\Delta\!T}$};
        
        \draw[rounded corners=5pt] (2.3,-.7) rectangle (3.7,.7);
        \node at (3,-.9) {\scriptsize{\bf{Environment}}};
        \node at (3,0) {$T_\env$};

        \draw[ultra thick, Gray, dashed] (-2.4,0) -- (-.8,0);
        \node at (-1.65,0) {\LARGE$\textcolor{red}{\times}$};
        \draw[<->, ultra thick, Blue] (.6,0) -- (2.7,0);
    }
    \caption{Schematic representation of the famine state. When the fuel reservoir is absent, the system can still exchange heat with the environment and reaches a unique equilibrium state maintained at the environment temperature $T_\env$.}
    \label{fig:famine-state}
\end{figure}

When an active system runs out of fuel, i.e. $\Delta E =0$, it cools down to a global thermal equilibrium state with $T_0=T_\env$. We refer to this lack of fuel/food as the \emph{famine state}; see \cref{fig:famine-state}. This state is effectively described by ordinary passive hydrodynamics, but with short-lived energy or heat fluctuations with the characteristic relaxation rate
\begin{equation}\label{eq:energy-relaxation-rate}
    \Gamma_\epsilon = \frac{ \ell^2 \kB^2 T_\env\gamma_\env}{c_v}~.
\end{equation}
This story is analogous to the recently developed hydrodynamic framework for approximate or weakly explicitly broken symmetries, featuring approximately conserved relaxed charges~\cite{Grossi:2021gqi, Delacretaz:2021qqu, Armas:2021vku, Armas:2022vpf, Armas:2023tyx, Armas:2023ouk}. In this instance, the approximate symmetry is time-translations, associated with the approximately conserved energy density. The passive statement of the local second law in \cref{eq:second-law-passive} is restored in the famine state when the source of activity is absent. To wit, when $T_\env$ is constant, \cref{eq:energy-balance,eq:second-law} together imply that $\dow_t s_{\text{fam}} + \dow_i s^i_{\text{fam}} \geq 0$, with $s_{\text{fam}} = s-\epsilon/T_\env$ and $s^i_{\text{fam}} = s^i-\epsilon^i/T_\env$. Note that this does not work when the environment temperature $T_\env$ is varying in space or time. The spatial or temporal gradients of $T_\env$ act as thermal engines that drive the system away from the famine state. We shall only consider a homogeneous and stationary environment in this paper.

Instead of the activity being supported by a chemical engine, we may also consider other engines of activity, such as an external thermal gradient, electric fields, or mechanical forcing. If we remain agnostic of the details of the engine itself, many such physical scenarios merely amount to reinterpretations of $r_\fuel$ and $\Delta E$. 
For example, consider a 2d active system driven by electric fields or mechanical forcing transverse to the plane~\cite{doi:10.1021/ja047697z, doi:10.1126/science.1230020, Amoretti:2022ovc, PhysRevE.67.040301}. In these cases, $r_\fuel$ and $\Delta E$ may be interpreted as the transverse components of the charge flux/momentum density and electric fields/acceleration respectively. 
The source of driving may also be inhomogeneous in time and/or space, modelled by a non-constant profile for $\Delta E$, e.g. $\Delta E\propto \sin(\Omega t)$ for an oscillatory activity engine with frequency $\Omega$. However, we will only consider homogeneous activity engines throughout this work for simplicity.
Lastly, we may also consider external electric fields or mechanical forcing within the dimensionality of the active system, but this additionally requires one to introduce momentum sinks to balance the momentum imparted by the external fields. We will comment on this case towards the end of this paper.

\subsection{Active KMS symmetry}
\label{sec:active-KMS}

Hydrodynamics describes the evolution of conserved charges in a physical system and is characterised by the constitutive relations for the conserved fluxes expressed in terms of the conserved densities, and possibly order parameters of spontaneously broken symmetries. In addition to symmetries, the construction of constitutive relations for a passive system is guided by physical requirements such as the local second law of thermodynamics~\cite{landau1959fluid} and the existence of local thermal equilibrium~\cite{Banerjee:2012iz, Jensen:2012jh}. However, active systems may freely dump entropy into the environment thereby violating the local second law. They also operate around non-equilibrium steady-states and typically do not admit stable thermal equilibrium states. To help us traverse this uncharted territory, we look towards \emph{Schwinger-Keldysh (SK) hydrodynamics}~\cite{Grozdanov:2013dba, Harder:2015nxa, Crossley:2015evo, Haehl:2015uoc, Haehl:2018lcu, Jensen:2017kzi, Glorioso:2018wxw}, which is a recently-developed effective field theory framework for hydrodynamics that, in principle, applies arbitrarily far from equilibrium.

The primary ingredient in SK hydrodynamics is the Kubo-Martin-Schwinger (KMS) symmetry. It is an incarnation of microscopic reversibility and ensures that the thermal correlators of hydrodynamic operators satisfy the fluctuation-dissipation theorems (FDTs) and Onsager's reciprocity relations~\cite{1931PhRv...37..405O, 1931PhRv...38.2265O, landau1980statistical, Wang:1998wg}. For two-point symmetric and retarded correlators, denoted $G^\rmS$ and $G^\rmR$ respectively, these statements read in Fourier space
\begin{align}\label{eq:FDT}
    G^\rmS_{\cO\cO'} &= \frac{1}{i\omega\beta_0} \Big( G^\rmR_{\cO\cO'} \mp G^{\rmR*}_{\cO\cO'} \Big)~, \nn\\
    G^\rmR_{\cO\cO'} &= \pm G^\rmR_{\cO'\cO}~,
\end{align}
where $\beta_0 = 1/(\kB T_0)$, and $\cO$, $\cO'$ represent the hydrodynamic operators of interest such as the conserved densities and fluxes. The upper/lower signs in \cref{eq:FDT} apply when the time-reversal eigenvalues of the two operators are the same/opposite. Analogous statements apply for higher-point functions~\cite{Wang:1998wg}. 

In the SK formalism, FDTs, Onsager's relations, and their higher-point generalisations, are all realised via a discrete symmetry of the effective field theory, known as the KMS symmetry. In simplest terms, we introduce a pair of external sources, denoted ${\sf s}_r$, ${\sf s}_a$, that can be used to compute the expectation values of the operators $\cO$ as well as their symmetric and retarded correlators, i.e.
\begin{align}
    \langle \cO\rangle 
    &= \frac{-i\delta}{\delta{\sf s}_a}\ln{\cal Z}~, \nn\\
    G^\rmS_{\cO\cO'} 
    &= \frac{-i\delta}{\delta{\sf s}'_a}\frac{-i\delta}{\delta{\sf s}_a}\ln{\cal Z}~, \nn\\
    G^\rmR_{\cO\cO'} 
    &= \frac{\delta}{\delta{\sf s}'_r}\frac{-i\delta}{\delta{\sf s}_a}\ln{\cal Z}~,
    \label{eq:corr-from-gf}
\end{align}
and similarly for higher-point correlation functions. Here $\cZ = \int\cD\psi\cD\psi_a\exp(i\int\df t\,L)$ is the SK generating functional of the theory, with the Lagrangian $L$, and $\psi$, $\psi_a$ collectively denote the physical and stochastic dynamical fields~\footnote{The physical dynamical fields $\psi$ are not given a label $r$ because they may generically differ from the $r$-type ``average'' fields used in the SK formalism; see \cref{app:SK}.}.
Heuristically, ${\sf s}_r$ can be understood as the true background fields and ${\sf s}_a$ as the associated thermal noise. For example, to obtain the correlation functions of conserved particle number/charge density and flux operators associated with an internal U(1) symmetry, we need to introduce the associated background gauge field and its noise partner. Whereas for correlation functions of conserved currents associated with spacetime symmetries, i.e. energy density, energy flux, momentum density, and stress tensor, we need to couple the theory to a background spacetime geometry and noise partners; see \cref{app:NC} for more details. We will see particular examples later in \cref{sec:simplediffusion,sec:active_nematics}.

The KMS symmetry can be stated as the invariance of the theory under a discrete transformation
\begin{align}\label{eq:KMS}
    {\sf s}_r &\KMSto \eta_\Theta\,{\sf s}_r~, \nn\\
    {\sf s}_a &\KMSto \eta_\Theta\,\hat{\sf s}_a\equiv \eta_\Theta\!\lb{\sf s}_a  
    + i\beta_0 \dow_t{\sf s}_r\rb~,
\end{align}
where $\eta_\Theta$ denotes the time-reversal eigenvalue. Throughout this paper, the right-hand sides of KMS transformation are understood to be evaluated at $(-t,\vec x)$. The KMS symmetry is realised on the dynamical fields as $\psi,\psi_a\KMSto \eta_\Theta\psi,\eta_\Theta\hat\psi_a$, where the explicit form of $\hat\psi_a$ depends on the model under consideration and will be discussed later in explicit examples. Depending on the application in mind, the KMS symmetry can analogously be defined for other kinds of reversibility symmetries involving combinations of spatial-parity and charge-conjugation, such as PT, CT, or CPT. More details on the KMS symmetry and its derivation from the full quantum KMS symmetry is provided in \cref{app:passive-sk}.

To describe an active system, we need two new ingredients in the SK framework: Firstly, we need a fuel source which generates the activity \cite{2018RPPh...81g6601J}. Secondly, we need a heat sink that balances out the continual influx of energy from the fuel source. These two ingredients are modelled using the external \emph{fuel} and \emph{environment sources} $\Phi^{\fuel,\env}_r$ and the associated noise partners $\Phi^{\fuel,\env}_a$. Physically, these may be interpreted as the pair of external sources coupled to the rate operators $r_{\fuel,\env}$ introduced in \cref{sec:fuelconsumption}, and for a homogeneous and non-stochastic fuel source and environment, take the value
\begin{align}\label{eq:bath-config}
    \Phi_r^\fuel = -\Delta E\,t, \quad 
    \Phi^\env_r = \kB T_\env\,t, \qquad 
    \Phi_a^\fuel = \Phi^{\env}_a = 0~.
\end{align} 
For inhomogeneous profiles of $\Delta E$ or $T_\env$, we can instead identify $\Phi_r^\fuel = -\int \df t\,\Delta E$ and $\Phi_r^\env = \int \df t\, T_\env$. 

To set up the active KMS symmetry, let us recall the hydrostatic principle which states that a passive system always flows towards global thermal equilibrium when coupled to time-independent non-stochastic background fields, i.e $\dow_t{\sf s}_r = {\sf s}_a = 0$. A crucial observation in this regard is that the external noise fields in such configurations remain zero under KMS transformation~\eqref{eq:KMS}. While this need not generically apply to the fuel/environment fields, we do expect the system to flow to global thermal equilibrium in the famine state, $\Delta E=T_0-T_\env=0$, i.e. $\Phi^\env_r=\kB T_0 t$, $\Phi^\fuel_{r},\Phi^{\fuel,\env}_a=0$. Requiring that the KMS transformation leaves the noise fields vanishing in and only in the famine state, fixes the KMS transformation to
\begin{align}\label{eq:activeKMS}
    \Phi^\fuel_r 
    &\KMSto -\Phi^\fuel_r~, \nn\\
    \Phi^\env_r 
    &\KMSto -\Phi^\env_r~, \nn\\
    \Phi^\fuel_a 
    &\KMSto - \hat\Phi^\fuel_a \equiv
    -\Phi^\fuel_a - i\beta_0 \dow_t\Phi^\fuel_r~, \nn\\
    \Phi^\env_a 
    &\KMSto - \hat\Phi^\env_a \equiv
    -\Phi^\env_a - i\beta_0 \dow_t\Phi^\env_r + i ~,
\end{align}
with time-reversal eigenvalues $-1$. Note the additional ``$i$'' shift in the KMS transformation of $\Phi^\env_a$. This is the \emph{active KMS transformation}, suitable for describing out-of-equilibrium systems coupled to a thermal bath. An appropriate extension to the quantum regime is provided in \cref{sec:activity-in-SK}.

The background field variations $\delta\Phi^\env_r = \Phi^\env_r - \kB T_\env t$, $\delta\Phi^\env_a = \Phi^\env_a$ satisfy the standard KMS symmetry \eqref{eq:KMS} in the famine state, hence the correlators in this state satisfy FDTs and Onsager's relations. Departing from the famine state, the background field configuration itself breaks the KMS symmetry  and the system has to settle in a non-equilibrium steady-state that causes violations of FDTs and Onsager's relations in \cref{eq:FDT}. Note that the effective field theory still realises the active KMS symmetry \eqref{eq:activeKMS}, but it relates states with $\Phi_a^{\fuel,\env} = 0$ to states with $\Phi_a^{\fuel,\env} \neq 0$. This is philosophically similar to how applying external magnetic fields to a rotationally-invariant theory leads to anisotropic low-energy observables and rotations relate states with different orientations of external magnetic fields. Except that in our case, the active KMS symmetry maps physical steady states without noise in an active system to auxiliary stochastic states with nonzero noise.

As a final comment, we note that while the environment fields $\Phi^\env_{r,a}$ are essential for our construction, the role of fuel fields $\Phi^\fuel_{r,a}$ may instead be played by any of the other background fields ${\sf s}_{r,a}$ relevant to the system under consideration. The only requirement for driving activity is that $\dow_t{\sf s}_r\neq 0$. This may be achieved, e.g., via a second heat bath with a temperature different from $T_{\env}$. Alternatively, one may consider background electric fields or mechanical driving, however, this introduces anisotropy in the system and also requires the introduction of a momentum sink to counter the momentum imparted by the external fields~\cite{Amoretti:2022ovc}. Lastly, apart from representing a fuel source such as ATP or food, the fuel fields $\Phi^\fuel_{r,a}$ can also be used for modelling internal batteries of motile particles, or, in two dimensions, electric fields or mechanical driving transverse to the plane of the system.

\subsection{Energy balance, unitarity, and the second law of thermodynamics}
\label{sec:second-law}

Just like the two copies of background sources, all the global symmetries in the SK framework are also doubled. The ``$r$-type symmetries'' ensure that the classical equations of motion are invariant under symmetry transformations, while the ``$a$-type symmetries'' are responsible for imposing the respective conservation equations. For instance, the symmetries relevant for energy balance are the doubled time-translations. The $r$-type time-translations act on all the fields as usual, i.e. $f(t)\to f(t+\chi^t_r)$, for a constant parameter $\chi^t_r$. Whereas, the $a$-type time-translations only act on the $a$-type noise fields and mix them with the physical fields. Collectively denoting the background sources as $\underline\sfs_{r,a}=(\sfs_{r,a},\ell\Phi^\fuel_{r,a},\ell\Phi^\env_{r,a})$, we have
\begin{align}\label{eq:a-type-translations}
    \ul\sfs_{a} \to \ul\sfs_{a} + \chi^t_a \dow_t\ul\sfs_r~,
\end{align}
for a constant parameter $\chi_a^t$. The transformations of the noise dynamical fields $\psi_a$ are more involved, depending on the particulars of the system under consideration, and will be discussed later. Using the standard Noether procedure, \cref{eq:a-type-translations} implies the balance of energy
\begin{align}\label{eq:energy-conservation}
    \dow_t{\epsilon} +  \partial_i\epsilon^i  
    = - \cO\dow_t{\sf s}_r - \ell r_\fuel\dow_t\Phi^\fuel_r - \ell  r_\env \dow_t\Phi^\env_r ~,
\end{align}
which extends \cref{eq:energy-balance} to account for the energy imparted or removed by time-dependent external fields.

The SK framework also features a set of constraints arising from the unitarity of the microscopic time-evolution operator~\cite{Crossley:2015evo, Glorioso:2018wxw}. These are summarised as
\begin{align}\label{eq:SK-conds}
    L\big|_{f_a\to 0} = 0~, \quad 
    L\big|_{f_a\to-f_a} = -L^*, \quad 
    \Im L\geq 0~,
\end{align}
where $f_a$ represents all the $a$-type fields. To implement the second condition later in the text, it is useful to introduce the SK-unitarity operator 
\begin{align}\label{eq:unitarity-dagger}
    (\ldots)^\dagger = (\ldots)^*\big|_{f_a\to-f_a}~,
\end{align}
so that $L^\dagger = -L$. In particular, $f_a^\dagger = -f_a^*$. 

The KMS symmetry, energy balance, and the unitarity constraints, together conspire to give rise to the local second law of thermodynamics for passive systems~\cite{Glorioso:2016gsa}. This is the statement that there exists an entropy density $s$ and associated flux $s^i$ such that entropy is locally produced as in \cref{eq:second-law-passive}. However for active systems, owing to the additional ``$i$'' term in the active KMS symmetry \eqref{eq:activeKMS}, one finds that the local second law modifies to \cref{eq:second-law}. We will see how this works in specific examples later, while a general derivation of the active second law from the active KMS symmetry is presented in \cref{app:2ndlaw}.

\subsection{Microscopic reversibility and detailed balance}
\label{sec:DB}

Let us take a quick detour to see how the (active) KMS symmetry relates to microscopic reversibility and detailed balance. Given that the system starts from an initial state $\psi(t_\i)=\uppsi_\i$ at time $t_\i$, the conditional probability distribution for it to follow a path $\psi(t)$ until a final time $t_\f$ is given by a path integral over the noise fields
\begin{equation}\label{eq:path-distribution}
    \bbP(\psi|\uppsi_\i,t_\i) = \frac{1}{\cN}\dsp\int{\cD\psi_a}
    \exp\!\lb i\int_{t_{\text i}}^{t_{\text f}}{\!\!\df t}\, L\big(\psi,\psi_a,\ul\sfs_r,\ul\sfs_a\big)\rb~,
\end{equation}
with boundary conditions $\psi_a(t_{\i,\f})=0$.
Using this, we can also obtain the conditional probability distribution for the system to transition to the final state $\psi(t_\f)=\uppsi_\f$ by integrating over all the paths
\begin{align}\label{eq:transition-probability}
    \bbP(\uppsi_\f,t_\f|\uppsi_\i,t_\i)
    = \int_{\uppsi_\i}^{\uppsi_\f}\cD\psi\, \bbP(\psi|\uppsi,t_\i)~.
\end{align}
The normalisation $\cN$ in \cref{eq:path-distribution} is fixed such that the probabilities add up, i.e. $\int\df\uppsi_\f\,\bbP(\uppsi_\f,t_\f|\uppsi_\i,t_\i)=1$. 

Under KMS transformation, $L\KMSto L-i\beta_0\dow_t\Omega$, where we have kept the possible temporal boundary term but the spatial boundary terms may be ignored. It is convenient to include a time-translation along with the time-reversal transformation while implementing KMS on a finite time-interval, $\Theta f(t) = \eta_\Theta f(-t+t_\i+t_\f)$, so that the interval maps to itself. Repeating the KMS transformation brings $L$ back to itself, which requires that $\Omega$ does not contain any $a$-type noise fields and is even under time-reversal. Note that the definition of $\Omega$ is ambiguous because $L$ can be redefined with arbitrary boundary terms. This may typically be fixed by requiring that $L$ is invariant under all the relevant continuous symmetries without leftover boundary terms~\footnote{In a nutshell, this means that the kinetic terms in the Lagrangian take the form $f(\psi)\dow_t\psi_a$ instead of $\psi_a\dow_tf(\psi)$. The terms of the latter kind are typically only invariant under $a$-type global symmetries up to boundary contributions; see e.g. \cref{eq:U1symmetry}.}. With this choice, $\Omega(\uppsi,t)=\Omega(\uppsi,\ul\sfs_r(t))$ can be interpreted as the grand canonical free energy distribution of states.

The KMS symmetry implies a relation between the probabilities of the original and time-reversed processes. Let us denote $\bbP_\Theta(\eta_\Theta\psi|\eta_\Theta\uppsi_\f,t_\i)$ as the conditional probability for the system to start from $\psi(t_\i)=\eta_\Theta\uppsi_\f$ and traversing the time-reversed path $\Theta\psi(t)=\eta_\Theta\psi(\Theta t)$, in the presence of time-reversed background sources $\eta_\Theta\ul\sfs_{r,a}(\Theta t)$. Relegating details to \cref{app:probs}, we find
\begin{align}
    \frac{\bbP_\Theta(\eta_\Theta\psi|\eta_\Theta\uppsi_\f,t_\i)}{\bbP(\psi|\uppsi_\i,t_\i)}
    = \e^{\beta_0\Delta\Omega -\beta_0W_\psi}~,
    \label{eq:microscopic-reversibility}
\end{align}
where $\Delta\Omega = \Omega(\uppsi_\f,t_\f)-\Omega(\uppsi_\i,t_\i)$ is the free energy differential between the end states, and 
\begin{align}\label{eq:work-done-along-path}
    \exp(-\beta_0W_\psi) 
    &= \left\langle \exp\!\lb -\beta_0\int\df t\,\df^dx\,\cW \rb\right\rangle_\psi~,
\end{align}
denotes the dissipative work done on the system during forward path $\psi(t)$, averaged over the thermal noise, i.e.
\begin{align}
    \cW
    &= - \cO\dow_t\sfs_r + \ell r_\fuel\Delta E + \ell r_\env\kB \Delta T \nn\\
    &= -\Big( \cO\dow_t\sfs_r + \ell r_\fuel\dow_t\Phi^\fuel_r 
    + \ell r_\env\dow_t\Phi^\env_a \Big)
    + \ell r_\env \kB T_0~.
\end{align}
defined as the total energy supplied by the sources, fuel, and environment in \cref{eq:energy-conservation}, minus the heat lost to the environment in \cref{eq:second-law}.
\Cref{eq:microscopic-reversibility} is the field-theoretic realisation of \emph{microscopic reversibility} from non-equilibrium statistical mechanics~\cite{Tolman:1924zz, tolman1979principles}. It states that, in the presence of activity, the likelihood of a process vs. its time-reverse is no longer governed by just the free energy differential $\Delta\Omega$ between the end states, but also the dissipative work $W_\psi$ performed on the system~\cite{Crooks1998,Basu2008}.

An equivalent statement for the transition probabilities can be obtained by summing over all paths, i.e.
\begin{align}\label{eq:detailed-imbalance}
    \frac{\bbP_\Theta(\eta_\Theta\uppsi_\i,t_\f|\eta_\Theta\uppsi_\f,t_\i)}{\bbP(\uppsi_\f,t_\f|\uppsi_\i,t_\i)}
    = \e^{\beta_0\Delta\Omega -\beta_0W}~,
\end{align}
where $W$ is now averaged over all paths, defined as 
\begin{align}\label{eq:work-done}
    \exp(-\beta_0W) 
    = \frac{\dsp\int_{\uppsi_\i}^{\uppsi_\f}\cD\psi \exp(-\beta_0W_\psi)\,\bbP(\psi|\uppsi_\f,t_\i)}
    {\dsp\int_{\uppsi_\i}^{\uppsi_\f}\cD\psi\,\bbP(\psi|\uppsi_\f,t_\i)}~.
\end{align}
\Cref{eq:detailed-imbalance} is the generalisation of the principle of \emph{detailed balance} in the presence of external work $W$. When all the background fields are time-independent and the fuel source is turned off, i.e. $\dow_t\sfs_r,\Delta E=0$, and the system temperature has equilibrated with the environment temperature, i.e. $T=T_\env$, we recover the original statement of detailed balance~\cite{de2013non, chandler1987introduction}.

A consequence of detailed balance is that, in the absence of external driving or activity, a transition is more likely to occur than its time-reversal if it decreases the free energy. Since this statement applies for arbitrary states, a passive system left to its own devices will ultimately settle into the state with least free energy, known as the principle of \emph{thermodynamic stability}. However, comparing free energies is not sufficient to determine the preferred state of an active system, and one must also account for the heat lost to the environment. In principle, one may use \cref{eq:detailed-imbalance} to determine the strength of activity required to induce an active phase transition to a state with higher free energy that is forbidden in passive systems. We leave such explorations for future work.


\section{Active superfluids}
\label{sec:simplediffusion}

To draw a qualitative picture of our framework, let us consider a toy model featuring a conserved number density $n$, together with the energy density $\epsilon$. The energy balance and charge conservation equations can be summarised as
\begin{align}\label{eq:diffusion-eqns}
    \dow_t{\epsilon} +  \partial_i\epsilon^i  
    &= E_i j^i 
    \blue{\,-\, \ell r_\fuel\dow_t\Phi^\fuel - \ell  r_\env \dow_t\Phi^\env}~,\nn\\
    \dow_t n
    + \dow_i j^i
    &= 0~,
\end{align}
where $j^i$ denotes the number flux. We have denoted the active contributions arising from the fuel and environment fields in \blue{{blue}} for emphasis. We have also introduced the background U(1) gauge field sources $A_t,A_i$ coupled to $n,j^i$ that contribute to the energy balance equation via the associated electric field $E_{i} = \dow_iA_t - \dow_t A_i$. Charge conservation may be understood as a consequence of an underlying U(1) symmetry that acts on the background fields as $A_t\to A_t+\dow_t\Lambda,A_i\to A_i+\dow_i\Lambda$. We may also introduce background ``clock fields'' $n_t,n_i$ coupled to $\epsilon,\epsilon^i$, however the resultant equations are quite involved due to non-linearities and we have relegated a full treatment to \cref{app:NC}. Note that the energy balance equation in \cref{eq:diffusion-eqns} differs from our previous expression in \cref{eq:energy-conservation} by a redefinition $\epsilon\to \epsilon-A_t n$, $\epsilon^i\to \epsilon^i-A_t j^i$ to make the equations gauge invariant.

Furthermore, our toy model consists of a complex scalar field $\Psi$ charged under the U(1) symmetry as $\Psi \to \Psi\e^{-i\Lambda}$. This can be used to study the superfluid phase where $\Psi$ attains a nonzero expectation value $\langle\Psi\rangle = \Psi_0$ and the U(1) symmetry is spontaneously broken. The massless fluctuations of $\Psi$ around the ground state are parametrised as $\Psi = \Psi_0 \e^{i\phi}$, where $\phi$ is the superfluid Goldstone field. Its dynamics is governed by the Josephson equation that will be determined in our formalism by varying the SK effective action, given in \cref{eq:Josephson}.

\subsection{Fields and symmetries}
\label{sec:SKstructure}

The primary dynamical ingredients in a SK-EFT featuring energy balance and charge conservation are: temperature $T$, chemical potential $\mu$, and the partner stochastic noise fields $X^t_a$, $\varphi_a$. To accommodate the possibility of spontaneous symmetry breaking, we also introduce the complex scalar field $\Psi_r\equiv\Psi$ and its stochastic partner $\Psi_a$. The background U(1) gauge field is doubled in the SK framework to $A_{r,a\,t},A_{r,a\,i}$. We may identify $A_{rt},A_{ri}$ as the classical gauge field $A_t,A_i$ and shall use the two notations interchangeably. The same holds for the doubled clock fields $n_{r,a\,t},n_{r,a\,i}$.

The effective theory is invariant under doubled U(1) global symmetries as well as doubled time-translations discussed around \cref{eq:a-type-translations}. The action of $r$-type time-translations is given by time-diffeomorphisms on all the fields as usual; see \cref{app:SK}. The remaining symmetries act on various dynamical and background fields as
\begin{align}\label{eq:U1symmetry}
    X^t_a &\to X^t_a -\chi^t_{a}~, \nn\\
    \varphi_a 
    &\to \varphi_a -\Lambda_a - X_a^t \dow_t \Lambda_r~, \nn\\
    \Psi_{r,a}
    &\to \e^{-i\Lambda_r} \Psi_{r,a}~, \nn\\
    A_t
    &\to A_t+\dow_t\Lambda_r~, \nn\\
    A_i
    &\to A_i+\dow_i\Lambda_r~, \nn\\
    A_{at} 
    &\to A_{at} + \dow_t\Lambda_{a} + \chi^t_a \dow_t A_t - A_t \dow_t \chi^t_a~, \nn\\
    A_{ai} 
    &\to A_{ai} + \dow_i\Lambda_{a}
    + \chi^t_a \dow_t A_i - A_t \dow_i \chi^t_a~, \nn\\
    n_{at} 
    &\to n_{at} + \chi^t_a \dow_t n_t - n_t \dow_t \chi^t_a~, \nn\\
    n_{ai} 
    &\to n_{ai} + \chi^t_a \dow_t n_i - n_t \dow_i \chi^t_a~,
\end{align}
while $T$, $\mu$, $n_t$, $n_i$ are invariant. A detailed derivation can be found of these symmetry transformations can be found in \cref{app:passive-sk,app:superfluid}.
In the following, it will be useful to define the quantities 
\begin{align}\label{eq:invariant-defs-noncov}
    N_{at} &= n_t \dow_t X^t_a + X^t_a\dow_t n_{t} + n_{at}~, \nn\\
    N_{ai} &= n_t \dow_i X^t_a + X^t_a\dow_t n_i + n_{ai}~, \nn\\
    B_{at} &= \dow_t\varphi_a + A_t \dow_t X^t_a + X^t_a\dow_t A_{t} + A_{at}~, \nn\\
    B_{ai} &= \dow_i\varphi_a + A_t \dow_i X^t_a + X^t_a\dow_t A_i + A_{ai}~, 
\end{align}
which are invariant under \cref{eq:U1symmetry}.
We also define the gauge-covariant derivatives of the complex scalar fields as $\Df_t = \dow_i\pm iA_{rt}$, $\Df_i = \dow_i\pm iA_{ri}$, with positive sign for $\Psi$,$\Psi_a$ and negative sign for $\Psi^*$, $\Psi^*_a$.

\begin{table}[t]
    \begin{tabular}{|c|c|}
        \hline
        \textbf{T-even} & \textbf{T-odd} \\
        \hline
        $x^i$, $\dow_i$ & $t$, $\dow_t$ \\ 
        $n$, $\epsilon$, $\tau^{ij}$
        & $j^i$, $\epsilon^i$, $\pi_i$ \\
        $A_{t}$, $A_{at}$, $n_{t}$, $n_{at}$, $h_{ij}$, $h_{aij}$
        & $A_{i}$, $A_{ai}$, $n_{i}$, $n_{ai}$, $v^i$, $v^i_a$ \\
        & $\Phi^\env$, $\Phi^\env_a$, $\Phi^\fuel$, $\Phi^\fuel_a$ \\
        $T$, $\mu$, $X^i_a$ & $X^t_a$, $\varphi_a$, $u^i$ \\
        $\Re\Psi$, $\Re\Psi_a$ & $\Im\Psi$, $\Im\Psi_a$, $\phi$, $\phi_a$ \\
        $Q_{ij}$, $\cQ_{ij}$ & \\
        \hline
    \end{tabular}
    \caption{Time-reversal even and odd quantities used in this work. The $r$-type background fields are identified with the unlabelled ones. The director field $q_i$ and its noise partner $\cq_{ai}$ may be time-reversal even or odd based on the system under consideration.}
    \label{tab:T-reversal}
\end{table}

The KMS transformation of the background fields is given in \cref{eq:KMS}, whereas it acts as a time-reversal transformation on the physical dynamical fields, with the time-reversal eigenvalues summarised in \cref{tab:T-reversal}. Note that the time-reversal acts oppositely on the real and imaginary parts of $\Psi$, so $\Psi\KMSto\Psi^*$.
The $a$-type noise dynamical fields are taken to transform under KMS as
\begin{alignat}{2} \label{eq:KMStransformations}
    X^t_a &\KMSto - \hat X^t_a \equiv 
    -X^t_a - i\lb \beta  - \beta_0  \rb~, \nn\\
    \varphi_a &\KMSto 
    - \hat\varphi_a \equiv -\varphi_a 
    - i\Big(\beta (\mu-A_t)  - \beta_0 \mu_0 \Big) ~, \nn\\
    \Psi_a &\KMSto -\hat\Psi_a^\dagger \equiv \Psi_a^* + i \beta  \Df_t\Psi^* - \beta \mu \Psi^*~, \nn\\
    \Psi_a^\dagger &\KMSto -\hat\Psi_a \equiv -\Psi_a - i \beta \Df_t\Psi
    - \beta \mu \Psi~,
\end{alignat}
where $\beta = 1/(\kB T)$. Here $\beta_0 = 1/(\kB T_0)$ and $\mu_0$ are the inverse global temperature chemical potential.
Note that these transformations do not respect complex-conjugation, but do respect the SK-unitarity operation \eqref{eq:unitarity-dagger}. 
We can define the hatted-versions of the quantities in \cref{eq:invariant-defs-noncov} by replacing the $a$-type field with their hatted versions, such that $f_a \KMSto \eta_\Theta \hat f_a$.

Finally, when activity is present, the $a$-type time-translation symmetry requires that the fuel and environment background fields from \cref{sec:active-KMS} appear in specific combinations
\begin{gather}
    \Delta E = -\dow_t\Phi^\fuel_r, \qquad
    \kB T_\env = \dow_t\Phi^\env_r~,\nn\\
    \Pi_{a}^{\fuel,\env} = \ell\Phi_a^{\fuel,\env} 
    + \ell X_a^t \dow_t\Phi^{\fuel,\env}_r~,
    \label{eq:cF-def}
\end{gather}
with the hatted versions
\begin{align}
    \hat\Pi_a^\fuel = \Pi_a^\fuel - i\ell \beta \Delta E ~, \quad 
    \hat\Pi_a^\env = \Pi_a^\env - i\ell \beta \kB\Delta T~,
\end{align}
so that $\Pi^{\fuel,\env}_a\KMSto -\hat\Pi^{\fuel,\env}_a$.

\subsection{Schwinger-Keldysh effective action}

We are now ready to construct the SK effective action. 
As the simplest first step, we invoke a physical assumption that the active system under consideration only has control over how the heat is dumped into the environment and not on how it is drawn from the fuel source. This entails, e.g., that the rate $r_\fuel$ of fuel consumption does not depend on the thermodynamic variables $T$, $\mu$, or their derivatives. In terms of the SK formalism, this means that the ``fuel part'' of the Lagrangian is fixed to the simple form
\begin{align}\label{eq:fuel-L}
    \cL_\fuel = \blue{i\kB T\gamma_\fuel \Pi_a^\fuel\hat\Pi_a^\fuel}~,
\end{align}
which respects all the SK symmetries in \cref{sec:SKstructure} as well as the unitarity constraints in \cref{eq:SK-conds}. The coefficient $\gamma_\fuel$ appearing here is the same as we saw in \cref{eq:losscurrent}.
Noting the composition of $\Pi^\fuel_a$ in \cref{eq:cF-def}, we see that the fuel fields only talk to the fluid via $T$ and $X^t_a$.
This simple fuel Lagrangian is sufficient to obtain all the universal features of active hydrodynamics agnostic of the details of the fuel burning process. One may easily generalise \cref{eq:fuel-L} if one wishes to simultaneously describe the fuel sector. By contrast, we shall allow the environment fields to non-trivially couple with all the hydrodynamic degrees of freedom.

Let us start with the ``fluid part'' of the SK Lagrangian depending only on the dynamical fields $T$, $\mu$, and their noise partners. For example, we may write down
\begin{align}\label{eq:lagrangian-passive-diffusion}
    \cL_{\text f}
    &= - \epsilon N_{at} + n B_{at} 
    + i\kB T^2\kappa N_{ai} \hat N_{a}^i
    + i\kB T\sigma B_{ai}\hat B_{a}^i \nn\\
    &\quad
    \blue{+\,i\kB T\gamma_\env \Pi_a^\env\hat\Pi_a^\env}~.
\end{align}
The terms in the first line are comprised of the SK model for energy diffusion~\cite{Chen-Lin:2018kfl} and charge diffusion~\cite{Jain:2020zhu}.
In addition to the densities $n$ and $\epsilon$, we have introduced the thermal conductivity $\kappa$, charge conductivity $\sigma$, and heat relaxation coefficient $\gamma_\env$. All the coefficients are functions of the thermodynamic variables $T$ and $\mu$. Generically, we may also introduce a thermo-electric conductivity $\sigma_\times$ coupling the energy and charge sectors that has been considered in \cref{app:diffusion}.

Next, we consider the ``superfluid part'' that contains the order parameter $\Psi$ and its noise partner in addition to the hydrodynamic fields. Consider
\begin{align}\label{eq:Psi-Langrangian}
    \cL_\Psi
    &= f_\Psi\Df^i\Psi \Big( \Df_t \Psi^* N_{ai}
    + i \Psi^*B_{ai} 
    - \Df_i\Psi_a^*
    \Big)
    - \frac{\dow V}{\dow\Psi^*}\Psi^*_a \nn\\
    &\quad
    + i\kB T \sigma_\Psi 
    \Big(\Psi_a^* \blue{\,-\,\frac{i\mu\lambda_{\phi\env}}{\kB T_\env}\Psi^*\Pi_a^\env}\Big)
    \Big(\hat \Psi_a \blue{\,+\,\frac{i\mu\lambda_{\phi\env}}{\kB T_\env}\Psi\hat\Pi_a^\env}\Big)
    \nn\\
    &\quad
    \blue{\,+\, i\frac{T}{T_\env} f_\Psi\lambda_{n\env} \Psi^*i\Df^i\Psi
    \Big(B_{ai}\hat\Pi_a^\env -  \hat B_{ai}\Pi_a^\env\Big) } \nn\\
    &\quad
    \blue{\,+\, i\frac{T}{T_\env}
    a_\env\lb \Psi^*\hat\Psi_a\Pi_a^\env - \Psi\Psi^*_a \hat\Pi_a^\env \rb}
    - (\ldots)^\dagger~.
\end{align}
The last term denotes the SK-unitarity-conjugate defined using \cref{eq:unitarity-dagger}. We have also introduced the superfluid density parameter $f_\Psi$, diffusion parameter $\sigma_\Psi$, and a few active parameters $\lambda_{\phi\env}$, $\lambda_{n\env}$, and $a_\env$ whose significance will be clear momentarily. $\Psi$ is subjected to a potential $V$, which is a function of $|\Psi|^2=\Psi^*\Psi$ and may take the representative form
\begin{align}\label{eq:potential}
    V = a |\Psi|^2 + \frac12 a_4|\Psi|^4~,
\end{align}
where $a$, $a_4$ are phenomenological parameters that determine the shape of the potential. All these coefficients may be functions of $T$ and $\mu$. 
A more exhaustive analysis of the allowed terms in the superfluid Lagrangian appears in \cref{app:superfluid}, while their description appears in \cref{tab:superfluids}.

\begin{table}[t]
    \centering
    \begin{tabular}{c|l}
        \hline\hline 
        \multicolumn{2}{c}{\textbf{Passive Transport}} \\
        \hline 
        $p$ & Thermodynamic pressure \\
        $\epsilon$ & Energy density \\
        $n$ & Particle/charge density \\
        $\kappa$ & Thermal conductivity \\
        $\sigma$ & Charge/particle conductivity \\
        \textcolor{gray}{$\sigma_\times$} 
        & \textcolor{gray}{Cross-conductivity between $\epsilon^i$ and $j^i$} \\
        \hline 
        $a$, $a_4$ & Parameters of superfluid potential $V$ \\
        $f_s \equiv 2\Psi_0^2f_\Psi$ & Superfluid density \\
        $\sigma_\phi \equiv 2\Psi_0^2\sigma_\Psi$ & Superfluid conductivity \\
        \textcolor{gray}{$\lambda_{n\phi}$} 
        & \textcolor{gray}{Cross-conductivity between $j^i$ and $\dow_t\phi$} \\
        \textcolor{gray}{$\lambda_{\epsilon\phi}$} 
        & \textcolor{gray}{Cross-conductivity between $\epsilon^i$ and $\dow_t\phi$} \\
        \hline\hline
        \multicolumn{2}{c}{\textbf{Active Transport}} \\
        \hline 
        $\gamma_\fuel$ & Controls rate of fuel consumption due to $\Delta E$ \\
        $\gamma_{\env}$ & Controls rate of entropy loss due to $\Delta T$ \\
        \hline
        $a_\env$ & Alters the equilibrium state $\Psi_0$ \\
        $\lambda_{\phi\env}$ & Screens $\mu$ in the Josephson equation for $\dow_t\phi$ \\
        $\lambda_{n\env}$ & Screens $f_s$ in the charge flux \\
        \textcolor{gray}{$\lambda_{\epsilon\env}$} 
        & \textcolor{gray}{Screens $f_s$ in the energy flux} \\
        \hline
    \end{tabular}
    \caption{Thermodynamic and transport coefficients of a simple (active) superfluid without momentum conservation. The gray coefficients have not been included in the main text, but have been detailed in the \cref{app:diffusion,app:superfluid}.}
    \label{tab:superfluids}
\end{table}

One may check that the theory is invariant under the global symmetries in \cref{eq:U1symmetry}.
The first two conditions in \cref{eq:SK-conds} are satisfied by construction, while the third one requires
\begin{gather}
    \kappa\geq 0~, \quad
    \sigma \geq 0~, \quad 
    \sigma_\Psi \geq 0~, \quad
    \gamma_{\env,\fuel} \geq 0~,
    \label{eq:ineq-1}
\end{gather}
which guarantees the positivity of $\Im\cL$. The KMS symmetry requires that $n$, $\epsilon$, and $f_\Psi$ are derived from the free energy density $\cF$ via the thermodynamic relations
\begin{align}\label{eq:thermodynamics-Psi}
    \df\cF &= - s\,\df T - n\,\df \mu 
    + f_\Psi\df(\Df^i\Psi^*\Df_i\Psi)
    + \frac{\dow V}{\dow|\Psi|^2}\df|\Psi|^2~, \nn\\
    \epsilon &= Ts+\mu n+\cF~, 
\end{align}
where $s$ is the entropy density. Noting these relations, one may verify that the SK Lagrangian is KMS-invariant up to a total derivative term $-i\dow_t(\beta\cF)$ that drops out from the effective action. Comparing with our discussion in \cref{sec:DB}, we have $\beta_0\Omega = \int\df^dx\,\beta\cF$.
Assuming $f_\Psi$ to be independent of $\Psi$, the free energy density takes the familiar Landau-Ginzburg form
\begin{align}
    \cF = -p + f_\Psi\Df^i\Psi^*\Df_i\Psi + V~,   
\end{align}
where $p$ is the thermodynamic pressure of the fluid. 

\subsection{Spontaneous symmetry breaking}
\label{sec:higgs-sf}

Extremising the SK effective action with respect to $\Psi_a$ yields the equation of motion for the order parameter
\begin{align}\label{eq:Psi-evolve}
    \Df_t\Psi
    &= 
    \frac{1}{\sigma_\Psi}\Df_i(f_\Psi\Df^i\Psi)
    - \frac{1}{\sigma_\Psi} \lb \frac{\dow V}{\dow|\Psi|^2}
    + \blue{\aleph a_\env} \rb \Psi \nn\\
    &\qquad
    + i\mu\lb 1 + \blue{\aleph\lambda_{\phi\env}} \rb\Psi~,
\end{align}
where $\aleph=\ell\Delta T/T_\env$ denotes the strength of activity.
In general, the equation of motion also contains an auxiliary noise part, which is classically set to zero by
extremising with respect to $T$, $\mu$, and $\Psi$, provided that the $a$-type background noise fields are turned off. This part is important in stochastic hydrodynamics when computing correlation functions of hydrodynamic operators. In the SK framework, however, correlation functions are computed directly using the effective action, so we do not need to concern ourselves with the explicit form of such stochastic corrections.

The qualitative behaviour of $\Psi$ depends on the form of the potential $V$ in \cref{eq:potential} and the coefficient $a_\env$. If the combination $a+\aleph a_\env>0$, the order parameter $\Psi$ is gapped with the gap-scale $(a+\blue{\aleph a_\env})/\sigma_\Psi$
and we are in the fluid phase. On the other hand, if $a+\aleph a_\env<0$, the potential is minimised at 
\begin{align}\label{eq:Psi-ground}
    \langle |\Psi|^2\rangle = \Psi_0^2 = - \frac{a +\blue{\aleph a_\env}}{a_4}~,
\end{align}
thereby spontaneously breaking the U(1) symmetry and leading us to the superfluid phase.

\begin{figure}
    \centering
    \tikz[thick]{
        \node at (0,0) {\includegraphics[width=0.48\linewidth]{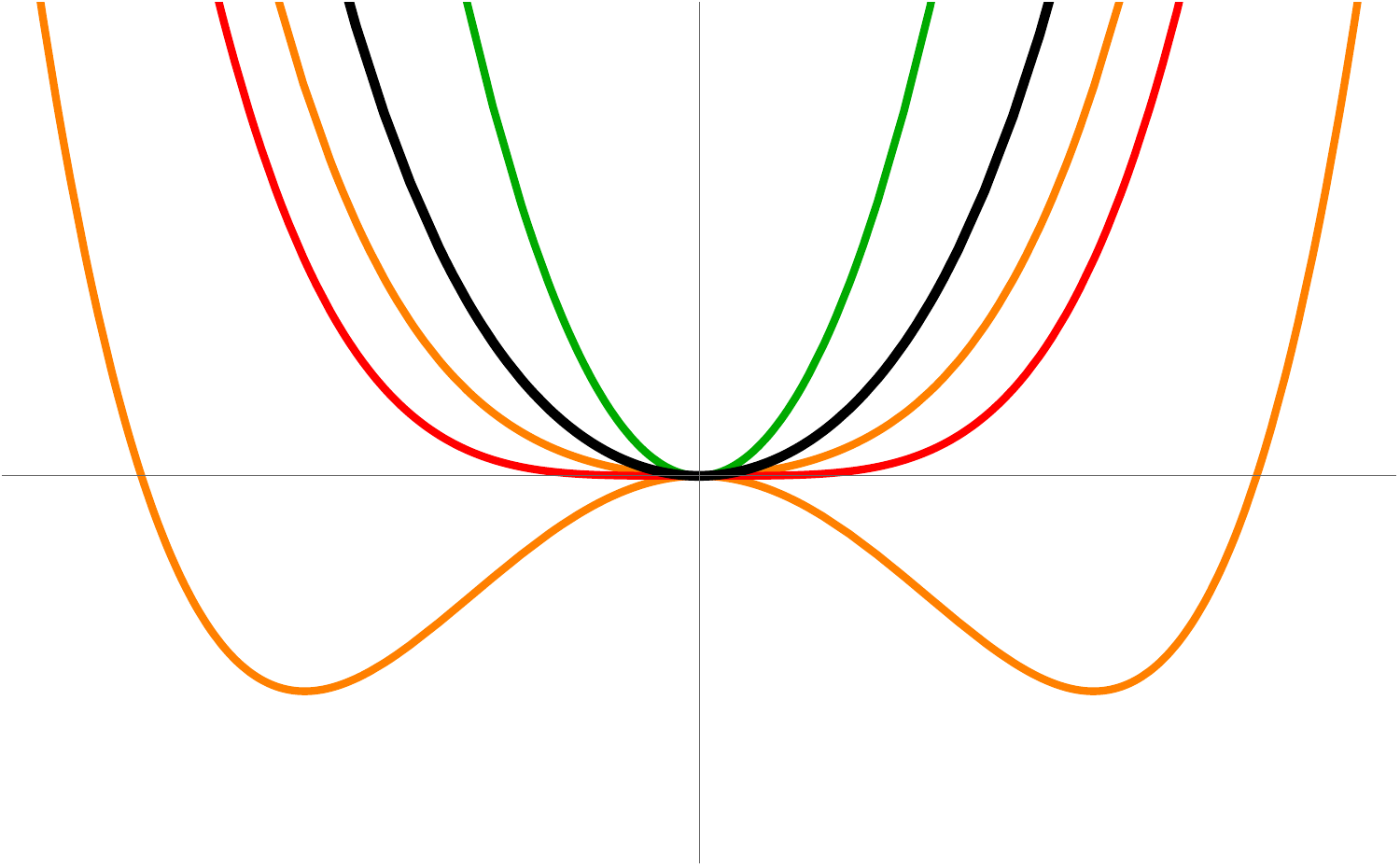}};
        \node at (0.45,1) [fill=white] {$\mathsmaller{a_\env>0}$};
        \node at (1.5,-.5) [fill=white] {$\mathsmaller{a_\env<0}$};
        \node at (0,-1) [fill=white] {$\boxed{a>0}$};
    }%
    \tikz[thick]{
        \node at (0,0) {\includegraphics[width=0.48\linewidth]{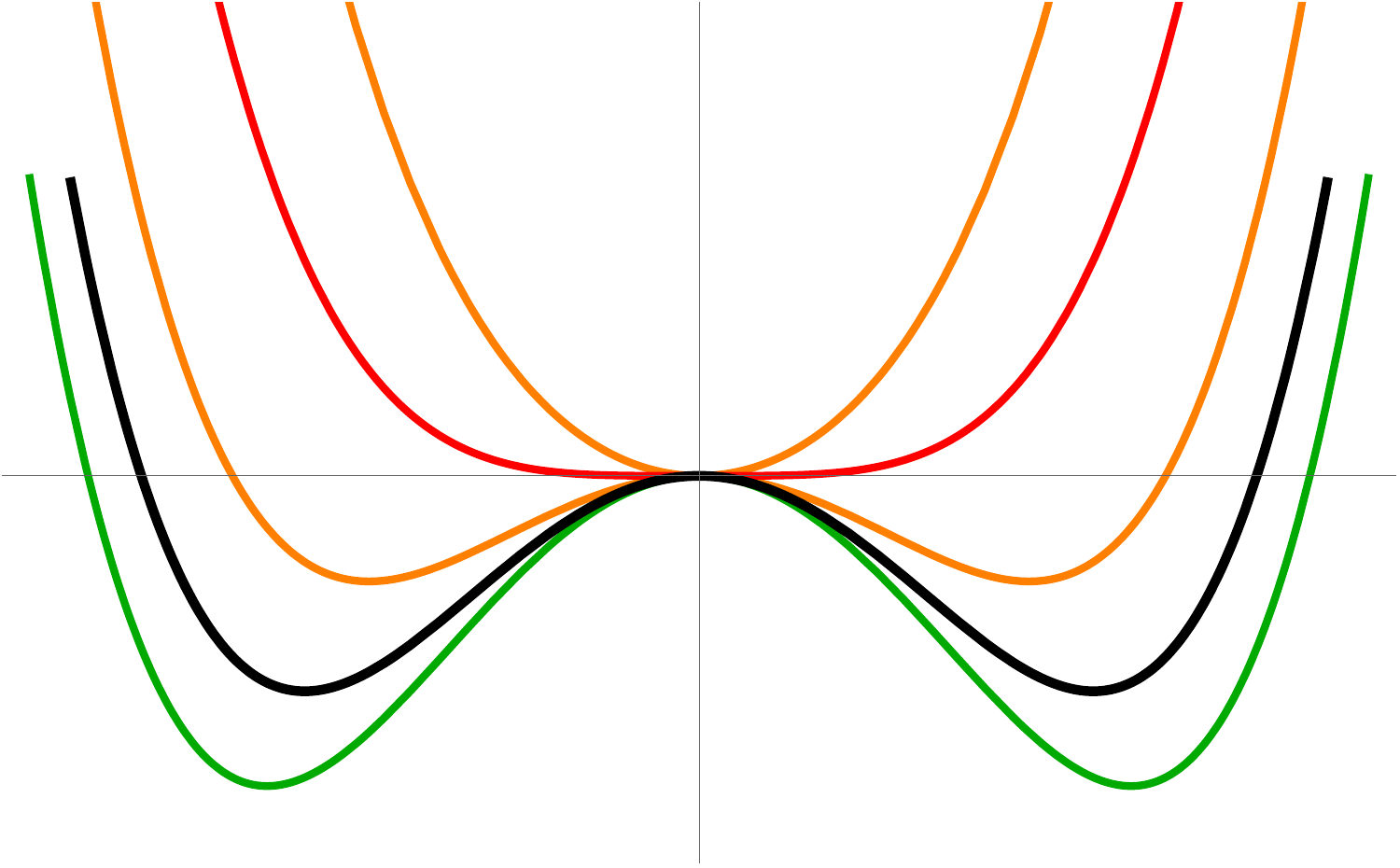}};
        \node at (1,1) [fill=white] {$\mathsmaller{a_\env>0}$};
        \node at (1.7,-1) [fill=white] {$\mathsmaller{a_\env<0}$};
        \node at (0,-1) [fill=white] {$\boxed{a<0}$};
    }
    \caption{The effective potential of $\Psi$ in an active superfluid. For $a>0$, the system exists in the spontaneously-unbroken or fluid phase in the absence of activity (black), while for $a<0$ it exists in the spontaneously-broken or fluid phase. For $a/a_\env>0$ (green), increasing activity retains the system in the same phase and drives it further from criticality (red), while for $a/a_\env<0$ (orange), increasing activity eventually flips the potential and induces a phase transition.}
    \label{fig:superfluid-active-transition}
\end{figure}

Depending on the signs of the parameters $a$ and $a_\env$, the activity may induce or destroy the superfluidity order. In particular, if $a>0$ and $a_\env<0$, the system exists in the fluid phase in the famine state and activity induces superfluidity beyond the critical scale
\begin{equation}\label{eq:criticalscale}
    T_0 = T_\env\lb 1 -\frac{a}{\ell a_\env} \rb~, \quad 
    \Delta E = \kB T_\env\sqrt{-\frac{\gamma_\env}{\gamma_\fuel}\frac{a}{\ell a_\env}}~.
\end{equation}
In contrast, if $a<0$ and $a_\env >0$, the system already exists in the superfluid phase in the famine state and activity beyond the critical scale \eqref{eq:criticalscale} destroys superfluidity. See \cref{fig:superfluid-active-transition}. In the remaining two cases, activity does not alter the phase of the system. We should emphasise that our framework of active hydrodynamics is only really reliable for ``small activity'', controlled by $\ell$, so that we do not stray too far from thermal equilibrium. In this sense, strictly speaking, our model can describe active phase transitions when the famine state is already close to criticality, i.e the dimensionless ratio $|a/a_\env|$ is sufficiently small.

The U(1) phase of $\Psi$ in the superfluid phase becomes a massless Goldstone mode. Parametrising the phase fluctuations as $\Psi = \Psi_0\e^{i\phi}$ in \cref{eq:Psi-evolve}, we can obtain the Josephson equation for $\phi$, i.e.
\begin{align}\label{eq:Josephson}
    \xi_t
    &= \lambda_\phi\mu
    + \frac{1}{\sigma_\phi} \dow_i\big(f_s\xi^i\big)~,
\end{align}
where $\xi_t = \dow_t\phi+A_t$, $\xi_i = \dow_i\phi+A_i$ are the superfluid potential and velocity respectively, and we have identified the superfluid density $f_s$ and diffusion parameter $\sigma_\phi$ as
\begin{align}\label{eq:sf-params}
    f_s = 2\Psi_0^2f_\Psi ~, \qquad 
    \sigma_\phi = 2\Psi_0^2\sigma_\Psi~.
\end{align}
The coefficient $\lambda_\phi$ in front of the chemical potential term in \cref{eq:Josephson} is 1 for passive superfluids, but the presence of activity improves it to
\begin{align}
    \lambda_\phi = 1 + \blue{\aleph\lambda_{\phi\env}}~,
\end{align}
controlled by the active coefficient $\lambda_{\phi\env}$. In other words, the superfluid Goldstone experiences a screened chemical potential $\lambda_\phi\mu$ in the presence of activity instead of the true thermodynamic chemical potential $\mu$. This behaviour is reminiscent of the pseudo-spontaneous symmetry breaking pattern found in superfluids in the presence of approximate U(1) symmetry~\cite{Armas:2021vku}.

The Lagrangian \eqref{eq:Psi-Langrangian} is useful for describing the dynamics of the full U(1) order parameter $\Psi$, together with the transition between the fluid and active superfluid phases. However, once we have settled into the superfluid phase sufficiently far from the phase transition point, we may integrate out $|\Psi|$ and its noise partner to arrive at the low-energy description exclusively for the massless Goldstone $\phi$ and the hydrodynamic degrees of freedom. In practice, this integrating-out procedure is quite technical, so a better strategy is to directly build the SK effective theory for $\phi$ from scratch. Since our construction is rooted in symmetries, the final effective description would be the same as that obtained after the integrating-out procedure, albeit up to renormalisation of coefficients and higher-derivative corrections. To this end, we introduce the noise field $\phi_a$ partner to the superfluid Goldstone, with the KMS transformation
\begin{align}
    \phi_a \KMSto -\hat\phi_a
    \equiv - \phi_a - i \beta (\xi_t - \mu)~.
\end{align}
In terms of this, we have
\begin{align}\label{eq:SK-Lagrangian-sf}
    \cL_{\text{sf}}
    &= f_s \xi^i \Big( \xi_t N_{ai} - B_{ai} - \dow_i\phi_a \Big)  \nn\\
    &\quad
    + i\kB T \sigma_\phi 
    \lb \phi_a + \blue{\frac{\mu\lambda_{\phi\env}}{\kB T_\env}\Pi^\env_a} \rb
    \lb \hat\phi_a + \blue{\frac{\mu\lambda_{\phi\env}}{\kB T_\env}\hat\Pi^\env_a} \rb \nn\\
    &\quad 
    \blue{-\,i\kB T f_s  \lambda_{n\env} \xi^i
    \lb B_{ai}\hat\Pi_a^\env - \hat B_{ai}\Pi_a^\env \rb}~,
\end{align}
that together with \cref{eq:lagrangian-passive-diffusion} describes the superfluid phase. All coefficients appearing here may be functions of $T$, $\mu$, and $\xi^2=\xi_i\xi^i$. One may check that this yields the correct Josephson equation in \cref{eq:Josephson}.
The low-energy thermodynamic relations are given as
\begin{align}\label{eq:thermodynamics}
    \df\cF &= - s\,\df T - n\,\df \mu + \half f_s\df\xi^2~, \nn\\
    \epsilon &= Ts+\mu n+\cF~,
\end{align}
which ensure the KMS symmetry. Assuming constant $\Psi_0$, it is straightforward to obtain \cref{eq:SK-Lagrangian-sf} from \cref{eq:Psi-Langrangian} by setting $\Psi = \Psi_0\e^{i\phi}$ and $\Psi_a = i\Psi_0\phi_a {\rm e}^{i\phi}$. More generally, $\Psi_0$ may be a function of $T$, $\mu$, and $\xi^2$, in which case a more careful derivation is warranted as given in \cref{app:superfluid}.

\subsection{Constitutive relations and the second law}
\label{eq:consti-diff}

The energy balance and charge conservation equations in \cref{eq:diffusion-eqns} are obtained by extremising the SK effective action with respect to $X_a^t$ and $\varphi_a$. The corresponding constitutive relations are given by
\begin{align}\label{eq:consti}
    \epsilon^i 
    &= - f_s\xi_t \xi^i
    - \kappa\,\dow^i T~, \nn\\
    j^i
    &= - \lambda_n f_s\xi^i
    - \sigma\lb T\dow^i\frac{\mu}{T} - E^i\rb~,
\end{align}
where we have identified 
\begin{gather}
    \lambda_n = 1 \blue{\,+\,\aleph\lambda_{n\env}}~.
    \label{eq:renormalised-dissipation}
\end{gather}
In the presence of activity, the superfluid density $f_s$ as observed by $\phi$ in \cref{eq:Josephson} is different from the superfluid density $\lambda_n f_s$ as appearing in $j^i$ in \cref{eq:consti}. The source of activity in \cref{eq:Josephson,eq:consti} is attributed to the heat exchange rates
\begin{align}\label{eq:sf-rates}
    r_\env 
    &= \ell \gamma_\env\kB\Delta T
    - f_s  \lambda_{n\env} \xi^i \lb T \dow_i\frac{\mu}{T} - E_i \rb 
    - \mu\lambda_{\phi\env} \dow_i(f_s\xi^i)~, \nn\\
    r_\fuel 
    &= \ell \gamma_\fuel \Delta E~,
\end{align}
obtained by varying the SK action with respect to $\Phi^\fuel_a$ and $\Phi^\env_a$ respectively.
We have recovered the rates in \cref{eq:losscurrent} for our model, together with the higher-derivative corrections.
Due to the simplifications in our toy model, we do not see any active corrections to the energy flux $\epsilon^i$, but in general they also receive corrections similar to those in the charge flux $j^i$. Furthermore, in writing \cref{eq:consti,eq:sf-rates}, we have already integrated out the magnitude $|\Psi|$ of the order parameter. This is why the active parameter $a_\env$ from \cref{eq:Psi-evolve} does not appear in \cref{eq:sf-rates} explicitly and is hidden within the renormalisations of other active coefficients. A more comprehensive discussion of these considerations is presented in \cref{app:superfluid}.

The SK structure outlined above conspires to give rise to the active modification to the second law of thermodynamics given in \cref{eq:second-law}. Using \cref{eq:thermodynamics}, we find that
\begin{align}\label{eq:second-law-diff}
    &\dow_t s + \dow_i s^i \blue{\,+\, \ell \kB r_\env} \nn\\
    &= \kappa\lb\frac1T\dow_iT \rb^2
    + \frac{\sigma}{T} \lb T\dow_i\frac{\mu}{T} - E_i\rb^2
    + \frac{1}{T\sigma_\phi}\big(\dow_i(f_s\xi^i)\big)^2 \nn\\
    &\qquad
     \blue{+\, \frac{\gamma_\env}{T} \ell^2 \kB^2\Delta T^2
    + \frac{\gamma_\fuel}{T} \ell^2 \Delta E^2}
    \geq 0~,
\end{align}
where the heat flux is given as
\begin{align}
    Ts^i
    &= \epsilon^i - \mu j^i + f_s\xi^i(\xi_t-\mu) \nn\\
    &= - \kappa\,\dow^i T
    + \mu\sigma\lb T\dow^i\frac{\mu}{T} - E^i\rb
    \blue{\,+\, \aleph\mu\lambda_{n\env} f_s\xi^i}~.
\end{align}
The positivity of the right-hand side of \cref{eq:second-law-diff} is guaranteed by the positivity constraints on the dissipative coefficients in \cref{eq:ineq-1}.

\subsection{Linearised mode spectrum and violation of fluctuation-dissipation theorem}
\label{eq:linear-diff}

Let us use the SK-EFT for active superfluids developed in the previous subsections to compute the linearised mode spectrum and symmetric and retarded two-point correlation functions. In particular, we will compute for the first time energy density correlation functions within the context of hydrodynamics of active matter. We will also see how activity gives rise to systematic violations of FDT in \cref{eq:FDT}. Consider the steady states of the hydrodynamic model
\begin{gather}
    T = T_0~, \qquad \mu = \mu_0~,  \qquad
    \phi = \lambda_\phi\mu_0 t~, \nn\\
    X^t_a = \varphi_a = \phi_a = 0~,
\end{gather}
with $T_0$ given in \cref{eq:eqb-temp}. In particular, note that $\phi\neq \mu_0 t$ as it would be in the thermal equilibrium state in the absence of activity. For illustrative purposes, let us turn off the cross-susceptibility, $\dow\epsilon/\dow\mu = 0$, and assume that all the transport coefficients are constants. We also focus on states with a particular value of the chemical potential satisfying $\lambda_\phi\mu_0=0$. Under these assumptions, the energy and charge fluctuations decouple from each other and give rise to a relaxed energy diffusion mode and a superfluid sound mode
\begin{align}\label{eq:mode-spectrun}
    \omega
    &= - i\Gamma_\epsilon -iD_\epsilon k^2 + \ldots~, \nn\\
    \omega 
    &= \pm v_s k - \frac{i}{2}(D_n+D_\phi)k^2 + \ldots~,
\end{align}
where we have identified the parameters
\begin{gather}
    D_\epsilon
    = \frac{\kappa}{c_v}~, \qquad 
    D_n 
    = \frac{\sigma}{\chi}~, \nn\\
    v_s = \sqrt{\frac{\lambda_n f_s}{\chi}}, \qquad 
    D_\phi = \frac{f_s}{\sigma_\phi}~,
\label{eq:D-defn}
\end{gather}
where $\chi = \dow n/\dow\mu$ denotes the charge susceptibility and $c_v = \dow\epsilon/\dow T$ the heat capacity. The energy relaxation rate $\Gamma_\epsilon$ has been given in \cref{eq:energy-relaxation-rate}. 

The retarded and symmetric correlation functions can be obtained by varying the SK generating functional with respect to the doubled background sources; see \cref{eq:corr-from-gf}. For example, for the charge density correlators we find
\begin{subequations}
    \begin{align} \label{eq:retardedcorrelator}
    G^\rmR_{nn}
    &= \frac{\lambda_n f_s k^2
    - \lb i\omega - D_\phi k^2 \rb \sigma k^2}{
    \lb i\omega - D_n k^2 \rb\lb i\omega - D_\phi k^2\rb + v_s^2 k^2
    }~,  \\
    G^\rmS_{nn}
    &= 2 \kB T_0\frac{
    |i\omega - D_\phi k^2|^2 \sigma\, k^2
    + \lambda_n^2 f_s^2/\sigma_\phi\, k^4
    }
    {\left|\lb i\omega - D_n k^2 \rb\lb i\omega - D_\phi k^2\rb + v_s^2 k^2\right|^2}~, \nn\\
    &= \frac{2 \kB T_0}{\omega}{\rm Im}\,G^\rmR_{nn}
    \nn\\
    &~~ 
    + \frac{2 \kB T_0 \chi D_\phi\lambda_{n\env} v_s^2 k^4}
    {\left|\lb i\omega - D_n k^2 \rb\lb i\omega - D_\phi k^2\rb + v_s^2 k^2\right|^2}
    \aleph~.  \label{eq:symmetriccorrelator}
\end{align}
\label{eq:FDT-violation}%
\end{subequations}
Note also that the symmetric correlator is strictly non-negative because of the constraint on $\sigma$ in \cref{eq:ineq-1}.
However, these correlators violate FDT when activity is present, i.e. $\aleph\neq 0$. For the energy density correlators we find
\begin{subequations}
    \begin{align} \label{eq:retardedcorrelator123}
    G^\rmR_{\epsilon \epsilon}
    &=  \frac{-c_v  T_0  \left(\Gamma_{\epsilon} + 
 D_{\epsilon}  k^2\right)}{i\omega -\Gamma_{\epsilon} - D_{\epsilon}  k^2 }~,  \\
    G^\rmS_{\epsilon \epsilon}
    &= \frac{2  \kB  c_v T_0^2 \left(\Gamma_{\epsilon}  +D_{\epsilon} 
  k^2 \right)}{\left|i\omega -\Gamma_{\epsilon} - D_{\epsilon}  k^2 \right|^2}  = \frac{2 \kB T_0}{\omega}{\rm Im}\,G^\rmR_{\epsilon \epsilon}  ~~ , \label{eq:symmetriccorrelator123} 
\end{align}
\label{eq:FDT-violation123}%
\end{subequations}
which obeys FDT as the active coefficients decouple from energy fluctuations. 
We can similarly work out the other correlators involving $j^i$ or $\epsilon^i$.

\subsection{Violation of fluctuation-dissipation theorem from fluctuating hydrodynamics} \label{sec:fluctuatinghydro}
To close this section, we show that the FDT violation \cref{eq:FDT-violation} is not dependent on the usage of doubled sources as is done with SK-EFT, but can also be obtained with an approach based on nonequilibrium fluctuating hydrodynamics \cite{sengersortiz}. To obtain the symmetric correlator, we generalise the constitutive equations of \cref{eq:consti} to
 \begin{align}\label{eq:consti1}
    \epsilon^i 
    &= - f_s\xi_t \xi^i
    - \kappa\,\dow^i T  + \upsilon^i_{\epsilon}~, \nn\\
    j^i
    &= - \lambda_n f_s\xi^i
    - \sigma\lb T\dow^i\frac{\mu}{T} - E^i\rb  + \upsilon^i_n~,
\end{align}
and take for \eqref{eq:sf-rates} instead
\begin{align}\label{eq:sf-rates123}
    r_\env 
    &= \ell \gamma_\env\kB\Delta T
    - f_s  \lambda_{n\env} \xi^i \lb T \dow_i\frac{\mu}{T} - E_i \rb + \upsilon_{\env}~, \nn\\
    r_\fuel 
    &= \ell \gamma_\fuel \Delta E  + \upsilon_{\fuel} ~,
\end{align}
and also generalise the Josephson equation of \cref{eq:Josephson} to
\begin{align}\label{eq:Josephson123}
    \xi_t
    &= \lambda_\phi\mu
    + \frac{1}{\sigma_\phi} \dow_i\big(f_s\xi^i\big)  +\upsilon_\phi~,
\end{align}
where $\upsilon_{a}$ with $a \in \{\epsilon, n, \env , \fuel  ,  \phi  \}$ are stochastic fluxes whose variances are given by
\begin{align}
    \langle \upsilon_a (x , t  ) \upsilon_a (x '  , t' )   \rangle     =  \Delta_a \delta (x - x ' )   \delta (t - t  ' )  ~~ . 
\end{align}
Let us again focus on charge conservation. To obtain the noise variances $\Delta_{n , \phi }$, we go to the famine state by taking $\Delta E =0 $. Then, linearizing and again assuming $\epsilon$ and $n$ are decoupled yields
\begin{align}
    &    G^\rmS_{nn} \Big|_{\Delta E =0 }  \nn 
    =     \Delta_{n}\left| \frac{D_{\phi} k^2-i \omega }{k^2 v_s^2+\left(D_n k^2-i \omega \right) \left(D_{\phi} k^2-i \omega \right)}\right| ^2 \\   &
     +     \Delta_{\phi} \left| \frac{f_s k^2}{k^2 v_s^2+\left(D_n k^2-i \omega \right) \left(D_{\phi} k^2-i \omega \right)}\right|^2 ~~ . 
\end{align}
When $\Delta E =0 $, FDT should be upheld. Therefore, it follows from comparison with \cref{eq:retardedcorrelator} that  
\begin{subequations}  \label{eq:noisevariance}
    \begin{align}
    \Delta_n   &  =  2 \sigma  \kB T_0   ~~ , \\ 
   \Delta_{\phi}  &  =   \frac{2  \kB T_0  }{\sigma_{\phi} } ~~ . 
\end{align}
\end{subequations}
Having tuned the noise in the famine state, we assume that upon departing from the famine state by turning on weak activity, the noise is not affected by this \cite{Basu2008,PhysRevE.80.011917}. Repeating the computation of the symmetric correlator with \cref{eq:noisevariance} and the activity turned on, we then obtain \cref{eq:symmetriccorrelator}. For energy we find
\begin{align}
\begin{split}
           &    G^\rmS_{\epsilon \epsilon } \Big|_{\Delta E =0 }  \nn 
    =     \frac{   \Delta_{\epsilon} k^2  + 
    \Delta_{\env}  \ell^2  (\kB  T_{\env})^2   }{ \omega ^2+\left(\Gamma_{\epsilon} +D_{\epsilon} 
  k^2  \right)^2 }   ~~ , 
   \end{split}
\end{align}
from which it follows that to match with \eqref{eq:symmetriccorrelator123} one must take
\begin{subequations}
\begin{align}
      \Delta_{\epsilon}  &  =  2    D_{\epsilon}   \kB c_v T_0^2 ~~ ,   \\ 
           \Delta_{\env}  &  =  2   \gamma_\env \kB    T_0  ~~ .  \end{align}    
\end{subequations}
Additionally taking
\begin{align} \label{eq:noisevariance12}
      \Delta_{\fuel}  &  =  2   \gamma_\fuel \kB    T_0  ~~ ,   \end{align}  
      \eqref{eq:noisevariance12} leads one to find outside of the famine state that
      \begin{align}
           &    G^\rmS_{\epsilon \epsilon }   \nn 
    = 2  \kB  T_0    \frac{    D_{\epsilon}   \kB c_v T_0   k^2  + 
  \gamma_{\env}  \ell^2  (\kB  T_{\env})^2   + 
       \gamma_{\fuel}  \ell^2  \Delta E^2    }{ \omega ^2+\left(\Gamma_{\epsilon} +D_{\epsilon} 
  k^2  \right)^2 }   ~~ , 
\end{align}
which matches with \eqref{eq:symmetriccorrelator123} if one uses \eqref{eq:eqb-temp} and \eqref{eq:energy-relaxation-rate}. The operational procedure outlined in this section can be used to compute energy correlation functions within fluctuating hydrodynamics applied to active matter.

\section{Active nematics}
\label{sec:active_nematics}
Having gained some insights using our simple toy model, we now proceed to apply our formalism of active hydrodynamics to a physically richer model of active matter, namely active nematics, with applications such as bacterial populations \cite{PhysRevLett.108.098102,PhysRevE.94.050602,PhysRevE.95.020601,PhysRevX.7.011029}, microtubule-motor protein mixtures \cite{Ndlec1997,Butt2010}, epithelial cells \cite{Epithelial1,Epithelial2,Epithelial3} and swarming sperm cells \cite{PhysRevE.92.032722}. A nematic liquid crystal is characterised by the presence of long-range orientational order in a physical system, thereby spontaneously breaking the global rotational symmetry. The order parameter for nematicity is a symmetric traceless tensor $Q_{ij} = \langle a_ia_j - 1/d\,\delta_{ij}\rangle_{\text{micro}}$, constructed by averaging over the orientations $a_i$ of all individual constituents with $a^ia_i=1$. Note that $Q_{ij}$ is invariant under the flip of individual orientations $a_i\to-a_i$, hence the nematic phase describes elongated rod-like constituents without a defined head or tail. We will mostly be interested in a uniaxial nematic, where the individual constituents align themselves along a single macroscopic director field $p_i$, with $p^ip_i=1$, i.e.
\begin{align}\label{eq:uniaxial}
    Q_{ij} = Q_0\lb p_ip_j - \frac1d \delta_{ij} \rb~,
\end{align}
with $Q_0$ representing the strength of alignment~\cite{Doostmohammadi2018}.

The theory of active nematic hydrodynamics is characterised by the associated energy, momentum, and mass conservation equations taking the form
\begin{align}\label{eq:nematic-conservation}
    \dow_t\epsilon + \dow_i\epsilon^i &= \blue{\,-\, \ell r_\fuel\dow_t\Phi^\fuel - \ell  r_\env \dow_t\Phi^\env}~, \nn\\
    \dow_t\pi^j + \dow_i\tau^{ij} &= \blue{\ell r_\fuel\dow^j\Phi^\fuel + \ell  r_\env \dow^j\Phi^\env}~, \nn\\
    \dow_t\rho + \dow_i\pi^i &= 0~,
\end{align}
where we have introduced the mass density $\rho$, momentum density $\pi^i = \rho u^i$, stress tensor $\tau^{ij}$, with $u^i$ being the fluid velocity. Note that, generically, the spatially inhomogeneous profiles of the fuel and environment fields $\Phi^{\fuel,\env}$ also impart momentum to the system in \cref{eq:nematic-conservation}, but these terms drop out for the homogeneous configuration in \cref{eq:bath-config}. The conservation equations together can be seen as determining the dynamics of the fluid temperature $T$, fluid velocity $u^i$, and mass chemical potential $\mu$. We still need an equation of motion for the nematic order parameter $Q_{ij}$, which will be obtained by extremising the SK effective action for the theory. 

\subsection{Fields and symmetries}

Since we have introduced conserved momentum into our setup, we need to add new degrees of freedom in the SK framework, i.e. fluid velocity $u^i$ and the associated noise field $X^i_a$, in addition to $T$, $\mu$, $X^t_a$, and $\varphi_a$ already introduced in \cref{sec:SKstructure}. Furthermore, to describe the nematic phase we need to introduce the order parameter $Q_{ij}$ and its noise partner $\cQ_{aij}$. The dynamical fields realise doubled U(1) symmetry, and doubled space- and time-translation symmetries. The action of $r$-type space- and time-translations is given as usual diffeomorphisms on all the fields. The remaining symmetries act as
\begin{align}
     X^t_a &\to X^t_a -\chi^t_{a}~, \nn\\
     X^i_a &\to X^i_a -\chi^i_{a}~, \nn\\
    \varphi_a 
    &\to \varphi_a -\Lambda_a - X_a^t \dow_t \Lambda_r - X_a^i \dow_i \Lambda_r~,
\end{align}
while leaving $T$, $\mu$, $u^i$, $Q_{ij}$, and $\cQ_{aij}$ invariant. We avoid introducing the associated background fields in the main text for simplicity, but a detailed treatment can be found in the appendices. We also restrict our attention to systems that feature Galilean boost symmetry. $N_{at}$, $N_{ai}$ defined in \cref{eq:invariant-defs-noncov} are Galilean-covariant, but $B_{at}$, $B_{ai}$ are not and we instead define
\begin{align}\label{eq:cB-def}
    \cB_{at} &= B_{at} + \half N_{at} \vec u^2~, \nn\\
    \cB_{ai} &= B_{ai} + \dow_t X_{ai} - u_i N_{at} + \half N_{ai} \vec u^2~, \nn\\
    \cH_{aij} &= 2\dow_{(i}X_{aj)} - 2 N_{a(i} u_{j)}~.
\end{align}
The chemical potential $\mu$ also needs to be improved to the Galilean-invariant mass chemical potential $\varpi = \mu + \half\vec u^2$. 
Due to new spacetime symmetries, the definitions in \cref{eq:cF-def} also need to be modified to
\begin{gather}
    \Delta E = - \lb \dow_t + u^i\dow_i \rb\! \Phi^\fuel_r, \qquad
    \kB T_\env = \lb \dow_t + u^i\dow_i \rb\!\Phi^\env_r~,\nn\\
    \Pi_{a}^{\fuel,\env} = \ell\Phi_a^{\fuel,\env} 
    + \ell \lb X_a^t \dow_t + X_a^i\dow_i \rb\!\Phi^{\fuel,\env}_r~.
    \label{eq:cF-mod}
\end{gather}
More details can be found in the appendices.

As with the superfluid model, the KMS transformation acts on the new physical dynamical fields $u^i$, $Q_{ij}$ as merely a time-reversal transformation, with the time-reversal eigenvalues given in \cref{tab:T-reversal}, while the noise fields are taken to transform as
\begin{align}
    X^i_a 
    &\KMSto \hat X^i_a \equiv X^i_a + i\beta u^i , \nn\\
    \cQ_{aij}
    &\KMSto \hat\cQ_{aij} 
    \equiv \cQ_{aij} 
    +  i\beta\frac{\df}{\df t} Q_{ij}  \nn\\
    &\qquad\qquad\qquad
    + i\beta\lb\dow_{[i} u_{k]} Q_{j}{}^k + \dow_{[j} u_{k]} Q_{i}{}^k \rb ~,
\end{align}
where $\df/\df t \equiv \dow_t + u^i\dow_i$ denotes the time-derivative long the fluid flow.
The derivation of these transformations can be found in \cref{app:passive-sk,app:activenematics}.

To model a nematic liquid crystal, $Q_{ij}$ and $\cQ_{aij}$ need to be traceless. We can achieve this by including the following Lagrange multiplier terms in the SK Lagrangian
\begin{align}
    \cL_{\tr} &= \tr(Q)\tr(\cQ_{a}) \nn\\
    &\qquad 
    + \half \tr(Q)^2
    \lb N_{at} + u^iN_{ai} + \half\cH_{ai}{}^i \rb~,
\end{align}
where the terms in the second line are necessitated by KMS. Therefore, $\tr(Q)$ and $\tr(\cQ_a)$ are set to zero onshell.

\subsection{Schwinger-Keldysh effective action}

We can use the ingredients outlined above to write down the SK effective action for an active nematic liquid crystal. The ``fuel part'' of the Lagrangian is still given by \cref{eq:fuel-L}, but with the modified definitions in \cref{eq:cF-mod}. The ``fluid part'' is given as
\begin{align}\label{eq:fluid-lagrangian}
    \cL_{\text f}
    &= - \varepsilon \lb N_{at} + u^i N_{ai} \rb
    + \rho\lb \cB_{at} + u^i \cB_{ai}\rb \nn\\
    &\!\!
    + \half \lb \rho\, u^i u^j - \cF\, \delta^{ij} \rb \cH_{aij}
    + i \kB T^2\kappa N_{ai}\hat N_a^i \nn\\
    &\!\!
    + \frac{i}{2} \kB T\eta\, \cH_{aij}\hat\cH^{ij}_a
    + \frac{i}{4} \kB T\!\lb \zeta - {\textstyle\frac{2}{d}}\eta\rb\!
    \cH_{ai}^{~~i}\hat\cH_{aj}{}^j \nn\\
    &\!\!
    \blue{\,+\, i \kB T\gamma_\env \Pi_a^\env\hat\Pi_a^\env 
    + \frac{i}{2}\frac{T}{T_\env} p_\env \delta^{ij}\Big( \cH_{aij}\hat\Pi^\env_a 
    - \hat \cH_{aij}\Pi^\env_a
    \Big)}~.
\end{align}
The passive contributions here are taken from~\cite{Jain:2020vgc}, where we have introduced the internal energy density $\varepsilon$, mass density $\rho$, free energy density $\cF$, thermal conductivity $\kappa$, shear viscosity $\eta$, and bulk viscosity $\zeta$.
Among the active contributions, the $\gamma_\env$ term is analogous to the one in \cref{eq:lagrangian-passive-diffusion}, while $p_\env$ is a new term that would give rise to an active correction to pressure. 

\begin{table}[t]
    \centering
    \begin{tabular}{c|l}
        \hline\hline 
        \multicolumn{2}{c}{\textbf{Passive Transport}} \\
        \hline 
        $p$ & Thermodynamic pressure \\
        $\varepsilon$ & Internal energy density \\
        $\rho$ & Mass density \\
        $\kappa$ & Thermal conductivity \\
        $\eta,\zeta$ & Shear and bulk viscosity\\
        $\gamma$ & Nematic-shear coupling \\
        \hline 
        $a$, $a_3$, $a_4$ & Parameters of nematic potential $V$ \\
        $K \equiv 2Q_0^2K_Q$ & Frank elasticity constant \\
        $\sigma_p \equiv 2Q_0^2\sigma_Q$ & Nematic diffusivity parameter \\
        \hline\hline
        \multicolumn{2}{c}{\textbf{Active Transport}} \\
        \hline 
        $\gamma_\fuel$ & Controls rate of fuel consumption due to $\Delta E$ \\
        $\gamma_{\env}$ & Controls rate of entropy loss due to $\Delta T$ \\
        \hline
        $a_\env$ & Alters the equilibrium state $Q_0$ \\
        $p_{\env}$ & Active correction to pressure \\
        $\lambda_{\env}$ & Gives rise to active-nematic stress $\tau^{ij}\sim Q^{ij}$ \\
        \hline
    \end{tabular}
    \caption{Thermodynamic and transport coefficients of an (active) nematic with Galilean boost symmetry.}
    \label{tab:nematics}
\end{table}

Moving onto the ``nematic part'', the simplest SK Lagrangian for the order parameter $Q_{ij}$ is written analogous to the superfluid model in \cref{eq:Psi-Langrangian}, but slightly more involved due to the tensorial structure of $Q_{ij}$. To wit
\begin{align} \label{eq:Qterms}
    \cL_Q
    &= K_Q\dow^k Q^{ij}\lb
    \frac{\df}{\df t}Q_{ij}N_{ak} 
    + \half \dow^l Q_{ij} \cH_{a kl} \rb \nn\\
    &\hspace{-1em}
    + 2K_Q\dow^i Q^{l[k} Q^{j]}_{~l}
    \Big( \dow_{k}\cH_{aij} 
    + \dow_k u_j N_{ai}
    + 2\dow_{(i} u_{k)} N_{aj}
    \Big) \nn\\
    &\hspace{-1em} 
    - \lb \frac{\dow V}{\dow Q_{ij}} + K_Q\dow^k Q^{ij} \dow_k \rb \cQ_{aij} \nn\\
    &\hspace{-1em}
    + i \kB T\sigma_Q \lb \cQ_a^{ij} - \half \gamma^{ijkl}\cH_{akl}\rb\!
    \lb \hat \cQ_{aij} - \half \gamma_{ij m n } \hat\cH^{m n }_{a}\rb \nn\\
    &\hspace{-1em}
    \blue{
    \,+\, \frac{i}{2} \kB T \lb \lambda_\env Q^{ij}
    + \frac{a_\env}{\kB T_\env} Q_{kl} \gamma^{klij}
    \rb \Big( \cH_{aij}\hat\Pi^\env_a 
    - \hat\cH_{aij}\Pi^\env_a
    \Big)} \nn\\ 
    &\hspace{-1em}
    \blue{\,-\, i\frac{T}{T_\env} a_\env Q^{ij}\Big( \cQ_{aij}\hat\Pi^\env_a 
    - \hat \cQ_{aij}\Pi^\env_a \Big)}~,
\end{align}
where we have introduced a single elastic constant $K_Q$, nematic conductivity $\sigma_Q$, and the nematic shear coupling tensor $\gamma_{ijkl}$ that is symmetric-traceless in the first two indices and symmetric in the last two. The potential $V$ may depend arbitrarily on $Q_{ij}$. For instance, we may take a simple form~\cite{Doostmohammadi2018}
\begin{align}\label{eq:potential-nem}
    V = \half a \tr(Q^2) - \frac13 a_3 \tr(Q^3) 
    + \frac{1}{4}a_4 \tr(Q^2)^2~,
\end{align}
where $a$, $a_3$, and $a_4$ are phenomenological parameters similar to \cref{eq:potential}. The form of the potential is sufficient to describe the nematic phase in $d\leq 3$ spatial dimensions, because higher-traces of $Q_{ij}$, i.e. $\tr(Q^n)$ for $n\geq 4$, are not independent. We will need to account for these traces in higher-dimensions. 
The active coefficients $a_\env$ and $\lambda_\env$ play a similar role to their namesakes from the active superfluid model.

The SK Lagrangian outlined above is invariant under all the global symmetries of the SK framework. The Lagrangian also satisfies the conditions in \cref{eq:SK-conds}, provided that we demand
\begin{gather}
    \kappa\geq 0~, \quad 
    \eta \geq 0~, \quad 
    \zeta \geq 0~, \quad 
    \sigma_Q \geq 0~, \quad
    \gamma_{\env,\fuel} \geq 0~.
    \label{eq:ineq-nem}
\end{gather}
Finally, the SK Lagrangian is KMS-invariant, provided that $\varepsilon$, $\rho$, and $K_Q$ are related to $\cF$ via the thermodynamic relations
\begin{align}\label{eq:thermoidentities}
    \varepsilon
    &= Ts + \rho\varpi + \cF~, \nn \\ 
    \df\cF 
    &= - s\,\df T - \rho\,\df\varpi  \nn\\
    &\qquad
    + \frac{1}{2} K_Q\df\Big(\dow^k Q_{ij}\dow_k Q^{ij} \Big)
    + \frac{\dow V}{\dow Q_{ij}}\df Q_{ij}~, \nn\\
    \df\varepsilon 
    &=   T \,\df s + \varpi\,\df  \rho\nn \\
    &\qquad
    + \frac{1}{2} K_Q\df\Big(\dow^k Q_{ij}\dow_k Q^{ij} \Big)
    + \frac{\dow V}{\dow Q_{ij}}\df Q_{ij}~ . 
\end{align}
With these in place, the SK Lagrangian is KMS-invariant up to a boundary term $-i\dow_t(\beta \cF)-i\dow_i(\beta \cF\,u^i)$.
Assuming $K_Q$ to be constant, we arrive at the Landau-de Gennes free energy \cite{Doostmohammadi2018,thampi2015intrinsic,de1993physics}
\begin{align}\label{eq:F-nematic}
    \cF = - p 
    + \frac{1}{2} K_Q\dow^k Q_{ij}\dow_k Q^{ij} 
    + V~.
\end{align}
The generalisation to multiple elastic constants is straightforward and has been discussed in the appendix.

\subsection{Spontaneous symmetry breaking}
\label{sec:nematic-ssb}

The equation of motion for the nematic order parameter $Q_{ij}$ can be obtained by extremising the SK effective action with respect to $\cQ_{aij}$. We find
\begin{equation}\label{eq:Q-EOM}
    \frac{\df}{\df t} Q_{ij}
    = \frac{1}{\sigma_Q } \Big(  \sfH_{ij} \blue{\,-\, \aleph a_\env Q_{ij}}\Big)
    + \sfS_{ijkl}\dow^k u^l~,
\end{equation}
where $\sfH_{ij}$ is the thermodynamic conjugate to $Q_{ij}$ and $\sfS_{ijkl}$ is the generalised advection tensor, defined as
\begin{align}
    \sfH_{ij} &= \dow_k\big(K_Q\dow^k Q_{ij}\big) 
    - \frac{\dow V}{\dow Q_{ij}} - \text{(trace)}~, \nn\\
    \sfS_{ijkl} &= Q_{k(i} \delta_{j)l} - Q_{l(i} \delta_{j)k} + \gamma_{ijkl}~.
\end{align}
Whether the system ends up in the nematic phase or the fluid phase depends on the potential $V$ in \cref{eq:potential-nem} and the activity parameter $a_\env$. The system admits a fluid phase for $a+\aleph a_\env>0$ with $\langle Q_{ij}\rangle = 0$. Whereas, when $a+\aleph a_\env < \frac{d-2}{8d}a_3^2/a_4$, the system admits a nematic phase with $\langle Q_{ij}\rangle = Q_0\lb p^0_i p^0_j - 1/d\,\delta_{ij} \rb$, where $p^0_i$ is a fixed unit vector and $Q_0$ is given by~\cite{de1993physics}
\begin{align}\label{eq:def-Q0}
    Q_0 
    &=
    \frac{a_3}{|a_3|}\sqrt{\frac{d}{(d-1)a_4}} \nn\\
    &\times
    \lb \sqrt{\frac{d-2}{8d} \frac{a_3^2}{a_4}}
    + \sqrt{\frac{d-2}{8d} \frac{a_3^2}{a_4} - a \blue{\,-\,\aleph a_\env}}\rb~.
\end{align}
Both the fluid and nematic phases are admitted in the overlapping regime $0<a+\aleph a_\env < \frac{d-2}{8d}a_3^2/a_4$. In thermal equilibrium, the system prefers the phase with lower free energy, with a first-order phase transition between them at the coexistence point $a_*=\frac{d-2}{9d}a_3^2/a_4$. We do not have the luxury to compare the free energies to determine thermodynamic stability of a non-equilibrium steady state in the presence of activity; see our discussion at the end of \cref{sec:DB}. However, for small activity, we can expect the first-order phase transition to happen somewhere in the vicinity of $a_*\approx \frac{d-2}{9d}a_3^2/a_4$.

Similar to our discussion for active superfluids around \cref{eq:criticalscale}, activity may induce or destroy the nematic phase transition when $a$ and $a_\env$ have opposite signs.

The fluctuations of $Q_{ij}$ in the fluid phase are gapped and the low-energy description is just given by the Lagrangian \eqref{eq:fluid-lagrangian}. In the nematic phase, however, we need to account for the Goldstone modes associated with the spontaneously broken rotation symmetry generators, parametrised in terms of the nematic director $p_i$ given in \cref{eq:uniaxial}. The associated equation of motion can be read off from \cref{eq:Q-EOM} as
\begin{align}\label{eq:p-eom}
    \frac{\df}{\df t}p_i
    &= \frac{\bar p_i^j}{\sigma_p} \dow_k\Big(K \dow^k p_j\Big) 
    + \lb p^{[k} \bar p^{l]}_i + \gamma p^{(k} \bar p^{l)}_i  \rb \dow_k u_l~,
\end{align}
where $\bar p_{ij} = \delta_{ij}-p_ip_j$ is the projector transverse $p_i$, and we have identified the Frank constant $K$, relaxation coefficient $\sigma_p$, and the shear coupling coefficient $\gamma$ as
\begin{gather}
    K = 2Q_0^2K_Q~, \qquad 
    \sigma_p = 2Q_0^2\sigma_Q~, \nn\\
    \gamma = \frac{2/Q_0}{d-1}p_ip_k \bar p_{jl}\gamma^{ijkl}~.
    \label{def:KtoKQ}
\end{gather}
The effects of activity are hidden within $Q_0$.

We can write down a low-energy effective theory for the director $p_i$ directly, akin to the superfluid model in \cref{eq:SK-Lagrangian-sf}. To this end, we need to define the KMS noise partner $\calp_{ai}$ to the director, satisfying $p^i\calp_{ai}=0$, with the KMS transformation
\begin{align}
    \calp_{ai} \KMSto \eta_\rmT\,\hat{\!\calp}_{ai} \equiv
    \eta_\rmT\! \lb \calp_{ai} 
    + i\beta \frac{\df}{\df t} p_i 
    + i\beta \dow_{[i} u_{j]} p^j \rb~.
\end{align}
Note that since the original nematic order parameter $Q_{ij}$ has a $p_i\to-p_i$ symmetry, we are free to choose either time-reversal eigenvalue for $p_i$, $\calp_{ai}$. In terms of these, the active nematic Lagrangian takes the simple form
\begin{align}\label{eq:nematic-action-q}
    \cL_{\text{nem}}
    &= - \lb \frac{\dow\cF}{\dow p_i} + \frac{\dow\cF}{\dow(\dow_k p_i)} \dow_k \rb \calp_{ai} \nn\\
    &\hspace{-1em}
    + \frac{\dow\cF}{\dow(\dow_k p_i)}
    \!\lb \frac{\df}{\df t} p_i N_{ak}
    + \half\dow^l p_i \cH_{akl} 
    \rb \nn\\
    &\hspace{-1em}
    + \frac{\dow\cF}{\dow(\dow_i p_{[k})} p^{j]}
    \lb \dow_k\cH_{aij}
    + \dow_k u_j N_{ai}
    + 2\dow_{(i} u_{k)} N_{aj}
    \rb \nn\\
    &\hspace{-1em}
    + i\kB T\sigma_p\! \lb \calp_a^i - \frac{\gamma}{2}\bar p^{ij} p^k \cH_{akj} \rb\!
    \lb \,\hat{\!\calp}_{ai} - \frac{\gamma}{2}\bar p_{ij} p_k \hat\cH^{kj}_a\rb \nn\\
    &\hspace{-1em}
    \blue{\,+\, \frac{i}{2}\kB T Q_0\lambda_\env p^i p^j
    \Big( \cH_{a\langle ij\rangle}\hat\Pi^\env_a 
    - \hat\cH_{a\langle ij\rangle}\Pi^\env_a
    \Big)}~,
\end{align}
with the normalisation conditions imposed by the Lagrange multiplier terms
\begin{align}
    \cL_{\text{norm}}
    &= (p^ip_i-1)(p^j\calp_{aj}) \nn\\
    &~~+ \frac14 (q^kq_k-1)^2
    \lb N_{at} + u^iN_{ai} + \half\cH_{ai}{}^i \rb~.
\end{align}
The low-energy thermodynamic relations are given as
\begin{align}
    \varepsilon
    &= Ts + \rho\varpi + \cF~, \nn \\ 
    \df\cF 
    &= - s\,\df T - \rho\,\df\varpi
    + \frac{1}{2} K \df\Big(\dow^ip^j\dow_ip_j \Big)~,
\end{align}
which ensure the KMS-invariance of the theory. As with the superfluid case, \cref{eq:nematic-action-q} can be obtained from \cref{eq:Qterms} quite simply under the assumption that $Q_0$ is constant, by identifying $Q_{ij}$ as in \cref{eq:uniaxial} and $\cQ_{aij}=2Q_0p_{(i}\calp_{aj)}$. For a non-constant $Q_0$, a more careful calculation needs to be performed to obtain the renormalisation of various coefficients appearing in \cref{eq:nematic-action-q}, which we leave for future work.

\subsection{Constitutive relations and the second law}

Extremising the SK effective action with respect to $X^t_a$, $X^i_a$, and $\varphi_a$, we recover the energy, momentum, and mass conservation equations, with the energy density $\epsilon = \varepsilon + \half \rho \vec u^2$, momentum density $\pi^i = \rho u^i$, and mass density $\rho$. The associated energy flux and stress tensor
\begin{align}\label{eq:consti-nematic}
    \epsilon^i 
    &= \lb \varepsilon + \half\rho \vec u^2
    - \cF \blue{\,+\, p_\env \ell  \kB  \Delta T}
    \rb u^i 
    \blue{\,+\, \lambda_\env Q^{ij} u_j \ell  \kB  \Delta T} \nn\\
    &~~
    - K_Q\dow^i Q^{kl}\dow_t Q_{kl} 
    - u_j\lb \gamma^{klij} - 2 \delta^{k[i} Q^{j]l} \rb \frac{\delta\cF}{\delta Q_{kl}}
    \nn\\
    &~~
    - 2\eta u_j\dow^{\langle i} u^{j\rangle}
    - \zeta u^i \dow_k u^k
    - \kappa\,\dow^i T
    {\color{Gray}\,-\, \dow_{k}\big( \cX^{[ik]j} u_j \big)} ~, \nn\\
    \tau^{ij}
    &= \rho\, u^i u^j 
    - \lb \cF \blue{\,-\, p_\env \ell  \kB  \Delta T} \rb \delta^{ij} 
    \blue{\,+\, \lambda_\env Q^{ij} \ell  \kB  \Delta T}
     \nn\\
    &~~
    + K_Q\dow^i Q^{kl} \dow^j Q_{kl} 
    - \lb \gamma^{klij} - 2 \delta^{k[i} Q^{j]l} \rb \frac{\delta\cF}{\delta Q_{kl}} \nn\\
    &~~
    - \eta \dow^{\langle i} u^{j\rangle}
    - \zeta \delta^{ij} \dow_k u^k
    {\color{Gray}\,-\, 2\dow_{k}\cX^{[ik]j}}~.
\end{align}
The grayed out terms are total-derivatives and drop out of the conservation equations whose form is discussed in \cref{app:activenematics}; the remaining contributions are sometimes referred to as the ``canonical'' constitutive relations. Note that while the full stress tensor is symmetric, the canonical part is not. 
The distinctive active feature in these constitutive relations $\sim\lambda_\env Q^{ij}$ term in the stress tensor that is absent in passive nematics~\cite{de1993physics,1995pcmp.book.....C}, whereas activity can allow such terms to appear \cite{PhysRevLett.103.058102,PhysRevLett.101.068102,2018RPPh...81g6601J}. The active contributions in blue may be attributed to the rates
\begin{align}
    r_\env 
    &= \ell\gamma_\env  \kB  \Delta T 
    + p_\env\dow_i u^i
    + iT a_\env Q^{ij}\hat \cQ_{aij} \nn\\
    &\qquad 
    + \lb \lambda_\env Q^{ij}
    + a_\env Q_{kl} \gamma^{klij}
    \rb \dow_i u_j~, \nn\\
    r_\fuel 
    &= \ell\gamma_\fuel \Delta E~.
\end{align}
We can also reduce these constitutive relations in terms of the director $p_i$ instead of the full $Q_{ij}$, but we do not perform this exercise here.

Using the thermodynamic relations in \cref{eq:thermoidentities}, we can verify that the constitutive relations satisfy the second law of thermodynamics
\begin{align}\label{eq:second-law-diff111}
    &\dow_t s + \dow_i s^i \blue{\,+\, \ell \kB r_\env} \nn\\
    &=  \kappa \left( \frac1T\partial_i  T \right)^2    
    + \frac{2 \eta}{T} (\partial_{\langle i} u_{j\rangle})^2 
    + \frac{\zeta}{T}( \partial_i u^i)^2 
    +  \frac{1}{T\sigma_Q}(\sfH_{ij})^2 \nn\\
    &\qquad 
    \blue{\,+\, \frac{\gamma_\env}{T}
    \ell^2 \kB^2  \Delta T^2  
     + \frac{\gamma_\fuel}{T} \ell^2  \Delta E^2}
     \geq 0~,
\end{align}
where the heat flux is given by
\begin{align}
    Ts^i
    &= \epsilon^i - \mathcal{F} u^i  - \lb\varpi - {\txp\frac{1}{2}} \vec u^2 \rb \rho  u^i  
    - u_j  \tau^{ji  }  \nn\\
    &\qquad 
    + K_Q \partial^i Q^{kl}  \frac{\df}{\df t} Q_{kl} \nn\\  
    &= Ts u^{i} + \kappa\,\partial^i T~,
\end{align}
The positivity of the right-hand side of \cref{eq:second-law-diff111} follows from the constraints in \cref{eq:ineq-nem}. 

Having formulated the SK effective action and equations of motion for our thermal theory of active nematics, one may proceed as \cref{eq:linear-diff} and perform a linearised analysis of the mode spectrum and correlation functions. This would also enable us to compute the violations of FDT analogous to \cref{eq:FDT-violation}. 

Interestingly, active nematics suffer from a linear instability to splay or bend deformations arising from the $\tau^{ij}\sim\lambda_\env Q^{ij}$ term in the constitutive relations~\cite{Ramaswamy_2007, PhysRevLett.89.058101}. This drives the system into a state of nematic turbulence~\cite{Alert_2020,turbulencealert}, invalidating any linearised results. In the turbulent regime, it is possible to describe an active nematic through numerical simulations, which is found to be dominated by the proliferation of
topological defect pairs \cite{morozovalexander,Doostmohammadi2018,PhysRevLett.98.118102,PhysRevE.76.031921,PhysRevLett.101.068102,PhysRevX.5.031003}. Although it is not possible to analytically extract observables in a generic nematic turbulent state at the moment, there may still be physical circumstances \aj{where the} instability is suppressed. In particular, the linear instability is suppressed for dry active nematics \cite{PhysRevE.90.062307,SRamaswamy_2003} and can also be suppressed by finite system size effects \cite{doi:10.2976/1.3054712,Ramaswamy_2007,duclos2018spontaneous,Doostmohammadi2017}. In such cases, it would be possible to extract the \aj{linearised mode spectrum and correlation functions,} as is done for the case of active superfluids in \cref{eq:linear-diff}. For nonequilibrium systems, it has been shown theoretically and experimentally through light-scattering experiments that the energy correlations display long-range behaviour characteristic of their non-equilibrium state \cite{sengersortiz,LI1994399}. We leave these explorations for future work; see~\cite{Armas:2025orz}.

\section{A buzz of activity}
\label{sec:buzz}

In this work we have developed a new hydrodynamic framework for active matter with local temperature fluctuations. The primary ingredient in this framework is the first law of thermodynamics, i.e. \emph{energy balance}: while the internal energy of an active system is not necessarily conserved, it must be balanced by the work done by the fuel source that drives activity and the heat lost to the environment in the process. This means that to appropriately model the thermal fluctuations of active matter, we must regard it as a driven open system and add both a fuel source and a distinct energy sink into the hydrodynamic framework, modelled in this work using the background scalar fields $\Phi^\fuel$ and $\Phi^\env$ respectively. 
In field-theoretic terms, these are seen as background sources coupled to the operators $r_\fuel$ and $r_\env$, measuring the rates of fuel consumption and heat loss respectively.
Another important ingredient that was used  in our framework is the existence of a \emph{famine state}: when an active system runs out of fuel, it must behave passively, i.e. it must obey the FDTs.

The SK-EFT formalism is perfectly suited for our purposes. It is a symmetry-based effective field theory approach to stochastic systems that, in principle, applies arbitrary far from equilibrium~\cite{Grozdanov:2013dba, Harder:2015nxa, Crossley:2015evo, Haehl:2015uoc, Haehl:2018lcu, Jensen:2017kzi, Glorioso:2018wxw}. In particular, the energy balance and (non-linear) FDT requirements are built into the SK-EFT formalism through time-translation symmetry and discrete KMS symmetry. While an active system by itself is not invariant under time-translations, the symmetry does apply when acting simultaneously on the fuel/environment components; see~\cite{Armas:2021vku, Armas:2022vpf, Armas:2023tyx} for a similar procedure for other symmetries. Furthermore, by requiring that the active system respects the original KMS symmetry in the famine state, we derived a new \emph{active KMS symmetry}, which makes the SK-EFT formalism suitable for modelling active matter in contact with a thermal bath (see \cref{sec:active-KMS}). We point out that it is possible to extract the lessons arising from SK-EFT that we learnt from this work and devise an operational procedure for implementing dynamical temperature, energy and stochastic noise in fluctuating hydrodynamics, as we explained in section \ref{sec:fluctuatinghydro}.

The time-translation symmetry, active KMS symmetry, and the unitarity constraints built into the SK-EFT framework, together yield the second law of thermodynamics appropriate for active systems. The active second law of thermodynamics dictates that only the total entropy, i.e. the sum of the entropy of the system and the entropy lost to the environment, is increasing. This is unlike passive systems, where entropy increases locally (see \cref{sec:second-law}). We also explored how the active KMS symmetry is related to microscopic reversibility and gives rise to an active correction to the principle of detailed balance in terms of the work performed by the external fields, as well as the fuel/environment sources (see \cref{sec:DB}).

We have applied our hydrodynamic framework for active matter to two interesting examples, namely active superfluids in \cref{sec:simplediffusion} and active nematics in \cref{sec:active_nematics}. In both these examples, we discussed how coupling to fuel/environment sources may generate entirely new active transport parameters in the hydrodynamic equations that are forbidden for passive systems. These active parameters are entirely fixed in terms of the fuel consumption and heat loss rate operators, $r_\fuel$ and $r_\env$. These coefficients characterise the explicit structure of FDT violation in frequency- and wavevector-dependent retarded and symmetric correlation functions. As a proof of concept, we have provided the first computation of energy correlation functions in active matter in \cref{eq:linear-diff}, which were previously inaccessible in other formalism for the hydrodynamics of active matter. Similar computations can be performed for other active phases including active nematics and the various phases that we describe below in this section. We expect that novel physical phenomena (so far unnoticed) associated with energy balance such as long range order may arise in active systems with additional broken symmetries.

Our hydrodynamic framework is particularly useful when the active system under consideration is operating sufficiently close to thermal equilibrium, meaning that the strength of activity $\aleph=\ell(T-T_\env)/T_\env$ is small, where $\ell$ is a small bookkeeping parameter. Assuming $\ell$ to scale on par with the spatial derivatives $\cO(\dow)$, which are treated as small in hydrodynamics, this allows us to systematically organise the active and passive corrections to the hydrodynamic constitutive relations order-by-order in derivatives. In \cref{sec:simplediffusion,sec:active_nematics}, we have focused on leading-order active corrections, while we will comment on a few potentially interesting subleading-order corrections later in this section.

Our hydrodynamic framework ceases to be applicable when the steady fuel consumption drives the system arbitrarily far away from equilibrium, as the Onsager relations that were used to constrain the action assume small deviations from equilibrium. Furthermore, stochastic fluctuations entirely decouple from dissipation as the system departs from  equilibrium. Nonetheless, hydrodynamic models have historically been known to apply well beyond their strict regime of applicability, so we may hope that our hydrodynamic model continues to provide useful insights even for active matter far from equilibrium.

While we have focused our attention to simple models of active superfluids and active nematics for concreteness, the active hydrodynamic framework developed in this work opens up possibilities for symmetry-based systematic modelling of active phenomena far beyond these specific examples. We outline a few of these potential avenues in the following.

\paragraph{Dissipation vs response vs fluctuation:} Ordinarily, as a result of FDT in a passive hydrodynamics, the transport coefficients controlling dissipation in the constitutive relations are the same coefficients controlling the strength of the retarded and symmetric correlators. However, this may no longer be the case in active hydrodynamics. Consider, for example, the fluid part of the superfluid effective action written in \cref{eq:lagrangian-passive-diffusion}. We may add to this new active terms of the kind
\begin{align}\label{eq:different-conductivities}
    - i\frac{T}{T_\env} \lb \sigma_{\rm t}^\env T\dow^i\frac{\mu}{T} 
    - \sigma_{\rm r}^\env E^i
    \rb \lb B_{ai}\hat\Pi^\env_a + \hat B_{ai}\Pi^\env_a \rb~,
\end{align}
which modify the conductivity contribution in the flux to $j^i\sim -\sigma_{\rm t}T\dow^i(\mu/T) + \sigma_{\rm r}E^i$, where $\sigma_{\rm t,\rm r}=\sigma+\aleph\sigma_{\rm t,\rm r}^\env$. Switching off the superfluid part, the three kinds of conductivities, $\sigma$, $\sigma_{\rm t}$, and $\sigma_{\rm r}$ show up differently in the retarded and symmetric correlators of density as
\begin{align}\label{eq:multi-conductivity-response}
    G^\rmR_{nn}
    &= \frac{-\sigma_{\rm r}\, k^2}{i\omega - \sigma_{\rm t}/\chi\, k^2}~, \quad
    G^\rmS_{nn}
    = \frac{2 \kB T_0\sigma\, k^2}
    {\left|i\omega - \sigma_{\rm t}/\chi\, k^2\right|^2}~.
\end{align}
The ``transport conductivity'' $\sigma_{\rm t}$ controls the poles of both the retarded and symmetric correlators, and thus the linearised mode spectrum. On the other hand, the ``response conductivity'' $\sigma_{\rm r}$ and the ``fluctuation conductivity'' $\sigma$ control the strength of retarded and symmetric correlators respectively. All three notions of conductivity may generically be different in an active system. We should note that for weak activity, $\ell\sim\cO(\dow)$, the terms in \cref{eq:different-conductivities} formally appear at $\cO(\dow^2)$ in the constitutive relations, which makes sense because these are active corrections on top of the already derivative suppressed dissipative corrections. Therefore, at least for weakly active systems, they are generically less important than the other active corrections considered in \cref{sec:simplediffusion}.

A similar strategy may be employed for adding active corrections to viscosities in a hydrodynamic model with conserved momentum, i.e.
\begin{equation}\label{eq:different-viscosity}
    - i\frac{T}{T_\env}
    \lb \eta^\env_{\rm t}\dow^{\langle i} u^{j\rangle} 
    + \half\zeta^\env_{\rm t}\delta^{ij} \dow_k u^k
    \rb
    \lb \cH_{aij}\hat\Pi_a^\env + \hat\cH_{aij}\Pi_a^\env \rb~,
\end{equation}
that corrects the viscosity terms in the stress tensor as $\tau^{ij}\sim -2\eta_{\rm t}\dow^{\langle i}u^{j\rangle} - \zeta_{\rm t} \delta^{ij}\dow_k u^k$, where $\eta_{\rm t} = \eta + \aleph\eta^\env_{\rm t}$ and $\zeta_{\rm t} = \eta + \aleph\zeta^\env_{\rm t}$ are the ``transport viscosities''. Unlike conductivities, however, spacetime symmetries generically force the ``response viscosities'' to be the same as  ``transport viscosities'', while these may generically differ from the ``fluctuation viscosities'' $\eta$ and $\zeta$ in an active system.

Staring at \cref{eq:different-conductivities,eq:different-viscosity}, we may notice that these are similar in form to the $\lambda_{n\env}$ term in \cref{eq:Psi-Langrangian} or the $\lambda_\env$ term in \cref{eq:Qterms}, except that the signs in the parenthesis are different due to different time-reversal eigenvalues.  This is indeed the common theme for generating active corrections in the constitutive relations, by writing down cross terms with the activity sources; see~\cite{2023arXiv230915142L, JULICHER20073, 2018RPPh...81g6601J}.

\paragraph{Sign-indefinite dissipation:} A consequence of the active corrections to the dissipative transport coefficients is that they may be sign-indefinite. Going back to the conductivity corrections in \cref{eq:different-conductivities}, we note that neither of $\sigma_{\rm t}^\env$ or $\sigma_{\rm r}^\env$ are required to be non-negative by the SK unitarity constraints in \cref{eq:SK-conds}; only $\sigma$ needs to be non-negative. This is essentially requiring that the symmetric correlator, which computes the variance of a stochastic random variable, must be non-negative, while no such requirement exists for the retarded correlator in the absence of FDT. This is to say that $\sigma_{\rm t}$ or $\sigma_{\rm r}$ may turn negative for sufficiently strong activity. Interestingly, since $\sigma_{\rm t}$ controls the pole of the correlation functions in \cref{eq:multi-conductivity-response}, this also means that such active systems will become unstable for sufficiently strong activity. While a purely diffusive system becomes unstable as soon as $\sigma_{\rm t}$ turns negative, this feature is sensitive to the details of the mode spectrum. For example, the superfluid mode spectrum in \cref{eq:mode-spectrun} remains stable until $\sigma_{\rm t}$ becomes sufficiently negative to overcome the Goldstone diffusion parameter $D_\phi=f_s/\sigma_\phi$. Similar considerations also apply for the transport viscosities $\eta_{\rm t}$ and $\zeta_{\rm t}$.

\paragraph{Odd viscosity and odd elasticity:} Another class of active phenomena that has gained recent traction are odd viscosity and odd elasticity, which can arise in two-dimensional parity-violating systems. For the former, let us consider the case where parity is broken but time-reversal symmetry is present. In that case, odd viscosity is prohibited for passive systems, however for active systems one can add a term to the SK-EFT Lagrangian
\begin{align}\label{eq:odd-viscosity}
    - i\frac{T}{T_\env}\tilde\eta\,\delta^{ik}\epsilon^{jl} \dow_{(k} u_{l)}
    \lb \cH_{aij}\hat\Pi_a^\env + \hat\cH_{aij}\Pi_a^\env \rb~,
\end{align}
where $\epsilon^{ij}$ is the 2d anti-symmetric Levi-Civita symbol.
This results in an odd-elasticity term in the stress tensor of the kind $\tau^{ij}\sim - 2\aleph\tilde\eta\,\delta^{k(i}\epsilon^{j)l} \dow_{(k} u_{l)}$, which is forbidden by the Onsager's relations in passive time-reversal symmetric hydrodynamics. 
Note that for weak activity, $\ell\sim\cO(\dow)$, the contribution in \cref{eq:odd-viscosity} appears at $\cO(\dow^2)$. In contrast, odd-elasticity appears already at $\cO(\dow)$ at weak activity. Directly importing the form in \cref{eq:odd-viscosity}, we may write
\begin{align}
    i\frac{T}{T_\env}K^o\,\delta^{ik}\epsilon^{jl} u_{kl}
    \lb \cH_{aij}\hat\Pi_a^\env - \hat\cH_{aij}\Pi_a^\env \rb~,
\end{align}
where $u_{ij}$ is the strain tensor and the sign inside the parenthesis is $+$ because $u_{ij}$ is even under time-reversal symmetry. This results in the odd-elasticity contribution to the stress tensor that take the form $\tau^{ij}\sim 2\aleph K^o\,\delta^{k(i}\epsilon^{j)l} u_{kl}$, introduced in~\cite{2019arXiv190207760S}.

\paragraph{Explicitly broken symmetries:} Just like the explicitly broken time-translation symmetry, many active systems of interest also explicitly break other internal and/or spacetime symmetries. Examples include: broken number/charge conservation in Malthusian active matter \cite{2012PhRvL.108h8102T,2010PhRvL.104t8101M,2010JSMTE..02..003M} where entities may replicate or die over time; broken spatial-translations and/or boosts when the system is subjected to physical barriers or friction~\cite{1998PhRvE..58.4828T}; broken rotations when subjected to electromagnetic fields or acceleration~\cite{{PhysRevE.67.040301}}; and also broken topological symmetries \cite{Shankar_2018,braverman,Armas:2023tyx}. Explicitly broken or approximate symmetries can be modelled by introducing the appropriate sink background fields in the hydrodynamic framework~\cite{Armas:2021vku, Armas:2023tyx}, just like $\Phi^\env$ plays the role of an energy sink in our construction. 

For instance, we introduce a momentum sink for broken spatial-translations in the form of a background scalar field $\Phi^I$, one for each spatial direction, with the noise partner $\Phi^I_a$ in the SK framework. They feature the regular KMS transformation \eqref{eq:KMS} with time-reversal eigenvalue $+1$, and take values $\delta^I_ix^i$ and $0$ respectively for a homogeneous momentum source. The SK-EFT Lagrangian now contains new terms such as 
\begin{align}\label{eq:mometum-sink}
    i\kB T\sigma_\Phi\Pi^I_a\hat\Pi_{Ia}
    - i\frac{T}{T_\env}
    \sigma^\env_{\Phi{\rm t}} \frac{\df\Phi_I}{\df t}\lb \Pi_{Ia}\hat\Pi^\env_a + \hat\Pi_{Ia}\Pi^\env_a \rb
    ~,
\end{align}
where $\Pi^I_a = \ell\Phi_a^I + \ell(X_a^t\dow_t + X_a^i\dow_i)\Phi^I$, and $\ell$ is taken to collectively control all forms of explicit symmetry breaking. Together, they give rise to a contribution on the right-hand side of the momentum conservation equation in \cref{eq:nematic-conservation}, which for a homogeneous background configuration takes the form $\dow_t\pi^i \sim -\ell^2 \sigma_{\Phi{\rm t}} u^i$, where $\ell\sigma_{\Phi{\rm t}} = \ell\sigma_\Phi + \aleph \sigma^\env_{\Phi{\rm t}}$.  Since $\sigma_{\Phi\rm t}$ must be strictly non-negative for passive systems, provided that none of the other momentum-imparting background fields are turned on, the only allowed solution is $u^i=0$. However, when activity is turned on, $\sigma_{\Phi\rm t}$ is no longer sign-definite and the momentum equation may admit other favourable solutions where $\sigma_{\Phi\rm t}$ goes to 0 instead. 

This effect is most evident for fluids without boost symmetry, in which case 
the coefficients may take the form $\sigma_{\Phi}= A + B \vec u^2$ and $\sigma^\env_{\Phi{\rm t}}=A_\env$, with $A,B>0$ and $A_\env<0$. When $\ell A + \aleph A_\env>0$, the system prefers the solution with $u^i=0$. Whereas, when $\ell A + \aleph A_\env<0$, the system spontaneously picks a state with $\vec u^2 = -(A + \aleph/\ell\, A_\env)/B$. This is precisely the structure underlying the Toner-Tu model of flocking~\cite{1998PhRvE..58.4828T}. Using the combination of broken translations, broken boosts, and activity, one may also generate all other terms in the Toner-Tu model; we leave an in-depth analysis for future work.

\paragraph{Active polar:} In \cref{sec:active_nematics}, we studied active nematic liquid crystals described by the order parameter $Q_{ij}$, whose low-energy dynamics reduces to the director $p_i$, together with the constraint $\vec p^2=1$ and $p_i\to-p_i$ symmetry. If we were to relax these, we may describe an active polar liquid crystal where the microscopic constituents feature distinct heads and tails~\cite{1998PhRvE..58.4828T,Toner2005,JULICHER20073,doi:10.2976/1.3054712,2012EPJST.202....1R}. The free energy density $\cF$ now contains a potential $V(\vec p^2) = \half a\,\vec p^2 + \frac14 a_4 \vec p^4$ that controls the transition between the ordered and disordered phases similar to our discussion in \cref{sec:nematic-ssb}. The explicit structure of the hydrodynamic description depends on the time-reversal eigenvalue of $p_i$, which the nematic phase is agnostic to on account of the $p_i\to-p_i$ symmetry. For example, $p_i$ is time-reversal-odd if it models the individual spins of microscopic constituents or time-reversal-even if it models the individual dipole moments. 

For time-reversal-even $p_i$, the free energy density $\cF$ may contain new terms at one-derivative order, e.g. $f_1\, p^i\dow_i\rho$, $\half f_2\,p^i\dow_i \vec p^2$, which contribute to SK-EFT Lagrangian in \cref{eq:nematic-action-q} accordingly. They generate a number of new terms in the equation for $\dow_t p_i$ in \cref{eq:p-eom}, i.e. $\lambda_2p_i\dow_k p^k$, $\lambda_3 \dow_i \vec p^2$, $\lambda_4 p_ip^k\dow_k\vec p^2$, $\lambda_5\dow_i\rho$, $\lambda_6p_ip^k\dow_k\rho$, with the coefficients $\lambda_{2,\ldots,6}$ fixed in terms of $f_{1,2}$. Importantly, the advective term like $\lambda_1p^k\dow_k p_i$ is forbidden in purely passive polar liquid crystals.  By contrast, all these terms are forbidden in a passive polar liquid crystal for time-reversal-odd $p_i$. These constraints may be overcome by introducing activity into the model, via new contributions in the SK-EFT Lagrangian
\begin{align}
    -i\frac{T}{T_\env}\lb \lambda_1 p^k \dow_k p^i + \ldots \rb \lb \calp_{ai} \hat\Pi^\env_a \mp \hat\calp_{ai}\Pi^\env_a \rb~,
\end{align}
where the upper/lower sign is applicable for even/odd $p_i$ under time-reversal. Note that even though activity can be used to generate every possible term in the equation for $\dow_tp_i$, casting both the even and odd cases on the same footing, the two scenarios remain qualitatively distinct. In particular, for weak activity, the terms allowed by passivity in either of these cases dominate over those only allowed by activity. 

Along the same lines, we noted that passive polar dynamics sets the advective coefficient $\lambda_1=0$ for both time-reversal even/odd $p_i$, and this must be generated purely from activity. This should be contrasted with the Toner-Tu model of flocking, where the role of the vector order parameter $p_i$ is played by the velocity $u^i$ of the individual constituents, as opposed to the fixed properties like spin or dipole moment. In this case, the passive dynamics is actually governed by the Navier-Stokes equations arising from momentum conservation and does allow for the advective term with $\lambda_1=1$ as seen from \cref{eq:consti-nematic}. We need activity to make this term different from 1.

\paragraph{Electrically-driven fluids:} Instead of a fuel source $\Phi^\fuel$, we may consider driving activity through one of the other background fields in the description, such as electric fields $E_i$ in a charged fluid~\cite{Amoretti:2022ovc, Brattan:2024dfv}. Since electric fields impart both energy $E_i j^i$ and momentum $E_i n$ to the fluid, we are forced to include both the energy sink $\Phi^\env_{r,a}$ and the momentum sink $\Phi^I_{r,a}$ introduced around \cref{eq:mometum-sink}. Taking $j^i = nu^i$, energy and momentum balance leads to the steady states
\begin{align}
    T = T_\env \lb 1 + \frac{n^2 E_iE^i}{\ell^4 \kB^2 T_\env^2 \gamma_\env\sigma_{\Phi\rm t}} \rb~, \qquad 
    u^i=\frac{nE^i}{\ell^2\sigma_{\Phi\rm t}}~.
\end{align}
Starting from this steady state, it is possible to retrace the discussion in this paper and formulate a hydrodynamic theory which includes electric field driving-induced contributions that are not bound by the local second law of thermodynamics. This example makes it clear that the framework we introduced in this paper can in general account for driven open systems, including systems for which the driving source does not originate from microscopic processes of burning fuel.

\paragraph{Active phase transitions:} In \cref{sec:higgs-sf,sec:nematic-ssb}, we discussed how the strength of activity may be used as a control parameter to induce phase transitions between the ordered and disordered states in our active hydrodynamic models that may be forbidden in passive systems. Active phase transitions are theoretically challenging because they do not have a notion of free energy minimisation and thus our usual theory of phase transitions based on Euclidean statistical field theory does not apply. The SK-EFT framework developed in this work perfectly sets the stage for a Wilsonian renormalisation group approach to study activity-induced phase transitions.  While we have not pursued this line of inquiry in this work, we anticipate that the models proposed in this work will help further our understanding of active phase transitions and we plan to return to these considerations in future work.

\paragraph{Kinetic theory, holography, and other driven open systems:} In this work, we have developed a systematic framework for constructing hydrodynamic models for active matter based on symmetries. However, such modelling typically features a number of undetermined transport coefficients that need to be fixed either through experiments or through an explicit microscopic calculation. At weak coupling, one can use the techniques of kinetic theory to derive the transport coefficients~\cite{pitaevskii2012physical, beris1994thermodynamics}. It will be interesting to revisit this approach for thermal active matter in the presence of fuel source and energy sink, as done previously for hydrodynamics with momentum sinks~\cite{Lucas:2017idv}.

However, our analytical tools are quite limited in the strong coupling regime where the hydrodynamic models are most reliable. Holography or the AdS/CFT correspondence provides an alternative route to derive the qualitative features of transport coefficients in a strongly coupled fluid, using a higher-dimensional gravitational theory~\cite{Maldacena:1997re, Policastro:2001yc, Bhattacharyya:2008jc}. However, the standard paradigm of holography only applies to relativistic systems, which has hindered its utilisation for studying active matter models that typically do not feature energy conservation. By contrast, the formalism of active hydrodynamics developed in this work systematically accounts for energy conservation and dynamical temperature, and while we have only applied this to non-relativistic systems, we do not anticipate any conceptual obstructions to extending these to relativistic systems. If so, it will be interesting to return to the prospect of active holography in another work. We also expect that the extension of this framework to relativistic driven open systems will offer new insights into other non-equilibrium systems beyond the holographic setup, including in the context of astrophysics.

\paragraph{Mechanosensitivity and complex environments:} Throughout this work we have assumed that there is no backreaction on the fuel burning process and hence that the rate $r_\fuel$ does not receive hydrodynamic corrections. Generically, however, the flow and transport properties of the fluid can affect the fuel consumption process and result in active systems becoming mechanosensitive \cite{J_licher_2018, JULICHER20073, PhysRevLett.92.078101, Callan-Jones_2011}. These effects can be taken into account by introducing generic hydrodynamic corrections in $r_\fuel$, similar to how we have derived corrections to $r_\env$ in this work. Mechanosensitivity is expected to be a general feature of certain classes of active systems, playing a role in the cytoskeleton and cell motility \cite{JULICHER20073}. We expect to return to this topic and single out generic features of this fuel consumption process in energy correlation functions.

A related problem of interest is to understand how active matter behaves in complex environments \cite{10.3389/fphy.2022.1005146}. This question is relevant not only for cell motility and its role in various biological processes, but also for understanding bacterial growth and collective motion under varying external conditions. The formalism we introduced here allows to model time and spatially modulated environment sources by giving a non-trivial profile to $\Phi^\env_r$. A concrete application is to study the collective motion of active matter when subjected to an external temperature gradient. We leave these interesting questions for future work.

\acknowledgments

We would like to thank Amin Doostmohammadi, Dominik Hahn, Ananyo Maitra, and Sriram Ramaswamy for helpful discussions. The authors are partly supported by the
Dutch Institute for Emergent Phenomena (DIEP) cluster at the University of Amsterdam and JA
via the DIEP programme Foundations and Applications of Emergence (FAEME). The work of
AJ was partly funded by the European Union’s Horizon 2020 research and innovation programme
under the Marie Skłodowska-Curie grant agreement NonEqbSK No. 101027527. Part of this project
was carried out during the ``Hydrodynamics at All Scales'' workshop at the Nordic Institute for
Theoretical Physics (NORDITA), Stockholm.

\clearpage
\appendix 

\onecolumngrid

\renewcommand{\thesection}{\Alph{section}}
\renewcommand{\thesubsection}{\thesection.\arabic{subsection}}
\renewcommand{\thesubsubsection}{\thesubsection.\arabic{subsubsection}}
\makeatletter
\renewcommand{\p@section}{}
\renewcommand{\p@subsection}{}
\renewcommand{\p@subsubsection}{}
\makeatother


\section{Chemical fuel and comparison to Julicher et al.}
\label{app:Julicher}

In this appendix we carefully review the work of \cite{2018RPPh...81g6601J} and show how it can fit into our work. We note, however, that this framework does not take into account environmental sinks and thus does not host non-equilibrium steady states. Let us start with the conservation equations for total energy, momentum, and number densities of chemical ingredients as outlined in~\cite{2018RPPh...81g6601J}, i.e.
\begin{align}
    \dow_t\epsilon_\tot + \dow_i\epsilon_\tot^i &= 0, \nn\\
    \dow_t\pi^j +\dow_i\tau^{ij} 
    &= 0, \nn\\
    \dow_tn_{\alpha} + \dow_ij^i_{\alpha} &= r_\alpha,
\end{align}
where $\alpha=0,\ldots,N$ labels different chemical species, characterised by constant mass per unit particle $m^\alpha$. 
The mass and momentum densities are given as $\rho=m^\alpha n_\alpha$ and $\pi^i=m^\alpha j^i_\alpha$. Given that the total mass is conserved, we have $m^\alpha r_\alpha = 0$, leading to
\begin{align}
    \dow_t\rho + \dow_i\pi^i &= 0.
\end{align}
We can decompose the rates into $r_\alpha = -r_I a^I_{\alpha}$, where independent rates $r_I$ are the independent rates of chemical reactions and $a^I_{\alpha}$ are stoichiometric coefficients. Thermodynamic relations can be summarised as
\begin{align}
    \df\epsilon_\tot 
    &= T\df s 
    + \lb\mu^\alpha - \half m^\alpha\vec u^2\rb \df n_\alpha 
    + u^i\df\pi_i,  \qquad
    p
    = Ts + \lb\mu^\alpha - \half m^\alpha\vec u^2\rb n_\alpha + u^i\pi_i
    -\epsilon_\tot.
\end{align}
Due to Galilean symmetry, we can also express this in terms of the total internal energy density $\varepsilon_\tot = \epsilon_\tot - \half \rho \vec u^2$, i.e.
\begin{align}
    \df\varepsilon_\tot
    &= T\df s 
    + \mu^\alpha \df n_\alpha, \qquad
    p
    = Ts 
    + \mu^\alpha n_\alpha
    - \varepsilon_\tot.
\end{align}
From here we can derive
\begin{align}
    \dow_t s + \dow_i s^i 
    &= - \frac{1}{T} \lb s^i - s\, u^i \rb \dow_iT
    - \frac1T\Big(\tau^{ij}-p\,\delta^{ij}-\rho u^iu^j\Big)\dow_iu_j 
    - \frac1T\Big(j^i_\alpha - n_\alpha u^i\Big)\dow_i\mu^\alpha
    + \frac{1}{T} r_I \Delta\mu^I \geq 0,
\end{align}
where we have identified the heat current
\begin{align}
    Ts^i = \epsilon^i_{\text{tot}} + \half \vec u^2\pi^i -\tau^{ij}u_j
- \mu^\alpha j^i_\alpha + p\,u^i,
\end{align}
and defined $\Delta\mu^I = a^I_{\alpha}\mu^\alpha$. From here we can read out the constitutive relations consistent with the second law of thermodynamics. In particular, we find the rates
\begin{align}
    r_I = M_{IJ} \Delta\mu^J,
\end{align}
for positive semi-definite matrix $M_{IJ}$. Using these, we can obtain temperature dynamics from energy conservation
\begin{align}
    \dow_t T 
    &=
    \frac{1}{c_v} 
    \lb \Delta\mu^I
    + T a^I_{\alpha}
    \frac{\dow s}{\dow n_\alpha}\bigg|_T
    \rb M_{IJ}\Delta\mu^J
    + \text{gradient terms}.
\end{align}
where $c_v = T\dow s/\dow T|_{n_\alpha}$ is the specific heat. This results in a steady increase of temperature when the activity parameters $\Delta\mu^I$ are turned on. On the other hand, when an energy sink is introduced as in \cref{sec:fuelconsumption}, we achieve a steady state at
\begin{align}
    T_0
    = T_\env \lb 1 + \frac{1}{\ell^2\kB^2T_\env^2\gamma_\env}
    \lb \Delta\mu^I
    + T a^I_{\alpha}
    \frac{\dow s}{\dow n_\alpha}\bigg|_T \rb
    M_{IJ}\Delta\mu^J\rb.
\end{align}

To make the mapping to the formalism presented in \cref{sec:fuelconsumption} more precise, note that only one linear combination of number densities, i.e. the mass density $\rho$, is conserved. Therefore, we can arbitrarily redefine $n_{\alpha} \to n_{\alpha} + g_\alpha(\varepsilon_\tot,n_\alpha)$ for arbitrary functions $g_\alpha(\rho,\varepsilon_\tot)$, such that $m^\alpha g_\alpha = 0$. The conservation equations take the same form after appropriately modifying the definitions of $j^i_\alpha$ and $r_\alpha$.  This also modifies the definitions of 
\begin{align}
    \frac{1}{T}\to \frac{1}{T}
    \lb 1 + \mu_\alpha\cJ_{\alpha\beta}\frac{\dow g_\beta}{\dow\varepsilon_\tot}\rb, \qquad 
    \frac{\mu_\alpha}{T}
    \to \frac{\mu_\beta}{T}\cJ_{\beta\alpha}.
\end{align}
where $\cJ_{\alpha\beta}$ is the inverse of $\delta_{\alpha\beta}+\dow g_\alpha/\dow n_\beta$. This hints at the inherent ambiguity in defining temperature in active systems. To fix this ambiguity, let us isolate the mass density contributions in the thermodynamic relations, leading to
\begin{align}
    \df\varepsilon_\tot
    &= T\df s 
    + \varpi_\tot \df \rho
    + \Delta \mu^I \df n_I, \qquad
    p
    = Ts + \varpi_\tot\rho
    + \Delta\mu^I n_I
    - \varepsilon_\tot.
\end{align}
where $n_I = \bar a_I^\alpha n_\alpha$ are independent non-conserved particle densities and $\varpi_\tot = \mu_\alpha m^\alpha/m^2$ is the total mass chemical potential. Here $\bar a_I^\alpha$ denotes the ``inverse stoichiometric coefficients'', satisfying the relations $\bar a_I^\alpha a^I_\beta = \delta^\alpha_\beta - m^\alpha m_\beta/m^2$ and $\bar a_I^\alpha a^J_\alpha = \delta_I^J$. In particular, using the last relation, we have that $\varpi_\tot=\varpi_\tot(p,T,\Delta\mu_I)$. To fix the aforementioned ambiguity, we require that the total energy density splits cleanly into a system part and a fuel part
\begin{align}
    \epsilon_\tot(s,n_\alpha,\pi_i)
    = \epsilon(s,\rho,\pi_i) + \rho\,\varepsilon_\fuel(n_I/\rho), \qquad
    \varepsilon_\tot(s,n_\alpha)
    = \varepsilon(s,\rho) + \rho\,\varepsilon_\fuel(n_I/\rho),
\end{align}
where we have defined $\epsilon=\varepsilon+\half\rho\vec u^2$. The parametrisation is chosen such that the mass chemical potential splits into
\begin{align}
    \varpi_\tot(p,T,\bar\mu_\alpha) = \varpi(p,T) + \varepsilon_\fuel 
    - \frac{n_I}{\rho}\Delta\mu^I,
\end{align}
along with the thermodynamic relations
\begin{gather}
    \df\varepsilon = T\df s + \varpi \df\rho, \qquad 
    p = Ts + \varpi\rho - \varepsilon, \qquad 
    \df\varepsilon_\fuel = \Delta\mu^I
    \df\frac{n_I}{\rho}.
\end{gather}
It is easy to check that the energy and entropy balance equations now become
\begin{align}
    \dow_t\epsilon 
    + \dow_i\epsilon^i
    &= 
    \ell r_\fuel \Delta E, \nn\\
    \dow_t s + \dow_i s^i 
    &= - \frac{1}{T} \lb s^i - s\, u^i \rb \dow_iT
    - \frac1T\Big(\tau^{ij}-p\,\delta^{ij}-\rho u^iu^j\Big)\dow_iu_j 
    + \frac{\ell}{T} r_\fuel \Delta E,
\end{align}
together with
\begin{align}
    \epsilon^i_\tot = \epsilon^i + \varepsilon_\fuel \pi^i 
    + \Delta\mu^I \lb j^i_I - n_I u^i\rb, \qquad
    Ts^i = \epsilon^i 
    + \half \vec u^2\pi^i -\tau^{ij}u_j
    - \varpi\pi^i + p\,u^i.
\end{align}
We isolate the constant $\ell\Delta E$ representing the strength of activity, by defining $\Delta\mu^I = \nu^I\ell\Delta E$. In terms of this, the net fuel burning rate is identified as
\begin{align}
    r_\fuel
    = r_I\nu^I
    - \lb j^i_I - n_I u^i\rb \dow_i\nu^I.
\end{align}
Note that momentum and mass conservation equations remain unaltered.
With this parametrisation, one find that $a^I_{\alpha}\dow s/\dow n_\alpha|_T=0$ and thus 
\begin{align}
    \gamma_\fuel
    = \nu^I M_{IJ}\nu^J.
\end{align}
We have hence recovered the formalism presented in the main text, without the energy sink. However, besides the inclusion of a sink mechanism, the formalism presented in the main text also poses two additional benefits: it is operationally simpler as one does not need to keep track of individual chemical species and it is easily generalisable to other kinds of activity sources such as electric of mechanical driving.

\section{Details of Schwinger-Keldysh formalism}
\label{app:SK}

In this appendix we give the details of the Schwinger-Keldysh (SK) formalism for active systems. The following discussion is a straightforward generalisation of non-relativistic SK hydrodynamics~\cite{Jain:2020vgc,Armas:2020mpr} to include background thermal bath fields; see also~\cite{Crossley:2015evo, Haehl:2018lcu, Jensen:2017kzi} for the original relativistic formulation. We begin with a brief review of non-relativistic spacetime geometries in \cref{app:NC}, which provide the appropriate background sources for conserved currents associated with spacetime symmetries. We then give a self-contained review of the SK formalism and KMS symmetry for passive hydrodynamics in \cref{app:passive-sk}, followed by the extension to include fuel and environment background fields in \cref{sec:activity-in-SK}. Finally, in \cref{app:2ndlaw,app:probs} we provide proofs of the (active) second law of thermodynamics and microscopic reversibility from the SK-EFT framework. The discussion in this appendix will be mostly formal; the actual construction of the SK-EFT effective actions has been relegated to \cref{app:examples}.

We will employ a covariant notation for spacetime coordinates $(x^\mu)=(t,x^i)$, with the spacetime indices running over $\mu,\nu,\ldots=0,1,2,\ldots$ and the spatial indices running over $i,j,\ldots=1,2,\ldots$. Later in the SK construction in \cref{app:passive-sk}, we will also introduce worldvolume coordinates $(\sigma^\alpha)$, with the indices running over $\alpha,\beta,\ldots=0,1,2,\ldots$. Importantly, despite the covariant notation, our construction is entirely non-relativistic.

\subsection{Non-relativistic geometries}
\label{app:NC}

We are interested in physical systems that feature global symmetries: time translations, spatial translations, spatial rotations, U(1) transformations, and possibly boosts. To keep track of the respective conserved currents, it is convenient to introduce the associated background fields. In non-relativistic setting without necessarily a boost symmetry, the appropriate structure is given by a curved \emph{Aristotelian background}~\cite{deBoer:2020xlc, Novak:2019wqg, Armas:2020mpr}. It features a clock-form $n_{\mu}$ coupled to the energy current $\epsilon^\mu$, a degenerate symmetric spatial metric $h_{\mu\nu}$ coupled collectively to the momentum density $\pi^\mu$ (s.t. $\pi^\mu n_\mu = 0$) and the symmetric stress tensor $\tau^{\mu\nu}$ (s.t. $\tau^{\mu\nu}n_\nu = 0$), and a U(1) gauge field $A_{\mu}$ coupled to the charge current $j^\mu$. Since $h_{\mu\nu}$ is degenerate, there exists a unique vector field $v^\mu$ such that $v^\mu h_{\mu\nu}=0$, normalised as $v^\mu n_{\mu}=1$, which represents the preferred observer with respect to whom the notions of space and time are defined. We can also define an ``inverse'' spatial metric $h^{\mu\nu}$, satisfying $h_\mu^\nu \equiv h^{\mu\rho} h_{\rho\nu}=\delta^\mu_\nu - v^\mu n_\nu$ and $n_\mu h^{\mu\nu}=0$. Tensorial indices can be raised and lowered by contractions with $h^{\mu\nu}$ and $h_{\mu\nu}$ respectively. However, note that these operations are not generically invertible, e.g. $X_\mu h^{\mu\nu}h_{\nu\rho}\neq X_\rho$ for some $X_\mu$.
We may introduce a spacetime connection on Aristotelian backgrounds given as
\begin{equation}\label{eq:connection}
    \Gamma^\lambda_{\mu\nu} = v^\lambda \partial_\mu n_\nu + \frac{1}{2} h^{\lambda\rho}\left( \partial_\mu h_{\nu\rho} + \partial_\nu h_{\mu\rho} - \partial_\rho h_{\mu\nu} \right)~,
\end{equation} 
used to define a covariant derivative $\nabla_\mu$ that acts on a mixed-index tensor as $\nabla_\mu X^\lambda_{~\nu} = \dow_\mu X^\lambda_{~\nu} + \Gamma^\lambda_{\mu\sigma}X^\sigma_{~\nu} - \Gamma^\sigma_{\mu\nu}X^\lambda_{~\sigma}$, and similarly for higher-rank tensors. Note that $\nabla_\mu n_\nu, \nabla_\mu h^{\nu\rho}=0$, but $\nabla_\mu v^\nu, \nabla_\mu h_{\nu\rho}\neq 0$.
It is also convenient to define a gradient operator $\nabla'_\mu = \nabla_\mu + F^n_{\mu\nu}v^\nu$, where $F^n_{\mu\nu} = 2\dow_{[\mu}n_{\nu]}$ and analogously $F_{\mu\nu} = 2\dow_{[\mu}A_{\nu]}$. This has the property that $\nabla'_\mu X^\mu = 1/\sqrt{\gamma}\,\dow_\mu(\sqrt{\gamma}\,X^\mu)$, with $\gamma = \det(n_\mu n_\nu + h_{\mu\nu})$. 

The action of global symmetries can simply be realised as background diffeomorphisms $x'^\mu(x)$ and gauge transformations $\Lambda(x)$ that act on various background fields as usual. In flat spacetime, the Aristotelian background fields take the trivial values $n_\mu,v^\mu = (1,\vec 0)$, $h_{\mu\nu},h^{\mu\nu} = ((0,\vec 0),(\vec 0,\delta_{ij}))$, $A_\mu=0$, which is only invariant under the global part of the transformations given as
\begin{align}\label{eq:arist-trans}
    t'(x) &= t + a^t~, \nn\\
    x'^i(x) &= \Lambda^i_{~j}\lb x^j + a^j\rb~, \nn\\
    \Lambda(x) &= \Lambda~,
\end{align}
where $a^t$, $a^i$ are the parameters of spacetime translations, $\Lambda^i_{~j}\in\SO(d)$ of rotations, and $\Lambda$ of global U(1) transformations. This set does not contain the boost symmetry, which needs to be supplied separately when relevant.

To model active systems, we will also introduce the fuel and environment background fields $\Phi_{\fuel,\env}$ coupled to the rate operators $r_{\fuel,\env}$ introduced in \cref{sec:active-formalism}

Given the effective action $S$ of the theory, the conserved currents can be defined via taking variations with respect to the background fields
\begin{equation}\label{eq:action-variation-basic}
    \delta S = \int_x 
    - \epsilon^\mu\delta n_\mu 
    + \biggl(v^{(\mu}\pi^{\nu)}+\frac{1}{2}\tau^{\mu\nu}\biggr)\delta h_{\mu\nu}
    + j^\mu\delta A_\mu 
    \blue{\,+\,\ell r_\env\Phi^\env 
    + \ell r_\fuel \Phi^\fuel}~,
\end{equation}
where $\int_x$ is short-hand notation for the integral measure $\int\df^{d+1}x\sqrt{\gamma}$. We will denote the active contributions arising from the external baths in \blue{blue} for emphasis.
 Requiring the symmetry transformations to leave the action $S$ invariant, we are led to the conservation equations
\begin{align}\label{eq:curved-conservation}
    \nabla_\mu' \epsilon^\mu 
    &= -v^\mu  f_\mu 
    - \tau^{\mu\nu} h_{\lambda\nu}\nabla_\mu v^\lambda~, \nn\\ 
    \nabla_\mu'\!\lb v^\mu \pi^\nu 
    + \tau^{\mu\nu} \rb
    &= h^{\nu\mu} f_\mu 
    - \pi_\mu h^{\nu\lambda}\nabla_\lambda v^\mu~, \nn\\ 
    \nabla'_\mu j^\mu &=0~,
\end{align}
where we have identified the forces $f_\mu = -F^n_{\mu\nu} \epsilon^\nu + F_{\mu\nu}j^\nu \blue{\,+\,\ell r_\env\dow_\mu\Phi_{\env}+\ell r_\fuel\dow_\mu\Phi_{\fuel}}$. See~\cite{deBoer:2020xlc, Novak:2019wqg, Armas:2020mpr} for more details on these conservation equations.

To describe a relativistic/Galilean system, we need to additionally impose the Lorentzian/Galilean boost symmetry on the background sources respectively. Specifically for Galilean systems, this is realised via the Milne transformations
\begin{align}\label{eq:Galilean-boost}
    v^\mu &\to v^\mu + b^\mu~, \nn\\
    h_{\mu\nu} &\to h_{\mu\nu} - 2n_{(\mu}b_{\nu)} + n_\mu n_\nu b^2~, \nn\\
    A_\mu &\to A_\mu + b_\mu - \half n_\mu b^2~,
\end{align}
where $b^\mu$ is the boost parameter satisfying $b^\mu n_\mu = 0$, and we have defined $b_\mu = h_{\mu\nu}b^\nu$, $b^2 = b^\mu b_\mu$. The background fields $n_\mu$ and $h^{\mu\nu}$ are Milne-invariant. The Aristotelian geometric structure with Milne transformations is called a Newton-Cartan geometry~\cite{Cartan:1923zea, Cartan:1924yea}. Specialising to the flat background described before \cref{eq:arist-trans}, these transformations act on the spatial coordinates as
\begin{align}
    x^i \to x^i - b^i t~,
\end{align}
as expected.
Requiring the action in \cref{eq:action-variation-basic} to be Milne-invariant sets the charge flux equal to the momentum density, i.e.
\begin{align}
    j^\mu h_{\mu\nu} = \pi_\nu~,
\end{align}
as we expect for a Galilean system. 
We do not concern ourselves with relativistic systems in this work. Note that the connection \eqref{eq:connection} is not Milne-invariant. In fact, no connection can be constructed out of the geometric structure at hand that is simultaneously Milne- and U(1)-invariant. However, one may write such a connection if provided with a Milne-invariant vector field, which in our case can be provided by the covariant fluid velocity $u^\mu$, normalised as $u^\mu n_\mu =1$. The Milne-invariant connection can then be written as~\cite{Jensen:2014ama}
\begin{align}\label{eq:gal-connection}
    \Gamma^{\text g}{}^\lambda_{\mu\nu} 
    &= v^\lambda \partial_\mu n_\nu + \frac{1}{2} h^{\lambda\rho}\left( \partial_\mu h_{\nu\rho} + \partial_\nu h_{\mu\rho} - \partial_\rho h_{\mu\nu} \right)
    + n_{(\mu}F_{\nu)\rho} h^{\rho\lambda}
    + \half \vec u^\lambda F^n_{\mu\nu}
    - \lb \vec u_{(\mu} - \half n_{(\mu} \vec u^2 \rb F^n_{\nu)\rho} h^{\rho\lambda}~,
\end{align}
with the associated covariant derivative operator $\Ng_\mu$ and the gradient operator $\nabla'^{\rm g}_\mu = \Ng_\mu + F^n_{\mu\nu}u^\nu$.
Note that this connection is not a purely geometric object, but may be used to make the Galilean symmetry manifest where required. More details can be found in~\cite{Jensen:2014aia, Jensen:2014ama, Jain:2020vgc, Christensen:2013lma, Bergshoeff:2014uea, Festuccia:2016awg, Bergshoeff:2017dqq}.

\subsection{Passive Schwinger-Keldysh formalism}
\label{app:passive-sk}

We will start with a lightning review of the SK formalism for passive non-relativistic hydrodynamics constructed in~\cite{Jain:2020vgc,Armas:2020mpr}. Readers familiar with the formalism may skim this discussion for the notation being used and proceed directly to \cref{sec:activity-in-SK}.

\paragraph{Fluid worldvolume, dynamical fields, and global symmetries:} SK field theories are defined on a closed-time contour, with leg ``1'' going forward in time and leg ``2'' returning backward in time to the initial state. Each leg of the contour is equipped with its own set of degrees of freedom and background fields. SK hydrodynamics is set up as a sigma-model on an auxiliary ``worldvolume'' with coordinates $\sigma^\alpha$, where the dynamical fields live. To describe a passive fluid featuring conserved energy, momentum, and charge, the dynamical field content consists of a pair of coordinate fields $X^\mu_{1,2}(\sigma)$ defining two copies of ``SK spacetime'' and a pair of U(1) phase fields $\varphi_{1,2}(\sigma)$. The subscripts label the respective legs of the contour.  Depending on the system under consideration, each leg of the contour may also feature certain additional fields, for instance order parameters associated with spontaneous symmetry breaking, that we shall return to for specific applications.

SK hydrodynamics realises the global spacetime and internal symmetries independently on the two spacetimes. Each global symmetry comes in pairs and acts on each SK spacetime independently, i.e.
\begin{align}
    X^t_{1,2} &\to X^t_{1,2} + a^t_{1,2}~, \nn\\
    X^i_{1,2} &\to \Lambda_{1,2}{}^i_{~j} \lb X^j_{1,2} + a^j_{1,2} - b^j_{1,2} X^t_{1,2} \rb~, \nn\\
    \varphi_{1,2} &\to \varphi_{1,2} - \Lambda_{1,2}~,
\end{align}
We may also replace Galilean boosts with Lorentz boosts for relativistic systems, or skip them altogether for boost-agnostic systems~\cite{deBoer:2020xlc, Novak:2019wqg, Armas:2020mpr}. It is convenient to gauge these symmetries by introducing the Aristotelian background fields $n_{1,2\mu}(X_{1,2})$, $h_{1,2\mu\nu}(X_{1,2})$, and $A_{1,2\mu}(X_{1,2})$ on each SK spacetime, reviewed in \cref{app:NC}, that transform under the global symmetries as usual. The action on the dynamical fields, on the other hand, is now given as simply
\begin{align}\label{eq:global-symm}
    X^\mu_{1,2}
    &\to X'^{\mu}_{1,2}(X_{1,2})~, \nn\\
    \varphi_{1,2}
    &\to \varphi_{1,2}
    - \Lambda_{1,2}(X_{1,2})~.
\end{align}
In practice, we can define objects that are invariant under global symmetries as simple pullbacks of background fields onto the fluid worldvolume, i.e.
\begin{align}
    \bbn_{1,2\alpha}
    &= n_{1,2\mu}(X_{1,2})\, 
    \dow_\alpha X^\mu_{1,2}~, \nn\\
    \bbh_{1,2\alpha\beta}
    &= h_{1,2\mu\nu}(X_{1,2})\, 
    \dow_\alpha X^\mu_{1,2}\,
    \dow_\beta X^\nu_{1,2}~, \nn\\
    \bbA_{1,2\alpha}
    &= A_{1,2\mu} (X_{1,2})\, 
    \dow_\alpha X^\mu_{1,2}
    + \dow_\alpha \varphi_{1,2}~.
    \label{eq:global-invariants-hydro}
\end{align}
We have dropped the explicit dependence on worldvolume coordinates for clarity.

It is useful to introduce an average-difference basis $f_r = (f_1+f_2)/2$, $f_a = (f_1-f_2)/\hbar$ for various dynamical and background fields. The average ``$r$'' combinations are understood as the ``physical'' macroscopic fields, while the difference ``$a$' combinations as the stochastic noise associated with them. 


\paragraph{Fluid worldvolume symmetries:} In addition to the global symmetries above, we also impose local diffeomorphisms $\sigma'^\alpha(\sigma)$ and U(1) gauge transformations $\lambda(\sigma)$ of the worldvolume acting on the dynamical fields as
\begin{align}
    X^\mu_{1,2}(\sigma) 
    &\to X'^\mu_{1,2}(\sigma'(\sigma))
    = X^\mu_{1,2}(\sigma), \nn\\
    \varphi_{1,2}(\sigma) 
    &\to \varphi'_{1,2}(\sigma'(\sigma)) 
    = \varphi_{1,2}(\sigma) + \lambda(\sigma).
\end{align}
All the global symmetry invariants in \cref{eq:global-invariants-hydro} are invariant under worldvolume gauge transformations, except for $\bbA_{r\alpha}$ that transforms as $\bbA_{r\alpha} \to \bbA_{r\alpha} + \dow_\alpha\lambda$. Worldvolume diffeomorphisms act on all global symmetry invariants as usual.

The effective theory is also endowed with a thermal vector $\bbbeta^\alpha(\sigma)$ and a chemical shift ${\mathbb\Lambda}_{\bbbeta}(\sigma)$, transforming as
\begin{align}
    \bbbeta^\alpha(\sigma)
    &\to \bbbeta'^\alpha(\sigma')
    = \bbbeta^\beta(\sigma)\dow_\beta\sigma'^\alpha(\sigma) ~, \nn\\ 
    {\mathbb\Lambda}_{\bbbeta}(\sigma)
    &\to {\mathbb\Lambda}'_{\bbbeta}(\sigma')
    = {\mathbb\Lambda}_{\bbbeta}(\sigma) - \bbbeta^\alpha\dow_\alpha\lambda(\sigma)~.
\end{align}
Without loss of generality, we may partially fix the worldvolume symmetries to set $\bbbeta^\alpha = (\beta_0,\vec 0)$ and ${\mathbb\Lambda}_{\bbbeta} = \beta_0\mu_0$, where $\beta_0=(\kB T_0)^{-1}$ is some reference global temperature and $\mu_0$ is some reference chemical potential.

\paragraph{Physical spacetime formulation:} The worldvolume picture of SK hydrodynamics with two copies of spacetimes is theoretically neat and appealing. However, for practical purposes, it is more transparent to move to a single physical spacetime formulation, defined via $x^\mu = X_r^\mu(\sigma)$. We can use pullbacks with respect to this map to define objects that are invariant under the fluid worldvolume diffeomorphisms.  Due to the non-linear nature of the theory, the relations between the average-difference quantities on the worldvolume and those on the spacetime are quite non-trivial. However, the relations simplify in the classical ($\hbar\to 0$) limit, i.e.
\begin{alignat}{2}\label{eq:physical-st-invariants}
    N_{r\mu} 
    &= \frac{\dow\sigma^\alpha}{\dow x^\mu}
    \bbn_{r\alpha}
    = n_{r\mu} + \cO(\hbar)~, &\qquad
    N_{a\mu} 
    &= \frac{\dow\sigma^\alpha}{\dow x^\mu}
    \bbn_{a\alpha}
    = n_{a\mu} + \lie_{X_a} n_{r\mu}
    + \cO(\hbar)~, \nn\\
    H_{r\mu\nu} 
    &= \frac{\dow\sigma^\alpha}{\dow x^\mu}
    \frac{\dow\sigma^\beta}{\dow x^\nu}
    \bbh_{r\alpha\beta}
    = h_{r\mu\nu} + \cO(\hbar)~, &\qquad
    H_{a\mu\nu} 
    &= \frac{\dow\sigma^\alpha}{\dow x^\mu}
    \frac{\dow\sigma^\beta}{\dow x^\nu}
    \bbh_{a\alpha\beta}
    = h_{a\mu\nu} + \lie_{X_a} h_{r\mu\nu}
    + \cO(\hbar)~, \nn\\
    B_{r\mu}
    &= \frac{\dow\sigma^\alpha}{\dow x^\mu}
    \bbA_{r\alpha}
    = A_{r\mu} + \dow_\mu\varphi_r + \cO(\hbar)~, &\qquad
    B_{a\mu}
    &= \frac{\dow\sigma^\alpha}{\dow x^\mu}
    \bbA_{a\alpha}
    = A_{a\mu} + \dow_\mu\varphi_a
    + \lie_{X_a} A_{r\mu} + \cO(\hbar)~,
\end{alignat}
up to quantum corrections, where $\lie_{X_a}$ denotes a Lie derivative along $X_a^\mu$. These quantities will be used as building blocks to construct the SK effective action. The fluid worldvolume gauge transformations become the diagonal spatial shift symmetry on the physical spacetime, acting only on $B_{r\mu}$ among these building blocks, i.e. $B_{r\mu} \to B_{r\mu} + \dow_\mu\lambda$. We can define the physical spacetime thermal vector $\beta^\mu$ and chemical shift $\Lambda_\beta$ as
\begin{align}
    \beta^\mu &= \bbbeta^\alpha \frac{\dow x^\mu}{\dow \sigma^\alpha(x)}~, \nn\\
    \Lambda_\beta  &= \beta^\mu \dow_{\mu}\phi_r + {\mathbb\Lambda}_{\bbbeta} ~,
    \label{eq:hydro-beta}
\end{align}
These objects are dynamical and can be used to define the local temperature $\kB T=1/\beta$, fluid velocity $u^\mu=\beta^\mu/\beta$, and chemical potential $\mu = (\Lambda_\beta + \beta^\mu A_{r\mu})/\beta$, where $\beta = \sqrt{\beta^\mu n_{r\mu}}$.
The compromise for passing onto the physical spacetime formulation is that the $r$-part of the physical spacetime diffeomorphisms \eqref{eq:global-symm} becomes non-manifest and takes the form $x^\mu \to x'^\mu(x)$, which acts on various fields on the physical spacetime as diffeomorphisms according to their tensor structure.

For convenience, we introduce the collective notation
\begin{alignat}{2}
    \text{Dynamical fields:}&\quad& \psi
    &= \lb~ \beta^\mu~,~ \Lambda_\beta~\rb~, \qquad 
    \psi_a = \lb~ X^\mu_{r,a}~,~\varphi_{r,a}~\rb~,\nn\\
    \text{Currents:}&\quad& \cO_{r,a}
    &= \lb~\epsilon^\mu_{r,a}~,~
    2v^{(\mu}_r\pi^{\nu)}_{r,a}+\tau^{\mu\nu}_{r,a}~,~
    j^\mu_{r,a}~\rb~, \nn\\
    \text{Background fields:}&\quad& {\sf s}_{r,a}
    &=\lb~ {-n}_{r,a\mu}~,~{\textstyle\half} h_{r,a\mu\nu}~,~A_{r,a\mu}~\rb~, \nn\\
    \text{Global symmetry invariants:}&\quad& \Psi_{r,a}
    &=\lb~ {-N}_{r,a\mu}~,~{\textstyle\half}H_{r,a\mu\nu}~,~B_{r,a\mu}~\rb~,
\end{alignat}
which will be useful in the forthcoming discussion. These sets are appropriately extended in the presence of additional degrees of freedom parametrising spontaneous symmetry breaking.

For later use, let us record the action of global symmetries \eqref{eq:global-symm} on various dynamical and background fields. As noted above, the $r$-type diffeomorphisms in \cref{eq:global-symm} act as usual on all the fields. Whereas, $a$-type diffeomorphisms only act on the $a$-type fields, i.e.
\begin{align}\label{eq:a-diffeo}
    \sfs_{a} \to \sfs_a + \lie_{\chi_a} \sfs_r~, \qquad 
    X^\mu_a \to X^\mu_a - \chi^\mu_a~,
\end{align}
while leaving $\varphi_a$ invariant, where $\chi_a^\mu$ denotes the parameter of $a$-type diffeomorphisms. The $r$-type and $a$-type gauge transformations act on the U(1) sector as
\begin{align}
    A_{r,a\mu} 
    &\to A_{r,a\mu} + \dow_\mu\Lambda_{r,a}~, \qquad 
    \varphi_{r} \to \varphi_r - \Lambda_r~, \qquad
    \varphi_a
    \to \varphi_a -\Lambda_a - X_a^\mu \dow_\mu \Lambda_r~.
\end{align}
One may check that these symmetries leave the quantities in \cref{eq:physical-st-invariants} invariant.

\paragraph{SK generating functional:} The fundamental object of interest in non-equilibrium field theory is the SK generating functional $\cZ[{\sf s}_r,{\sf s}_a]$, which can be used to probe various non-equilibrium operators via the variational formulae such as those in \cref{eq:corr-from-gf}. In SK hydrodynamics, the generating functional is obtained by performing a path integral of the effective action $S[\psi,\psi_a,\sfs_r,\sfs_a]$, as in
\begin{equation}\label{eq:Z-def}
    {\cal Z}[{\sf s}_r,{\sf s}_a]
    = \int\cD\psi\,\cD\psi_a
    \exp\!\Big(iS[\psi,\psi_a,\sfs_r,\sfs_a]\Big)~.
\end{equation}
In practice, the effective action may be construct using the building blocks $\Psi_{r,a}$, $u^\mu$, $T$, and $\mu$, arranged order-by-order in derivatives.
Varying the effective action with respect to the two types of sources, we can read out the associated operators
\begin{equation}
    \delta S = \int_x \cO_r\cdot \delta{\sf s}_a 
    + \cO_a\cdot \delta{\sf s}_r~.
\end{equation}
The $\cO_r$ operators (obtained by varying with respect to ${\sf s}_a$) can be understood as physical, while the $\cO_a$ operators (obtained by varying with respect to ${\sf s}_r$) as their stochastic noise counterparts. The classical constitutive relations are given by $\cO = \cO_r|_{f_a\to 0}$. One may check that the associated conservation equations \eqref{eq:curved-conservation} are obtained by extremising the action with respect to $\psi_a$ dynamical fields, while varying with respect to $\psi$ fields yields the equivalent equations for the stochastic noise.

The SK generating functional is required to satisfy the following three conditions
\begin{equation}
    \cZ[{\sf s}_r,{\sf s}_a = 0] = 1, \qquad 
    \cZ[{\sf s}_r,-{\sf s}_a] = \cZ^*[{\sf s}_r,{\sf s}_a]~, \qquad 
    \Re\cZ[{\sf s}_r,{\sf s}_a] \leq 0~,
    \label{eq:SK-conditions-pf}
\end{equation}
arising from generic properties of quantum field theories defined on a closed-time contour.
More details regarding the underlying physics can be found in the review of~\cite{Glorioso:2018wxw}. These conditions can naturally be extended to the effective action as given in \cref{eq:SK-conds}.
We can arrange $S$ as a series in powers of $\Psi_a$, in which case the three conditions mean that: $S$ must at least be linear in $\Psi_a$, the terms with even-powers of $\Psi_a$ must be imaginary, and these imaginary terms must be arranged into a quadratic form with non-negative coefficients.

\paragraph{Dynamical KMS symmetry:} While the SK conditions \eqref{eq:SK-conditions-pf} are satisfied for arbitrary non-equilibrium field theories defined on a closed-time contour, the SK generating functional $\cZ[{\sf s}_r,{\sf s}_a]$ for a thermal system is also required to satisfy a discrete dynamical Kubo-Martin-Schwinger (KMS) symmetry.
Let us assume that the microscopic theory underlying our hydrodynamic features a discrete symmetry $\Theta$ that includes the time-reversal transformation T. Depending on the physical system under consideration, this could be just T, or some combination involving spatial-parity charge-conjugation transformation such as PT, CT, CPT. The KMS symmetry is defined as the invariance of the Schwinger-Keldysh generating functional $\cZ[{\sf s}_r,{\sf s}_a]$ under a transformation of the background fields~\footnote{We use the convention $\Theta f(x+z) = \eta^\Theta_f f(\Theta x + z^*)$, where $\eta^\Theta_f$ is the $\Theta$-eigenvalue of $f$, and we have defined $\Theta t = -t$ and $\Theta x^i =\pm x^i$ depending on whether $\Theta$ includes the spatial parity transformation P. Similarly, $\Theta \dow f(x+z) = \eta^\Theta_{\dow f} \dow f(\Theta x + z^*)$.}
\begin{gather}
    {\sf s}_{1}(x)
    \KMSto \Theta{\sf s}_{1}(x)~, \qquad
    {\sf s}_{2}(x)
    \KMSto \Theta{\sf s}_{2}(x + i\hbar\,\Theta\beta_0)~,
    \label{eq:KMS-full}
\end{gather}
where the $\Theta{\sf s}_2$ is evaluated on the complex spacetime arguments $x^\mu + i\hbar\,\Theta\beta^\mu_0$, with $\beta^\mu_0 = u_0^\mu/(\kB T_0)$ being the thermal vector associated with the inertial equilibrium observer. More details can be found in e.g.~\cite{Bellac:2011kqa,Glorioso:2018wxw}. For classical non-equilibrium field theories valid at small frequencies $\omega \ll \kB T_0/\hbar$, the classical truncation of the KMS symmetry is more relevant, given by
\begin{align}
    {\sf s}_{r}
    &\KMSto \Theta{\sf s}_{r} + \cO(\hbar)~, \qquad 
    {\sf s}_{a}
    \KMSto 
    \Theta\hat{\sf s}_a \equiv
    \Theta\lb{\sf s}_{a}
    + i\beta_0^\mu\dow_\mu {\sf s}_{r}\rb  + \cO(\hbar)~.
    \label{eq:KMS-back}
\end{align}

The action of KMS symmetry on the dynamical fields is naturally defined on the the fluid worldvolume as
\begin{alignat}{2}
    X_1^\mu(\sigma) 
    &\KMSto 
    \Theta X_1^\mu(\sigma), &\qquad
    X_2^\mu(\sigma) 
    &\KMSto 
    \Theta X_2^\mu(\sigma + i\hbar\,\Theta\bbbeta)
    - i\hbar\, \Theta\beta_0^\mu~, \nn\\
  \varphi_1(\sigma) &\KMSto 
  \Theta\varphi_1(\sigma), &\qquad
  \varphi_2(\sigma) &\KMSto 
  \Theta \varphi_2(\sigma + i\hbar\,\Theta\bbbeta)~.
  \label{eq:KMS-dynamicalfields}
\end{alignat}
Importantly, the extra constant contribution in the transformation $X_2^\mu$ is taken so that KMS symmetry preserve the equilibrium configuration $X_{1,2}^\mu(\sigma) = \delta^\mu_\alpha \sigma^\alpha$, $\varphi_{1,2}(\sigma) = 0$. 
In the physical spacetime formulation in the classical limit, these give rise to
\begin{alignat}{2}
    \beta^\mu &\KMSto \Theta\beta^\mu  + \cO(\hbar)~, &\qquad 
    X^\mu_a &\KMSto 
    \Theta\hat X^\mu_a \equiv 
    \Theta\lb X^\mu_a
    + i(\beta^\mu - \beta^\mu_0) \rb + \cO(\hbar)~, \nn\\
    \varphi_r &\KMSto \Theta\varphi_r + \cO(\hbar)~, &\qquad 
    \varphi_a &\KMSto 
    \Theta\hat\varphi_a \equiv 
    \Theta\lb \varphi_a + i\lie_\beta\varphi_r \rb + \cO(\hbar)~.
    \label{eq:KMS-classical-hydro}
\end{alignat}
These transformation properties induce the following dynamical KMS transformation on the building blocks of the effective action
\begin{equation}
    \Psi_r \KMSto \Theta\Psi_r + \cO(\hbar)~, \qquad 
    \Psi_a \KMSto 
    \Theta\hat\Psi_a \equiv 
    \Theta\hat\!\lb \Psi_a + i\lie_\beta \Psi_r \rb + \cO(\hbar)~, \qquad 
    \psi \KMSto \Theta\psi + \cO(\hbar)~.
    \label{eq:KMS-Phira_hydro}
\end{equation}

\paragraph{Galilean hydrodynamics:} The SK setup so far has been boost-agnostic, i.e. it applies to systems that may or may not possess a boost symmetry. To describe a Galilean system, we also need to impose the Milne boost symmetry \eqref{eq:Galilean-boost} on the two sets of spacetime background fields $(n_{1,2\mu},h_{1,2\mu\nu},A_{1,2\mu})$ independently~\cite{Jain:2020vgc}. In the classical limit, $\Psi_r$ fields only transform under the diagonal Milne boosts that act in the same way as \cref{eq:Galilean-boost}. Milne boosts act quite complicatedly on the $\Psi_a$ fields, though we can write down the combination
\begin{align}\label{eq:gal-H}
    \cH_{a\mu\nu}
    &= H_{a\mu\nu} + 2N_{r(\mu} B_{a\nu)}
    - 2N_{a(\mu} \lb \vec u_{\nu)}
    - \half N_{r\nu)} \vec u^2 \rb~,
\end{align}
that is Milne-invariant together with $N_{a\mu}$. Here $\vec u^\mu = u^\mu - v^\mu$ denote the spatial components of the fluid velocity, with $\vec u_\mu = H_{r\mu\nu}\vec u^\nu$, and $\vec u^2 = \vec u^\mu \vec u_\mu$ is the fluid velocity squared. We can also use this to define a Galilean-invariant version $B_{a\mu}$, i.e. $\cB_{a\mu} = \cH_{a\mu\nu}u^\nu
- \half N_{r\mu}\cH_{a\rho\sigma}u^\rho u^\sigma$ , used in \cref{eq:cB-def}.
Furthermore, the chemical potential $\mu$ defined in \cref{eq:hydro-beta} is not Galilean-invariant and we instead need to improve the definition as
\begin{align}
    \beta \varpi &=
    \beta \lb\mu + \half\vec u^2\rb~,
\end{align}
known as the mass chemical potential.
The definitions of fluid velocity and temperature are already Galilean-invariant.

\vspace{1em}

This finishes our lightening review of the formal aspects of passive non-relativistic SK hydrodynamics. Next, we will discuss how to introduce activity into the framework.
Later in \cref{app:examples}, we will see these concepts applied to particular examples.

\subsection{Active ingredients}
\label{sec:activity-in-SK}

To extend the SK framework discussed above to active hydrodynamics, we need to introduce the background fuel and environment fields on each spacetime $\Phi^{\fuel,\env}_{1,2}(X_{1,2})$ from \cref{sec:active-KMS}. The pullbacks onto the worldvolume are defined simply as
\begin{equation}
    \mathbb{\Phi}^{\fuel,\env}_{1,2}
    = \Phi^{\fuel,\env}_{1,2}(X_{1,2})~,
\end{equation}
that are evaluated on the worldvolume coordinates $\sigma^\alpha$. We can define the average-difference basis of these worldvolume fields $\mathbb\Phi^{\fuel,\env}_{r,a}$,
which can be further pulled onto the physical spacetime defined by identifying $x^\mu = X^\mu_r$, denoted as $\Pi^{\fuel,\env}_{r,a}$. In the classical limit, the objects $\Pi^{\fuel,\env}_{r,a}$ are related to the background fields $\Phi^{\fuel,\env}_{r,a}$ as
\begin{alignat}{2}
    \Pi^{\fuel,\env}_{r}
    &= \ell\Phi^{\fuel,\env}_{r} + \cO(\hbar)~, &\qquad
    \Pi^{\fuel,\env}_{a}
    &= \ell\Phi^{\fuel,\env}_{a}
    + \ell\lie_{X_a}\Phi^{\fuel,\env}_{r}
    + \cO(\hbar)~.
    \label{eq:cF-def-general}
\end{alignat}
We have included a factor of $\ell$ in these definitions for later convenience.
The SK effective action $S[\psi,\psi_a,\ul\sfs_r,\ul\sfs_a]$ is constructed similar to the passive case, where $\ul\sfs_{r,a}=(\sfs_{r,a},\Phi^\fuel_{r,a},\Phi^\env_{r,a})$ contains the fuel/environment background fields as well, and is still required to satisfy the conditions in \cref{eq:SK-conds}. The SK generating functional $\cZ[\ul\sfs_r,\ul\sfs_a]$ is defined similarly to \cref{eq:Z-def}.

We impose the regular KMS symmetry \eqref{eq:KMS-full} on the fuel fields
\begin{gather}
    \Phi^\fuel_{1}(x)
    \KMSto \Theta\Phi^\fuel_{1}(x)~, \qquad
    \Phi^\fuel_{2}(x)
    \KMSto \Theta\Phi^\fuel_{2}(x + i\hbar\,\Theta\beta_0)~.
    \label{eq:KMS-fuel1}
\end{gather}
The qualitative point of departure from passive hydrodynamics is the new active KMS symmetry imposed on the environment fields, i.e.
\begin{gather}
    \Phi^\env_{1}(x)
    \KMSto \Theta\Phi^\env_{1}(x)~, \qquad
    \Phi^\env_{2}(x)
    \KMSto \Theta\Phi^\env_{2}(x + i\hbar\,\Theta\beta_0) - i\hbar~.
    \label{eq:active-KMS}
\end{gather}
The activity arises because of the additional $-i\hbar$ term in the KMS transformation of $\Phi^\env_{2}$. This term is similar to the additional term introduced in the KMS transformation of $X_2^\mu$ in \cref{eq:KMS-dynamicalfields}, which was required to ensure that the equilibrium configuration $X_{1,2}^\mu(\sigma) = \delta^\mu_\alpha \sigma^\alpha$ is preserved by KMS. Whereas, the additional term here means that the environment field configuration $\Phi^\env_{1,2} = T_\env t$ is only preserved by KMS when $T_0=T_\env$, i.e. the system temperature is same as the environment temperature. Recall from \cref{eq:eqb-temp} that $T_0-T_\env \propto \Delta E^2$ in an active steady state, meaning that the environment field configuration is never preserved by KMS when the fuel is available to provide heat. Accordingly, the fuel field configuration $\Phi^\fuel_{1,2} = -\Delta E\, t$ is also only preserved by KMS when the fuel chemical differential is absent, i.e. $\Delta E=0$. This also means that retarded and symmetric correlation functions computed in a state with $T_0\neq T_\env$ or $\Delta E\neq 0$ will not satisfy FDTs. \Cref{eq:active-KMS} implies a KMS transformation for the $\cF^{\fuel,\env}_{r,a}$ fields as
\begin{align}
    \Pi^{\fuel}_r &\KMSto \Theta\Pi^{\fuel}_r + \cO(\hbar)~, \qquad 
    \Pi^\fuel_a \KMSto \Theta\hat\Pi^\fuel_a \equiv 
    \Theta \lb \Pi^\fuel_a + i\lie_\beta\Pi^\fuel_r \rb + \cO(\hbar)~, \nn\\
    \Pi^{\env}_r &\KMSto \Theta\Pi^{\env}_r  + \cO(\hbar)~, \qquad 
    \Pi^\env_a \KMSto \Theta\hat\Pi^\env_a \equiv 
    \Theta\lb \Pi^\env_a + i\lie_\beta\Pi^\env_r - i \rb + \cO(\hbar)~.
\end{align}

We should emphasise that the full Schwinger-Keldysh generating functional $\cZ[\ul\sfs_r,\ul\sfs_a]$ for active hydrodynamics respects the KMS symmetry. However, this symmetry is ``spurious'' in the sense that it non-trivially relates a fuel and environment configuration $\Phi^\fuel_2 = -\Delta E\, t$, $\Phi_2^\env =  \kB  T_\env t$ to another configuration $\Phi^\fuel_2 = -\Delta E\, t - i\hbar \Delta E/T_0$, $\Phi^\env_2 =   \kB  T_\env t - i\hbar \Delta T/T_0$, shifted by imaginary terms that are nonzero in the presence of activity. 

\subsection{Second law of thermodynamics}
\label{app:2ndlaw}

In this appendix, we discuss how the local second law of thermodynamics emerges from the SK formalism and, in particular, how the derivation gets modified in the presence of activity. The derivation presented here differs slightly from the one given originally in~\cite{Glorioso:2016gsa}, but essentially follows the same line of reasoning. Comparing the $a$-type time-translations in \cref{eq:energy-conservation} against the KMS transformations \cref{eq:KMS,eq:activeKMS}, we make a crucial observation that the inhomogeneous shift piece in the KMS transformations of $\sfs_a$ and $\Phi^\fuel_a$ can be undone by an imaginary $a$-type time-translation with $\chi^t_a = -i\beta_0$. This does not apply to our environment field $\Phi^\env_a$, due to the additional ``$i$'' term in the active KMS transformation. Let us denote the combined transformation of the noise dynamical fields under KMS and this imaginary $a$-type time-translation as $\psi_a\to\Theta(\psi_a+ig_r)=\Theta(X_a^\mu+i\beta^\mu,\varphi_a+i\lie_\beta\varphi_r)$. The SK Lagrangian, $L=\int\df^dx\,\cL(\psi,\psi_a,\ul\sfs_r,\ul\sfs_a)$, is invariant under this combined transformation, up to a  total derivative term, leading to
\begin{align}\label{eq:KMS-TT}
    \Theta\cL\big(\Theta\psi,\Theta(\psi_a+ig_r),\Theta\ul\sfs_r,\Theta\ul\sfs_a\big)
    \big|_{\Phi^\env_a\to\Phi^\env_a-i} 
    &= \cL(\psi,\psi_a,\ul\sfs_r,\ul\sfs_a) 
    + i\dow_t\cN^t(\psi,\ul\sfs_r) + i\dow_i\cN^i(\psi,\ul\sfs_r)~.
\end{align}
Note that repeating the KMS symmetry takes $\cL$ back to itself, meaning that $\cN^t$, $\cN^i$ do not contain any $a$-type noise fields and have time-reversal eigenvalue $+1$, $-1$. Also, note that the total-derivative terms need to be imaginary to preserve $\cL^\dagger=-\cL$ from \cref{eq:SK-conds}.

Since \cref{eq:KMS-TT} is valid for any field arguments, we can evaluate it at $\Phi^\env_a\mapsto\Phi^\env_a + i$ and $\psi_a\mapsto\psi_a-ig_r$, and use the second condition in \cref{eq:SK-conds}, to find
\begin{align}
    \Im&\Big[\Theta\cL(\Theta\psi,\Theta\psi_a,\Theta\ul\sfs_r,\Theta\ul\sfs_a)\Big] \nn\\
    &= \Im\Big[\cL\big(\psi,\psi_a{-}ig_r,\ul\sfs_r,\ul\sfs_a\big)\big|_{\Phi^\env_a\to\Phi^\env_a+i}
    + i\dow_t\cN^t(\psi,\ul\sfs_r) + i\dow_i\cN^i(\psi,\ul\sfs_r)
    \Big] \nn\\
    &= \half\Im
    \Big[
    \cL\big(\psi,\psi_a{-}ig_r,\ul\sfs_r,\ul\sfs_a\big)\big|_{\Phi^\env_a\to\Phi^\env_a+i}
    - \cL^*\big(\psi,\psi_a{+}ig_r,\ul\sfs_r,\ul\sfs_a\big)\big|_{\Phi^\env_a\to\Phi^\env_a-i}
    \Big]
    + \dow_t\cN^t(\psi,\ul\sfs_r)+\dow_i\cN^i(\psi,\ul\sfs_r) \nn\\
    &= \half\Im
    \Big[
    \cL\big(\psi,\psi_a{-}ig_r,\ul\sfs_r,\ul\sfs_a\big)\big|_{\Phi^\env_a\to\Phi^\env_a+i}
    + \cL\big(\psi,-\psi_a{-}ig_r,\ul\sfs_r,-\ul\sfs_a\big)\big|_{\Phi^\env_a\to\Phi^\env_a-i}
    \Big]
    + \dow_t\cN^t(\psi,\ul\sfs_r)+\dow_i\cN^i(\psi,\ul\sfs_r)~.
    \label{eq:Delta-from-SK}
\end{align}
Note that the expression in the first line is merely the time-reversed $\Im\cL$ evaluated on time-reversed field arguments. The third condition in \cref{eq:SK-conds} requires $\Im\cL$ to be non-negative for all possible field configurations, so this expression must also be non-negative. As for the last line in \cref{eq:Delta-from-SK}, we can set $\sfs_a,\Phi^\fuel_a\mapsto0$ and Taylor expand the two $\cL$ terms in their remaining $a$-type arguments $\psi_a\pm ig_r$ and $\Phi^\env_a\pm i$. The 0th order pieces in these expansions vanish due to the first condition in \cref{eq:SK-conds}, i.e. $\cL$ vanishes when all $a$-type arguments are turned off. The 1st order pieces are quite important, but they turn out to be independent of $\psi_a$ and $\Phi^\env_a$, taking the form
\begin{align}
    \half \Im\Big[\ldots\Big]_{(1\text{st order})} 
    &= \half\Im\lB
    (\psi_a-ig_r)\frac{\delta\cL}{\delta\psi_a}
    + (\Phi^\env_a+i)\frac{\delta\cL}{\delta\Phi_a^\env}
    + (-\psi_a-ig_r)\frac{\delta\cL}{\delta\psi_a}
    + (-\Phi^\env_a+i)\frac{\delta\cL}{\delta\Phi_a^\env}
    \rB 
    + \dow_t\cN_{(1)}^t + \dow_i\cN_{(1)}^i
    \nn\\
    &= - g_r \frac{\delta\cL}{\delta\psi_a} 
    + \frac{\delta\cL}{\delta\Phi_a^\env}
    + \dow_t\cN_{(1)}^t + \dow_i\cN_{(1)}^i \nn\\
    &= \ell r_\env
    + \dow_t\cN_{(1)}^t + \dow_i\cN_{(1)}^i~,
\end{align}
where we have identified possible total-derivative terms left over after integration-by-parts. In the last line, we have used the classical equations of motion $\delta\cL/\delta\psi_a=0$, and identified $\ell r_\env = \delta\cL/\delta\Phi^\env_a$. This is how $r_\env$ makes its way into the second law statement. The fields $\psi_a$ and $\Phi^\env_a$ appear first in the 2nd order pieces. Denoting $\cO(\psi_a^m,(\Phi_a^\env)^n)$ terms in the Lagrangian as
$\cL\sim(-i)^{m+n+1}\fL_{(m,n)}\psi_a^m(\ell\Phi_a^\env)^n$, where $\fL_{(m,n)}$ may be a differential operator, we have
\begin{align}
    \half \Im\Big[\ldots\Big]_{(2\text{nd order})} 
    &= \half\fL_{(2,0)}\lb\psi_a^2-g_r^2\rb
    + \ell\fL_{(1,1)}\lb\psi_a\Phi^\env_a+g_r\times 1\rb
    + \half\ell^2\fL_{(0,2)}\lb(\Phi^\env_a)^2-1\times 1\rb~.
    \label{eq:second-order}
\end{align}
Similarly, we can find $k$th order terms as
\begin{align}
    \half \Im\Big[\ldots\Big]_{(k\text{th order})} 
    &= \sum_{m=0}^k \frac{1}{m!(k-m)!}\fL_{(m,k-m)}
    \Im\Big((-i)^{k+1}(\psi_a-ig_r)^m(\ell\Phi^\env_a+i\ell)^{k-m}\Big)~.
    \label{eq:n-order}
\end{align}

If we assume Gaussian noise, meaning that the Lagrangian is at most quadratic in noise fields, the expansion ends at the 2nd order. In this case, we may evaluate \cref{eq:Delta-from-SK} at $\sfs_a,\Phi^\fuel_a\mapsto0$, $\psi_a\mapsto g_r$, and $\Phi^\env_a\mapsto -1$, which kills the 2nd order contributions from \cref{eq:second-order}, and we arrive at the desired statement of the second law of thermodynamics
\begin{align}
    \dow_t s^t + \dow_is^i + \ell \kB r_\env = \Delta \geq 0~,
\end{align}
with 
\begin{align}
    \frac{1}{\kB} s^t = \cN^t + \cN^t_{(1)}~, \qquad 
    \frac{1}{\kB} s^i = \cN^i + \cN^i_{(1)}~, \qquad 
    \frac{1}{\kB}\Delta = \Im\Big[\Theta\cL\big(\Theta\psi,\Theta g_r,\Theta\ul\sfs_r,\{0,0,-\ell\}\big)\Big]\geq 0~.
\end{align}
This argument also works in the presence of non-Gaussian noise, but we need to find the appropriate substitution values for $\psi_a$ and $\Phi^\env_a$ that sets the 2nd and higher order terms in \cref{eq:n-order} to zero, possibly up to total-derivative terms that correct the entropy density and flux. In practise, we can solve for $\psi_a$ and $\Phi^\env_a$ order-by-order in derivatives.
Typically, the potential corrections from the $k$th order terms only affect the constitutive relations at $\cO(\dow^{k-1})|_{k>2}$, and thus we can safely ignore these for our purposes. A more comprehensive non-perturbative derivation of the second law from the SK formalism can be found in~\cite{Glorioso:2016gsa}.

\subsection{Microscopic reversibility}
\label{app:probs}

In this appendix, we give the details of the derivation of microscopic reversibility in \cref{eq:microscopic-reversibility} from the SK path integral. The conditional probability distribution for the system to traverse the time-reversed path $\Theta\psi(t)=\eta_\Theta\psi(\Theta t)$, given that it starts at $\Theta\psi(t_\i)=\eta_\Theta\uppsi_\f$, in the presence of time-reversed background sources $\eta_\Theta\ul\sfs_{r,a}(-t)$, is given by the path integral
\begin{equation}
    \bbP_\Theta(\Theta\psi|\eta_\Theta\uppsi_\f,t_\i) 
    = \frac{1}{\cN}\dsp\int{\cD\psi_a}
    \exp\!\lb i\int_{t_{\text i}}^{t_{\text f}}{\!\!\df t}\, L\big(\Theta\psi,\Theta\psi_a,\Theta\ul\sfs_r,\Theta\ul\sfs_a\big)\rb~.
\end{equation}
To relate this to the probability distribution of the forward path, let us consider the following manipulations of the SK path integral
\begin{align}    
    \int\cD\psi_a&\exp\!\bigg[ i\int_{t_{\text i}}^{t_{\text f}}\df t\,
    L\big(\Theta\psi,\Theta\psi_a;\Theta\ul\sfs_r,\Theta\!\!\!\hat{\,\,\,\ul\sfs_a}\big)
    \bigg] \nn\\
    &= 
    \int\cD\psi_a\exp\!\bigg[ i\int_{t_{\text i}}^{t_{\text f}}\df t\,
    L\big(\Theta\psi,\Theta\hat\psi_a;\Theta\ul\sfs_r,
    \Theta\!\!\!\hat{\,\,\,\ul\sfs_a}\big)
    \bigg] \nn\\
    &= 
    \int \cD\psi_a\exp\!\bigg[ i\int_{t_{\text i}}^{t_{\text f}}\df t\,
    \Theta \Big( 
    L\big(\psi,\psi_a;\ul\sfs_r,\ul\sfs_a\big)
    + \beta_0 \dow_t \Omega(\psi,\ul\sfs_r)\Big)\bigg]~, \nn\\
    &= 
    \e^{\beta_0\Delta\Omega}
    \int\cD\psi_a\exp\!\bigg[ i\int_{t_{\text i}}^{t_{\text f}}\df t\,
    L\big(\psi,\psi_a;\ul\sfs_r,\ul\sfs_a\big)\bigg]~.
    \label{eq:path-integral-ft}
\end{align}
In the first step, we have changed the path integration variables, $\psi_a\to\hat\psi_a$, which is legal provided that $\hat\psi_a(t_{\i,\f})=0$. The precise physical interpretation of this, of course, depends on the model under consideration and the KMS transformation properties of the $\psi_a$ fields. In hydrodynamic models, this typically means that the initial and final states are at the same temperature and chemical potentials, see \cref{eq:KMStransformations}, and $\beta_0$ is equal to the initial/final inverse temperature. In the second step, we have used the KMS symmetry of the Lagrangian, while in the final step we have performed the coordinate relabelling $t\to-t+t_{\rm i}+t_{\rm f}$. We have almost arrived at a relation between the original and time-reversed path probabilities, except that the path integrals in the first and last lines are evaluated at different values of the noise sources. To bring them in the same form, consider evaluating \cref{eq:path-integral-ft} at $\sfs_a \mapsto \sfs_a - i\beta_0\dow_t\sfs_r$, $\Phi^\fuel_a \mapsto \Phi^\fuel_a - i\beta_0\dow_t\Phi^\fuel_r$, and $\Phi^\env_a \mapsto \Phi^\env_a - i\beta_0\dow_t\Phi^\env_r + i$. This converts the noise sources in the first line to just $\Theta\ul\sfs_a$, while those in the last line will get shifted as
\begin{align}
    \int\cD\psi_a&\exp\!\bigg[ i\int_{t_{\text i}}^{t_{\text f}}\df t\,
    L\big(\Theta\psi,\Theta\psi_a;\Theta\ul\sfs_r,\Theta{\ul\sfs_a}\big)
    \bigg] \nn\\
    &= \e^{\beta_0\Delta\Omega}
    \int\cD\psi_a\exp\!\bigg[ i\int_{t_{\text i}}^{t_{\text f}}\df t\,
    L\big(\psi,\psi_a;\ul\sfs_r,\sfs_a+i\beta_0\ul{\sf w}_r\big)\bigg] \nn\\
    &= \e^{\beta_0\Delta\Omega}
    \exp\!\lB \beta_0\int_{t_\i}^{t_\f}\df t\,\df^dx\,
    \ul{\sf w}_r\cdot \frac{i\delta}{\delta\ul\sfs_a} \rB
    \int\cD\psi_a\exp\!\bigg[ i\int_{t_{\text i}}^{t_{\text f}}\df t\,
    L(\psi,\psi_a;\ul\sfs_r,\ul\sfs_a)\bigg]~,
    \label{eq:all-source-shift}
\end{align}
where we have introduced the notation $\ul{\sf w}_r=-(\dow_t\sfs_r,\dow_t\Phi^\env_r,\dow_t\Phi^\env_r-\kB T_0)$. In the second step, we have represented the shifts in the $a$-type sources as Taylor series expansion in variational derivatives. Note that the $a$-type sources couple to the respective operators $\ul\cO = (\cO,\ell r_\fuel,\ell r_\env)$, so we can replace the variational derivatives with the appropriate operators insertions, leading to
\begin{align}
    \int\cD\psi_a&\exp\!\bigg[ i\int_{t_{\text i}}^{t_{\text f}}\df t\,
    L\big(\Theta\psi,\Theta\psi_a;\Theta\ul\sfs_r,\Theta{\ul\sfs_a}\big)
    \bigg] \nn\\
    &= \e^{\beta_0\Delta\Omega}
    \left\langle\exp\!\lB
    -\beta_0 \int\df t\,\df^dx\,\ul\cO\cdot \ul{\sf w}_r\rB\right\rangle_\psi
    \int\cD\psi_a\exp\!\bigg[ i\int_{t_{\text i}}^{t_{\text f}}\df t\,
    L(\psi,\psi_a;\ul\sfs_r,\ul\sfs_a)\bigg] \nn\\
    &= \e^{\beta_0\Delta\Omega-\beta_0 W_\psi}
    \int\cD\psi_a\exp\!\bigg[ i\int_{t_{\text i}}^{t_{\text f}}\df t\,
    L(\psi,\psi_a;\ul\sfs_r,\ul\sfs_a)\bigg]~.
    \label{eq:final-path-manipulation}
\end{align}
The notation $\langle\ldots\rangle_\psi$ denotes that the averaging is only done over $\psi_a$ fields for fixed $\psi$. This yields the microscopic reversibility relation in \cref{eq:microscopic-reversibility}.
Furthermore, by integrating \cref{eq:final-path-manipulation} over all paths $\int_{\uppsi_\i}^{\uppsi_\f}\cD\psi$, we recover the active detailed balance relation in \cref{eq:detailed-imbalance}.

Instead of integrating over all paths, let us only integrate \cref{eq:final-path-manipulation} over the paths and final states that require a fixed amount of work, $W_\psi=W_0$, averaged over all the initial states drawn from the free energy distribution
\begin{align}\label{eq:eqb-distr}
    w(\uppsi,t) &= \exp\bigg(\beta_0F(t)-\beta_0\Omega(\uppsi,t)\bigg)~, \qquad
    F(t) = \frac{-1}{\beta_0}\log\int\df\uppsi\exp\bigg(-\beta_0\Omega(\uppsi,t)\bigg)~.
\end{align}
This recovers \emph{Crook's fluctuation theorem} for non-equilibrium processes~\cite{Crooks:1999ttd}
\begin{align}\label{eq:crooks}
    \frac{\bbP_\Theta[-W_0]}{\bbP[W_0]} 
    = \exp\lb \beta_0 \Delta F -\beta_0W_0\rb~,
\end{align}
where $\Delta F = F(t_\f)-F(t_\i)$ and $\bbP[W_0]$ is the total probability of the forward trajectories that require work $W_0$, given by
\begin{align}
    \bbP(W_0) = \int\df\uppsi_\i\df\uppsi_\f\, w(\uppsi_\i,t_\i) 
    \,\delta(W_\psi-W_0)\,\bbP(\psi|\uppsi_\i,t_\i)~.
\end{align}
Lastly, we can integrate \cref{eq:crooks} over all $W_0$ to obtain the Jarzynski equality~\cite{PhysRevLett.78.2690, 1997PhRvE..56.5018J}, relating the average work done during a process to the differential of total free energy
\begin{align}
    \Big\langle \exp\lb -\beta_0W\rb\Big\rangle
    = \exp(-\beta_0 \Delta F)~,
\end{align}
where the averaging is understood over all processes and all final states drawn from the $w(\uppsi_\i,t_\i)$ distribution. This equality can also be obtained directly from the active detailed balance condition in \cref{eq:detailed-imbalance} by integrating over the initial states from the $w(\uppsi_\i,t_\i)$ distribution.


\section{Details of Schwinger-Keldysh effective actions}
\label{app:examples}

In this appendix, we use the SK formalism from \cref{app:SK} to construct effective action descriptions for various active systems that appear in the bulk of this paper. In particular, we discuss active diffusion in \cref{app:diffusion}, active superfluids in \cref{app:superfluid}, active fluids in \cref{app:activehydro}, and active nematics in \cref{app:activenematics}.
As in the main text, we assume the coupling between the fluid and fuel sectors to be minimal for simplicity, physically meaning that the fluid only uses the heat provided by the burning of fuel but does not back-react on the burning process itself. For concreteness, we will also choose $\Theta=\rmT$ to set up the KMS symmetry and only work in the classical limit. For parity-preserving systems, this is equivalent to the choice $\Theta={\rm PT}$ as well.

We will organise the effective field theory in a double perturbative expansion in small derivatives, controlled by $\dow_\mu$, as well as small activity, controlled by $\ell$. For concreteness, we take $\ell\sim\cO(\dow)$ and we take the derivative ordering of various background and dynamical fields as follows
\begin{gather}
    n_\mu,~ h_{\mu\nu},~ A_\mu
    \sim\cO(\dow^0)~, \qquad 
    n_{a\mu},~ h_{a\mu\nu},~ A_{a\mu}
    \sim\cO(\dow^1)~, \nn\\
    \blue{ \Phi^{\fuel,\env} \sim \cO(\dow^{-1})~, \qquad 
    \Phi^{\fuel,\env}_a \sim \cO(\dow^0)}~, \nn\\
    u^\mu,~ T,~\mu \sim \cO(\dow^0)~, \qquad
    X^\mu_a,~ \varphi_a \sim \cO(\dow^0)~.
\end{gather}
In the presence of spontaneous symmetry breaking, we will also need to choose a derivative ordering for the associated order parameters and Goldstone fields, which we will address when we get there.
We will ignore $\cO(\dow^3)$ and higher-order terms in the SK effective action. In terms of the constitutive relations, this means that we will ignore $\cO(\dow^2)$ and higher contributions to the conserved currents and rate operators.

\subsection{Active diffusion}
\label{app:diffusion}

Let us start with a simple toy model of active diffusion without conserved momentum, where the only relevant ingredients are the energy and charge conservation. We take the momentum sector to be trivial by setting $u^i = 0$, $X^i_a = 0$, $h_{\mu\nu} = ((0,0),(0,\delta_{ij}))$, and $h_{a\mu\nu} = 0$. This further implies $v^\mu = (1/n_t,0)$ and $h^{\mu\nu} = ((n_kn^k/n_t^2,-n^j/n_t),(-n^i/n_t,\delta^{ij}))$.
The relevant conservation equations are
\begin{align}
    \dow_\mu\!\lb n_t \epsilon^\mu\rb
    &= F^n_{t\mu}\epsilon^\mu - F_{t\mu}j^\mu 
    \blue{\,-\, \ell r_\env \dow_t\Phi^\env
    - \ell r_\fuel \dow_t\Phi^\fuel}~, \nn\\
    \dow_\mu\!\lb n_t j^\mu\rb &=0~,
\end{align}
which generalise \cref{eq:diffusion-eqns} in the presence of $n_\mu$.
Note that $\sqrt{\gamma} = n_t$ and $\int_x = \int \df t\,\df^{d}x\,n_t$ when $h_{\mu\nu}$ is flat. 

Truncating the effective theory to at most quadratic order in $a$-type fields, and assuming minimal coupling to the fuel sector, we can write down a simple effective Lagrangian for active diffusion
\begin{align}\label{eq:SKaction-active-diffusion}
    \cL
    &= - \epsilon\, u^\mu N_{a\mu} 
    + n\,u^\mu B_{a\mu} 
    \nn\\
    &\qquad
    + i\kB T
    \begin{pmatrix}
         -N_{a\mu} \\ B_{a\mu}  \\ \blue{\beta_\env\Pi^\env_a}
    \end{pmatrix}^{\!\!\intercal}
    \begin{pmatrix}
        T\kappa h^{\mu\nu} & \sigma_\times  h^{\mu\nu}
        & \blue{0} \\
        \sigma_\times h^{\nu\mu} & \sigma h^{\mu\nu}
        & \blue{0} \\
        \blue{0} & \blue{0} & \blue{\gamma_\env/\beta_\env^2}
    \end{pmatrix}
    \begin{pmatrix}
         -\hat N_{a\mu} \\ \hat B_{a\mu}  \\ \blue{\beta_\env\hat\Pi^\env_a}
    \end{pmatrix}
    \blue{\,+\, i\kB T \gamma_\fuel\, \Pi^\fuel_a\hat\Pi^\fuel_a}~,
\end{align}
where $\beta_\env = 1/(\kB T_\env)$.
We have introduced the energy density $\epsilon$, charge density $n$, thermal conductivity $\kappa$, charge conductivity $\sigma$, thermo-electric conductivity $\sigma_\times$, and active coefficients $\gamma_{\env,\fuel}$. All coefficients appearing here are functions of $T$ and $\mu$.  

The action \eqref{eq:SKaction-active-diffusion} is manifestly invariant under the spacetime global symmetries and worldvolume gauge symmetries. It also trivially obeys the first two SK conditions in \cref{eq:SK-conds}, while the third one requires
\begin{equation}\label{eq:ineq-sf-app}
    T\kappa \geq \sigma_\times^2/\sigma~, \qquad 
    \sigma \geq 0~, \qquad 
    \blue{\gamma_{\env,\fuel} \geq 0}~,
\end{equation}
which is sufficient to set ${\rm Im}\,S\geq 0$ order-by-order in derivatives, i.e.
\begin{align}
    \beta\Im\cL
    &=  \lb T\kappa - \frac{\sigma_\times^2}{\sigma} \rb 
    \Big(N_{a\mu} + \cO(\dow^3) \Big)^2 
    +   \sigma \lb B_{a\mu} - \frac{\sigma_\times}{2\sigma}N_{a\mu} + \cO(\dow^3) \rb^2 
    \blue{\,+\, \gamma_\env\lb\Pi^\env_a\rb^2
    +  \gamma_\fuel \lb\Pi^\fuel_a\rb^2}
    \geq 0~,
\end{align}
where $\ldots$ denote higher-derivative corrections.
The squares of vector objects above are understood appropriately contracted with $h^{\mu\nu}$. The terms in the first line in \cref{eq:SKaction-active-diffusion} are KMS-invariant provided that the charge density $n$ and energy density $\epsilon$ satisfy the thermodynamic relation \eqref{eq:thermodynamics}; we may check that the KMS variation of these terms is given as
\begin{align}
    -\epsilon\, u^\mu\, &i\lie_\beta N_{r\mu} 
    + n\, u^\mu \,i\lie_\beta B_{r\mu} 
    = i\Big( -\epsilon\,u^\mu\dow_\mu \beta 
    + n\,u^\mu\dow_\mu  ( \beta \mu ) \Big)
    = \frac{i}{n_t} \dow_t (\beta p ) ~,
\end{align}
which drops out from the effective action as a boundary term.

The classical constitutive relations may be obtained by varying the effective action with respect to the associated $a$-type background fields and setting all the $a$-type fields to zero. We find
\begin{align}
    \epsilon^\mu 
    &= \epsilon\,u^\mu 
    - T\kappa\, \sfV_\epsilon^\mu
    - \sigma_\times\sfV^\mu_n, \nn\\
    j^\mu 
    &= n\,u^\mu 
    - \sigma\,\sfV^\mu_n
    - \sigma_\times\sfV^\mu_\epsilon~, \nn\\
    r_\env 
    &= \blue{\gamma_\env \ell\kB\Delta T}~, \nn\\
    r_\fuel 
    &= \blue{\gamma_\fuel \ell\Delta E}~, 
\end{align}
where we have introduced the notation
\begin{align}\label{eq:V-def}
    \sfV_{n\mu} = T\dow_\mu\frac{\mu}{T} - F_{\mu\nu}u^\nu~, \qquad 
    \sfV_{\epsilon\mu} = \frac1T\dow_\mu T + F^n_{\mu\nu}u^\nu~.
\end{align}
Turning off $F^n_{\mu\nu}$ and $\sigma_\times$, we recover the ``fluid part'' of the constitutive relations presented in \cref{eq:consti}. The ``superfluid part'' will be discussed in the next sub-appendix.

\subsection{Active superfluids}
\label{app:superfluid}

\paragraph{SK effective action:} To model an active superfluid, we need to introduce the associated order parameter for each SK spacetime, i.e. a pair of complex scalar fields $\bbPsi_{1,2}(\sigma)$. We take these to transform under the worldvolume U(1) symmetry, i.e. $\bbPsi_{1,2} \to {\rm e}^{-i\lambda}\bbPsi_{1,2}$. In the physical spacetime formulation, the associated average-difference basis can be used to define
\begin{align}
    \Psi_{r,a} = {\rm e}^{i\varphi_r}\bbPsi_{r,a} + \cO(\hbar)~,
\end{align}
which we met in \cref{sec:simplediffusion}. These derived fields are invariant under the worldvolume U(1) symmetry but instead transform under the diagonal part of the spacetime U(1) symmetry. The action of KMS is defined as
\begin{alignat}{2}
    \bbPsi_1(\sigma) &\KMSto  
    \Theta\bbPsi_1^*(\sigma), &\qquad
    \bbPsi_2(\sigma) &\KMSto 
    \Theta\bbPsi^*_2(\sigma + i\hbar\,\Theta\bbbeta)~, \nn\\
    \bbPsi^*_1(\sigma) &\KMSto  
    \Theta\bbPsi_1(\sigma), &\qquad
    \bbPsi^*_2(\sigma) &\KMSto 
    \Theta\bbPsi_2(\sigma + i\hbar\,\Theta\bbbeta)~,
\end{alignat}
with the time-reversal eigenvalues $+1$.
In the classical limit, this leads to
\begin{align}\label{eq:hatPsi}
    \Psi_r &\KMSto \Theta\Psi_r^\dagger + \cO(\hbar)~, \qquad 
    \Psi_a \to -\Theta\hat\Psi^\dagger_a \equiv 
    \Theta\lb\Psi_a^* + i\beta^\mu\Df_\mu\Psi_r^* 
    - \beta\mu \Psi_r^*\rb + \cO(\hbar)~,\nn\\
    \Psi^\dagger_r &\KMSto \Theta\Psi_r + \cO(\hbar)~, \qquad 
    \Psi_a^\dagger \to -\Theta\hat\Psi_a \equiv
    -\Theta\lb\Psi_a + i\beta^\mu\Df_\mu\Psi_r 
    + \beta\mu \Psi_r \rb + \cO(\hbar)~,
\end{align}
where the dagger operation has been defined in \cref{eq:unitarity-dagger} and is comprised of a complex conjugation together with a sign-flip of the $a$-type fields. We shall identify $\Psi_r$ with $\Psi$.

The derivative counting for the order parameter is a bit subtle. To begin with, we take $\Psi\sim\cO(\dow^0)$, $\Psi_a\sim\cO(\dow^1)$. However, the U(1) phase of $\Psi$ is a massless Goldstone mode whose spacetime derivatives need not be small, so we take $\Im(\Psi^*\Df_\mu\Psi)\sim\cO(\dow^0)$ but keep $\Re(\Psi^*\Df_\mu\Psi)\sim\cO(\dow^1)$. Lastly, since $\Im(\Psi^* u^\mu\Df_\mu\Psi) = \mu|\Psi|^2 + \ldots$ due the Josephson equation \eqref{eq:Josephson}, we will take the difference $(u^\mu\Df_\mu-i\mu)\Psi\sim\cO(\dow^1)$, which also ensures that $\hat\Psi_a\sim\cO(\dow^1)$ in \cref{eq:hatPsi}.

Using the new ingredients outlined above, the SK effective Lagrangian for an active superfluid can be constructed by generalising \cref{eq:SKaction-active-diffusion} as
\begin{align}\label{eq:SKaction-sf-general}
    \cL
    &= - \Big( \epsilon\, u^\mu - f_\Psi\Xi^\mu_\epsilon\Big) N_{a\mu} 
    + \Big( n\,u^\mu - f_\Psi\Xi^\mu \Big) B_{a\mu} 
    - 2f_\Psi\Re\!\lB\Df^\mu\Psi^* \Df_\mu\Psi_a  \rB
    - 2\frac{\dow V}{\dow|\Psi|^2} \Re\!\lB\Psi^*\Psi_a\rB
    \nn\\
    &\qquad 
    + i\kB T\, \fV^\intercal_a
    \begin{pmatrix}
        T\kappa^{\mu\nu} & \sigma^{\mu\nu}_\times 
        & \blue{-\lambda_{\epsilon\env}f_\Psi \Xi^\mu_\epsilon}
        & -\sigma_\Psi\lambda_{\epsilon\phi}\Df^\mu\Psi^*
        & -\sigma_\Psi\lambda_{\epsilon\phi}\Df^\mu\Psi
        \\
        \sigma^{\nu\mu}_\times & \sigma^{\mu\nu}
        & \blue{-\lambda_{n\env} f_\Psi\Xi^\mu}
        & -\sigma_\Psi\lambda_{n\phi}\Df^\mu\Psi^*
        & -\sigma_\Psi\lambda_{n\phi}\Df^\mu\Psi \\
        \blue{\lambda_{\epsilon\env} f_\Psi \Xi^\mu_\epsilon} 
        & \blue{\lambda_{n\env} f_\Psi\Xi^\mu} 
        & \blue{\gamma_\env/\beta_\env^2} 
        & \blue{a_\env\Psi^*} & \blue{a_\env\Psi} \\
        \sigma_\Psi\lambda_{\epsilon\phi}\Df^\nu\Psi^* 
        & \sigma_\Psi\lambda_{n\phi}\Df^\nu\Psi^* 
        & \blue{-a_\env\Psi^*} & 0 & \sigma_\Psi \\
        \sigma_\Psi\lambda_{\epsilon\phi}\Df^\nu\Psi
        & \sigma_\Psi\lambda_{n\phi}\Df^\nu\Psi
        & \blue{-a_\env\Psi} & \sigma_\Psi & 0
    \end{pmatrix}
    \hat\fV_a \nn\\
    &\qquad
    + \blue{i\kB T \gamma_\fuel \Pi^\fuel_a\hat\Pi^\fuel_a}~,
\end{align}
where for compactness we have identified $\Xi^\mu = 2\Im[\Psi^*\Df^\mu\Psi]$, $\Xi_\epsilon^\mu = 2\Re[u^\lambda\Df_\lambda\Psi^*\Df^\mu\Psi]$. The vector $\fV_a$ given by
\begin{align}
    \fV_a = \begin{pmatrix}
         -N_{a\mu} \\ B_{a\mu}  \\ \blue{\beta_\env\Pi^\env_a} \\ 
         \Psi_a - i\gamma_{\epsilon\phi} \Df^\mu\Psi N_{a\mu} + i\gamma_{n\phi} \Df^\mu\Psi B_{a\mu} 
         \blue{\,+\, i\beta_\env\mu\lambda_{\phi\env} \Psi\Pi_a^\env} \\ 
         -\Psi_a^\dagger 
         + i\gamma_{\epsilon\phi} \Df^\mu\Psi^* N_{a\mu} - i\gamma_{n\phi} \Df^\mu\Psi^* B_{a\mu} 
         \blue{\,-\, i\beta_\env\mu\lambda_{\phi\env} \Psi^*\Pi_a^\env}
    \end{pmatrix}~,
\end{align}
and the hatted version $\hat\fV_a$ is given as usual by converting all the $a$-type fields with their hatted versions.
The Lagrangian has been designed to identically satisfy the first two conditions in \cref{eq:SK-conds}, as well as the KMS-symmetry. In particular, using the thermodynamic relation in \cref{eq:thermodynamics-Psi}, whereby $\Df^i\Psi^*\Df_i\Psi$ is replaced with the covariant version $h^{\mu\nu}\Df_\mu\Psi^*\Df_\nu\Psi$, we can verify that the KMS variation of the terms in the first two lines add up to a total derivative
\begin{align}
    -&\Big( \epsilon\, u^\mu - f_\Psi\Xi^\mu_\epsilon \Big) i\lie_\beta N_{r\mu} 
    + \Big( n\,u^\mu - f_\Psi\Xi^\mu \Big) i\lie_\beta B_{r\mu} \nn\\
    &\qquad
    - 2i f_\Psi\Re\!\lB \Df^\mu\Psi^* \Df_\mu\lb \beta^\mu\Df_\mu\Psi - i\beta\mu \Psi\rb\rB 
    - 2i\frac{\dow V}{\dow|\Psi|^2}
    \Re\!\lB \Psi^*\lb \beta^\mu\Df_\mu\Psi - i\beta\mu \Psi \rb\rB \nn\\
    &= 
    - i\lb T\epsilon \lie_\beta \frac{1}{T}
    - Tn\lie_\beta\frac{\mu}{T}
    + f_\Psi\lie_\beta\lb \Df^\mu\Psi^* \Df_\mu\Psi \rb
    + \frac{\dow V}{\dow|\Psi|^2}\lie_\beta |\Psi|^2\rb \nn\\
    &= 
    - \frac{i}{n_t}\dow_t(\beta\cF)~.
\end{align}
Note that some entries in the coefficient matrix in \cref{eq:SKaction-sf-general} are antisymmetric because of the time-reversal transformation contained within the KMS transformation. We have chosen to represent the possible symmetric cross-couplings between the Goldstone and other sectors in the definition of $\fV_a$ instead, which will turn out to be convenient later. The thermal/charge conductivity matrices are allowed to admit anisotropic pieces for a superfluid, e.g.
\begin{align}
    \sigma^{\mu\nu} = \sigma\,h^{\mu\nu} + \sigma_{\text{anis}} \Df^{(\mu}\Psi^*\Df^{\nu)}\Psi~,
\end{align}
and similarly for $\kappa^{\mu\nu}$ and $\sigma_\times^{\mu\nu}$. The non-negativity of the imaginary part of $\cL$ imposes 
\begin{align}
    \begin{pmatrix}
        T\kappa^{\mu\nu} & \sigma^{\mu\nu}_\times \\
        \sigma^{\nu\mu}_\times & \sigma^{\mu\nu}
    \end{pmatrix}
    \geq 0~, \qquad
    \sigma_\Psi \geq 0~, \qquad
    \blue{\gamma_{\env,\fuel} \geq 0~},
\end{align}
which upgrades the inequality constraints in \cref{eq:ineq-sf-app}. The non-negativity of a symmetric matrix means that all its eigenvalues are non-negative.

\paragraph{Equations of motion and constitutive relations:}
The equation of motion for the order parameter arising from this model takes the form
\begin{align}\label{eq:Psi-eom-cov}
    u^\mu\Df_\mu\Psi
    &=
    \frac{1}{\sigma_\Psi}
    \Big(\sfH_\Psi \blue{\,-\,\aleph a_\env\Psi} \Big)
    + i\lambda_\phi\mu\Psi 
    - \lb \lambda_{\epsilon\phi}+i\gamma_{\epsilon\phi}\rb \sfV_{\epsilon}^\mu \Df_\mu\Psi
    - \lb \lambda_{n\phi}+i\gamma_{n\phi}\rb \sfV_{n}^\mu \Df_\mu\Psi \nn\\
    &= i\lambda_\phi\mu\Psi
    - i\lb \gamma_{n\phi} \sfV_{n}^\mu 
    + \gamma_{\epsilon\phi} \sfV_{\epsilon}^\mu  \rb \Df_\mu\Psi
    + \frac{1}{\sigma_\Psi}\sfS_\Psi~,
\end{align}
where $\aleph = (T-T_\env)/T_\env$ and we have introduced the notation
\begin{gather}
    \sfH_\Psi = \Df'_\mu(f_\Psi\Df^\mu\Psi) - \frac{\dow V}{\dow|\Psi|^2} \Psi~, \qquad 
    \lambda_\phi = 1 \blue{\,+\,\aleph\lambda_{\phi\env}}~, \nn\\
    \sfS_\Psi = \sfH_\Psi \blue{\,-\,\aleph a_\env\Psi}
    - \sigma_\Psi \Df^{\mu}\Psi\lb\lambda_{\epsilon\phi}\sfV_{\epsilon\mu}
    + \lambda_{n\phi}\sfV_{n\mu}\rb~.
\end{gather}
More discussion can be found in \cref{sec:higgs-sf}. The constitutive relations are given by
\begin{align}
    \epsilon^\mu 
    &= \epsilon\, u^\mu
    - \blue{\lambda_\epsilon}f_\Psi \Xi^\mu_\epsilon
    - T\kappa^{\mu\nu}\sfV_{\epsilon\nu}
    - \sigma^{\mu\nu}_\times \sfV_{n\nu} 
    + \lambda_{\epsilon\phi}\Re[\sfS_\Psi \Df^\mu\Psi^*]
    - \gamma_{\epsilon\phi}\Im[\sfH_\Psi \Df^\mu\Psi^*]~, \nn\\
    j^\mu 
    &= n\,u^\mu 
    - \blue{\lambda_n}f_\Psi \Xi^\mu
    - \sigma^{\mu\nu} \sfV_{n\nu}
    - \sigma^{\mu\nu}_\times \sfV_{\epsilon\nu}
    + \lambda_{n\phi} \Re[\sfS_\Psi \Df^\mu\Psi^*]
    - \gamma_{n\phi}\Im[\sfH_\Psi \Df^\mu\Psi^*]~, \nn\\
    r_\env 
    &= \blue{\ell\gamma_\env  \kB  \Delta T
    - \beta_\env\lambda_{\epsilon\env} f_\Psi\Xi^\mu_\epsilon \sfV_{\epsilon\mu}
    - \beta_\env\lambda_{n\env} f_\Psi\Xi^\mu \sfV_{n\mu} 
    - \frac{\beta_\env a_\env}{\sigma_\Psi}\Re[\sfS_\Psi\Psi^*]
    - \beta_\env\mu\lambda_{\phi\env} \Im[\sfH_\Psi\Psi^*]}~, \nn\\
    r_\fuel 
    &= \blue{\ell\gamma_\fuel \Delta E}~.
\end{align}
where we have identified
\begin{align}
    \lambda_\epsilon = 1 \blue{\,+\,\aleph\lambda_{\epsilon\env}}~, \qquad
    \lambda_n = 1 \blue{\,+\,\aleph\lambda_{n\env}}~.
\end{align}

\paragraph{Integrating out the order parameter:} It is instructive to integrate out the magnitude of the order parameter and express the theory exclusively in terms of the Goldstone $\phi$ identified via $\Psi = |\Psi|\e^{i\phi}$. The equation of motion for the magnitude is given as
\begin{align}
    u^\mu\dow_\mu|\Psi|
    &=
    - \frac{1}{\sigma_\Psi}\lb 
    \frac{\dow V}{\dow|\Psi|^2} 
    + f_\Psi\xi^\mu\xi_\mu
    \blue{\,+\,\aleph a_\env} \rb |\Psi|
    + \gamma_{n\phi} \xi^\mu \sfV_{n\mu}|\Psi| 
    + \gamma_{\epsilon\phi} \xi^\mu \sfV_{\epsilon\mu} |\Psi|
    + \cO(\dow^2)~,
\end{align}
where $\xi_\mu=\dow_\mu\phi + A_\mu$. Assuming that $|\Psi| = \Psi_0$ is the zero of $\dow V/\dow|\Psi|^2 + f_\Psi\xi^\mu\xi_\mu
\blue{\,+\,\aleph a_\env}$, we can solve this equation at leading order in derivatives to find
\begin{align}\label{eq:eliminate-Psi}
    \frac{\dow V}{\dow|\Psi|^2} 
    + f_\Psi\xi^\mu\xi_\mu
    &= \sigma_\Psi\sfg_\epsilon \xi^\mu \sfV_{\epsilon\mu} 
    + \sigma_\Psi\sfg_n \xi^\mu \sfV_{n\mu}
    + \frac{1}{2\Psi_0^2}\sfg_\xi \nabla'_\mu(\Psi_0^2f_\Psi\xi^\mu)
    \blue{\,-\,\aleph a_\env}
    + \cO(\dow^2) \nn\\
    \implies
    |\Psi|
    &= \Psi_0 + \frac{1}{2\Psi_0}\frac{\sigma_\Psi\sfg_\epsilon \xi^\mu \sfV_{\epsilon\mu} 
    + \sigma_\Psi\sfg_n \xi^\mu \sfV_{n\mu}
    + \frac{1}{2\Psi_0^2}\sfg_\xi \nabla'_\mu(\Psi_0^2f_\Psi\xi^\mu)
    \blue{\,-\,\aleph a_\env}}{V''(\Psi_0^2)
    + f'_\Psi(\Psi_0^2)\xi^\mu\xi_\mu}
    + \cO(\dow^2)
\end{align}
where we have used the definitions in \cref{eq:sf-params} and treated $\Psi_0$ as a function of $s$, $n$, and $\xi^2=\xi^\mu\xi_\mu$ (as opposed to $T$, $\mu$, and $\xi^2$) to define
\begin{align}
    \sfg_\epsilon = \gamma_{\epsilon\phi} - 2\mu\frac{\dow\ln\Psi_0}{\dow \xi^2}~, \qquad
    \sfg_n = \gamma_{n\phi} - 2\frac{\dow\ln\Psi_0}{\dow \xi^2}~, \qquad
    \sfg_\xi = - 2\Psi_0^2\sigma_\Psi \frac{\dow\ln\Psi_0}{\dow n}~.
\end{align}
After $|\Psi|$ has been integrated out, these parameters renormalise various thermodynamic variables and transport coefficients in the low-energy description as
\begin{align}\label{eq:renormalisation}
    T\big|_\ren &= T + \frac{\dow\Psi_0^2}{\dow s}
    \lb \sigma_\Psi\sfg_\epsilon \xi^\mu \sfV_{\epsilon\mu} 
    + \sigma_\Psi\sfg_n \xi^\mu \sfV_{n\mu}
    + \frac{1}{2\Psi_0^2}\sfg_\xi \nabla'_\mu(\Psi_0^2 f_\Psi\xi^\mu)
    \blue{\,-\,\aleph a_\env} \rb~, \nn\\
    \mu\big|_\ren &= \mu + \frac{\dow\Psi_0^2}{\dow n}
    \lb \sigma_\Psi\sfg_\epsilon \xi^\mu \sfV_{\epsilon\mu} 
    + \sigma_\Psi\sfg_n \xi^\mu \sfV_{n\mu}
    + \frac{1}{2\Psi_0^2}\sfg_\xi \nabla'_\mu(\Psi_0^2 f_\Psi\xi^\mu)
    \blue{\,-\,\aleph a_\env}
    \rb ~, \nn\\
    f_s\big|_\ren &= \Psi_0^2 f_\Psi + 2\frac{\dow\Psi_0^2}{\dow\xi^2}
    \lb \sigma_\Psi\sfg_\epsilon \xi^\mu \sfV_{\epsilon\mu} 
    + \sigma_\Psi\sfg_n \xi^\mu \sfV_{n\mu}
    + \frac{1}{2\Psi_0^2}\sfg_\xi \nabla'_\mu(\Psi_0^2 f_\Psi\xi^\mu)
    \blue{\,-\,\aleph a_\env}
    \rb~, \nn\\
    \kappa^{\mu\nu}\big|_\ren
    &= \kappa^{\mu\nu}
    + \frac2T \Psi_0^2\sigma_\Psi
    \frac{(\sfg_\epsilon + \lambda_{\epsilon\phi}\sfg_\xi)^2}{1 + \sfg_\xi^2}
    \xi^\mu\xi^\nu~, \nn\\
    \sigma^{\mu\nu}\big|_\ren
    &= \sigma^{\mu\nu}
    + 2\Psi_0^2\sigma_\Psi
    \frac{(\sfg_n + \lambda_{n\phi}\sfg_\xi)^2}{1 + \sfg_\xi^2}  
    \xi^\mu\xi^\nu~, \nn\\
    \sigma^{\mu\nu}_\times\big|_\ren
    &= \sigma^{\mu\nu}_\times
    + 2\Psi_0^2\sigma_\Psi
    \frac{(\sfg_\epsilon + \lambda_{\epsilon\phi}\sfg_\xi)
    (\sfg_n + \lambda_{n\phi}\sfg_\xi)}{1 + \sfg_\xi^2}  
    \xi^\mu\xi^\nu , \nn\\
    \sigma_\phi\big|_\ren
    &= \frac{2\Psi_0^2\sigma_\Psi}{1+\sfg_\xi^2}~, \qquad 
    \lambda_{\epsilon\phi}\big|_\ren
    = \lambda_{\epsilon\phi} - \sfg_\epsilon \sfg_\xi~, \qquad 
    \lambda_{n\phi}\big|_\ren
    = \lambda_{n\phi} - \sfg_n \sfg_\xi~, \nn\\
    \blue{\lambda_{\phi\env} \big|_\ren}
    &= \blue{\lambda_{\phi\env} - \frac{a_\env\sfg_\xi}{\mu\sigma_\Psi}~, \qquad
    \lambda_{\epsilon\env}\big|_\ren
    = \lambda_{\epsilon\env} - \frac{a_\env\sfg_\epsilon}{\mu f_\Psi}~, \qquad
    \lambda_{n\env}\big|_\ren
    = \lambda_{n\env} - \frac{a_\env\sfg_n}{f_\Psi}}~.
\end{align}

Eliminating $|\Psi|$ using \cref{eq:eliminate-Psi} and performing the renormalisations in \cref{eq:renormalisation}, we can obtain the Josephson equation for the Goldstone phase 
\begin{align}
    u^\mu \xi_\mu
    &=
    \lambda_\phi\mu
    + \frac{1}{\sigma_\phi}\nabla'_\mu(f_s\xi^\mu)
    - \lambda_{\epsilon\phi} \xi^\mu\sfV_{\epsilon\mu}
    - \lambda_{n\phi} \xi^\mu\sfV_{n\mu}
    + \cO(\dow^2) \nn\\
    &\equiv \lambda_\phi\mu + \frac{1}{\sigma_\phi}\sfS_\phi~.
\end{align}
We have dropped the renormalisation labels ``ren'' for clarity. Similarly, we find the constitutive relations
\begin{align}
    \epsilon^\mu 
    &= \epsilon\, u^\mu
    - \blue{\lambda_\epsilon} f_s u^\nu\xi_\nu\xi^\mu
    - T\kappa^{\mu\nu} \sfV_{\epsilon\nu}
    - \sigma^{\mu\nu}_\times \sfV_{n\nu}
    + \lambda_{\epsilon\phi}\xi^\mu\sfS_\phi
    + \cO(\dow^2)
    ~, \nn\\
    j^\mu 
    &= n\,u^\mu 
    - \blue{\lambda_n} f_s\xi^\mu
    - \sigma^{\mu\nu} \sfV_{n\nu}
    - \sigma^{\mu\nu}_\times \sfV_{\epsilon\nu}
    + \lambda_{n\phi} \xi^\mu \sfS_\phi
    + \cO(\dow^2) ~, \nn\\
    r_\env 
    &= \blue{\ell\gamma_\env\kB\Delta T
    - \beta_\env\lambda_{\epsilon\env} f_s u^\nu\xi_\nu\xi^\mu \sfV_{\epsilon\mu}
    - \beta_\env\lambda_{n\env} f_s\xi^\mu \sfV_{n\mu}
    - \beta_\env\mu\lambda_{\phi\env} \dow_\mu(f_s\xi^\mu)
    + \cO(\dow^2)}~, \nn\\
    r_\fuel 
    &= \blue{\ell\gamma_\fuel \Delta E}~.
\end{align}

\paragraph{Goldstone effective action:} If we are interested in the low-energy description of the superfluid phase sufficiently far away from the critical point, instead of introducing the pair of order parameters $\bbPsi_{1,2}(\sigma)$, we may directly introduce the phase fields $\bbphi_{1,2}(\sigma)$ on the SK worldvolume, transforming under the worldvolume U(1) symmetry as $\bbphi_{1,2} \to \bbphi_{1,2}-\lambda$. These can be used to define the physical spacetime fields
\begin{align}
    \phi_{r} = \varphi_{r} + \bbphi_r + \cO(\hbar)~, \qquad
    \phi_{a} = \bbphi_a + \cO(\hbar)~,
\end{align}
where $\phi_r$ shifts under the diagonal U(1) spacetime symmetry while $\phi_a$ is entirely invariant. The KMS transformation is defined simply as
\begin{alignat}{2}
    \bbphi_1(\sigma) &\KMSto  
    \Theta\bbphi_1(\sigma), &\qquad
    \bbphi_2(\sigma) &\KMSto 
    \Theta\bbphi_2(\sigma + i\hbar\,\Theta\bbbeta)~,
\end{alignat}
with time-reversal eigenvalues $-1$. In the classical limit, this gives
\begin{alignat}{2}
    \phi_r &\KMSto \Theta\phi_r + \cO(\hbar)~, &\qquad 
    \phi_a &\KMSto 
    \Theta\hat\phi_a \equiv 
    \Theta\lb \phi_a + i\beta^\mu\xi_\mu
    - i \beta \mu 
    \rb + \cO(\hbar)~.
\end{alignat}

We can write down the low-energy effective action as
\begin{align}\label{eq:SKaction-sf-phi}
    \cL
    &= - \Big( \epsilon\, u^\mu - f_s u^\nu\xi_\nu\xi^\mu \Big) N_{a\mu}
    + \Big( n\,u^\mu - f_s\xi^\mu \Big) B_{a\mu} 
    - f_s\xi^\mu \dow_\mu\phi_a
    \nn\\
    &\qquad
    + i\kB T
    \begin{pmatrix}
         -N_{a\mu} \\ B_{a\mu}  \\ \blue{\beta_\env\Pi^\env_a} \\ 
         \phi_a \blue{\,+\,\beta_\env\mu_\env \Pi_a^\env}
    \end{pmatrix}^{\!\!\intercal}
    \begin{pmatrix}
        T\kappa^{\mu\nu} & \sigma^{\mu\nu}_\times 
        & \blue{-\lambda_{\epsilon\env}f_s u^\lambda\xi_\lambda\xi^\mu} 
        & -\sigma_\phi\lambda_{\epsilon\phi}\xi^\mu \\
        \sigma^{\nu\mu}_\times & \sigma^{\mu\nu}
        & \blue{-\lambda_{n\env} f_s\xi^\mu}
        & -\sigma_\phi\lambda_{n\phi}\xi^\mu \\
        \blue{\lambda_{\epsilon\env}f_s u^\lambda\xi_\lambda\xi^\nu} 
        & \blue{\lambda_{n\env} f_s\xi^\nu} & \blue{\gamma_\env/\beta_\env^2} & \blue{0} \\
        \sigma_\phi\lambda_{\epsilon\phi}\xi^\nu
        & \sigma_\phi\lambda_{n\phi}\xi^\nu
        & \blue{0} & \sigma_\phi
    \end{pmatrix}
    \begin{pmatrix}
         -\hat N_{a\mu} \\ \hat B_{a\mu}  \\ 
         \blue{\beta_\env\hat\Pi^\env_a} \\ 
         \hat\phi_a \blue{\,+\,\beta_\env\mu_\env\hat\Pi_a^\env}
    \end{pmatrix}
    \nn\\
    &\qquad 
    + \blue{i\kB T \gamma_\fuel\, \Pi^\fuel_a\hat\Pi^\fuel_a}~,
\end{align}
which is much simpler than the one in \cref{eq:SKaction-sf-general} and gives rise to \cref{eq:SK-Lagrangian-sf} in the main text. 

\subsection{Active fluids}
\label{app:activehydro}

We now construct the SK effective action for active Galilean hydrodynamics including momentum conservation; the relevant discussion for passive Galilean hydrodynamics can be found in~\cite{Jain:2020vgc}.
Truncating the effective theory to at most quadratic order in $a$-type fields, we can write down a simple effective action for active Galilean hydrodynamics 
\begin{align}\label{eq:SKaction-active}
    S 
    &= \int_x 
    - \varepsilon\, u^\mu N_{a\mu} 
    + \half\lb \rho\,u^\mu u^\nu 
    + p\, h^{\mu\nu}\rb \! \cH_{a\mu\nu}
    \nn\\
    &\qquad
    + i  \kB T 
    \begin{pmatrix}
        -N_{a\mu} \\
         \half\cH_{a\mu\nu} \\ 
         \blue{\beta_\env\Pi^\env_a}
    \end{pmatrix}^{\!\!\intercal}
    \begin{pmatrix}
        T\kappa h^{\mu\rho} & 0 & \blue{0} \\
        0 & 2\eta\,h^{\mu\langle\rho}h^{\sigma\rangle\nu}
        + \zeta h^{\mu\nu}h^{\rho\sigma}
        & \blue{p_\env h^{\mu\nu}} \\
        \blue{0} & \blue{-p_\env h^{\rho\sigma}}
        & \blue{\gamma_\env/\beta_\env^2}
    \end{pmatrix}
    \begin{pmatrix} 
        -\hat N_{a\rho} \\
        \half\hat\cH_{a\rho\sigma} \\
        \blue{\beta_\env\hat\Pi^\env_a}
    \end{pmatrix} 
    + \blue{i\kB T \gamma_\fuel \Pi^\fuel_a\hat\Pi^\fuel_a}
    ~.
\end{align}
Note that we have used the Milne-invariant quantities $\cH_{a\mu\nu}$ from \cref{eq:gal-H}. Angular brackets denote a symmetric-traceless contribution. The effect of activity is slightly more pronounced in this case compared to the simple diffusion model, because of the $p_\env$ term that will act as an active correction to pressure.

As with the diffusion model, the action \eqref{eq:SKaction-active} is consistent with the SK conditions in \cref{eq:SK-conds}, provided that we impose the inequality constraints on the leading-order diagonal dissipative coefficients
\begin{equation}
    \kappa \geq 0~, \qquad 
    \eta \geq 0~, \qquad 
    \zeta \geq0~, \qquad\qquad 
    \blue{\sigma_\fuel \geq 0~, \qquad 
    \sigma_\env \geq 0~.}
\end{equation}
The thermodynamic terms in the first line in \cref{eq:SKaction-active} are KMS-invariant up to a total derivative term, i.e.
\begin{align}
    - \varepsilon\, & u^\mu i\lie_\beta N_{r\mu} 
    + \lb \rho\,u^\mu u^\nu 
    + p\, h^{\mu\nu}\rb \! 
    \lb \half i\lie_\beta H_{r\mu\nu}
    + N_{r\mu} i\lie_\beta B_{r\nu}
    - i\lie_\beta N_{r\mu}
    \lb \vec u_{\nu} - \half N_{r\nu} \vec u^2 \rb \rb
    \nn\\
    &= 
    i\lb - T(\varepsilon+p) \lie_\beta \frac1T
    + T\rho \lie_\beta \frac{\varpi}{T}
    + \frac{p}{\sqrt{\gamma}} \lie_\beta \sqrt{\gamma}\rb
    \nn\\
    &= i\nabla'_\mu\big(p\,\beta^\mu\big)~,
\end{align}
where we have used (the non-nematic versions of) the thermodynamic relations in \cref{eq:thermoidentities} with $\cF=-p$.
The terms in the second line in \cref{eq:SKaction-active} are manifestly KMS-invariant.

We can vary the effective action \eqref{eq:SKaction-active} with respect to the $a$-type background fields to read off the associated constitutive relations
\begin{align}
    \epsilon^\mu 
    &= \lb \varepsilon + \half\rho \vec u^2 \rb u^\mu + (p+\blue{p_\env\aleph})\vec u^\mu 
    - T\kappa\, \sfV^\mu_{\epsilon}
    - \eta\, \sigma^{\mu\nu} \vec u_\nu
    - \zeta\,\vec u^{\mu} \nabla'_\lambda u^\lambda~, \nn\\
    \tau^{\mu\nu}
    &= \rho\,\vec u^\mu \vec u^\nu + (p+\blue{p_\env\aleph})h^{\mu\nu}
    - \eta\, \sigma^{\mu\nu}
    - \zeta\,h^{\mu\nu} \nabla'_\lambda u^\lambda~,
    \nn\\
    j^\mu 
    &= \rho\,u^\mu~, \qquad 
    \pi^\mu = \rho\vec u^\mu~, \nn\\
    r_\env &= \blue{\ell\gamma_\env \kB \Delta T
    + \beta_\env p_\env \nabla'_\lambda u^\lambda}~, \nn\\
    r_\fuel &= \blue{\ell\gamma_\fuel \Delta E}~,
\end{align}
where we have used $\sfV_{\epsilon\mu}$ from \cref{eq:V-def} and further identified the fluid shear and vorticity tensors
\begin{align}
    \sigma^{\mu\nu} 
    &= 2 h^{\sigma(\mu}\nabla^{\text g}_\sigma u^{\nu)}
    - \frac2d h^{\mu\nu} \nabla'^{\rm g}_\lambda u^\lambda
    = 2 h^{\sigma(\mu}\nabla_\sigma u^{\nu)}
    + 2 \vec u^{(\mu} h^{\nu)\rho} F^n_{\rho\sigma} u^\sigma
    - \frac2d h^{\mu\nu} \nabla'_\lambda u^\lambda ~, \nn\\
    \omega^{\mu\nu}
    &= 2h^{\sigma[\mu}\Ng_{\sigma}u^{\nu]}
    = 2h^{\rho[\mu}h^{\nu]\sigma}\partial_\rho \vec u_\sigma
    + F^{\mu\nu}
    - \half F_n^{\mu\nu} \vec u^2~.
\end{align}
Note that the shear tensor is Galilean-invariant, as has been manifested using the Galilean-covariant derivative defined using the connection in \cref{eq:gal-connection}.

\subsection{Active nematics}
\label{app:activenematics}

To model an active nematic liquid crystal in the SK framework, we need to introduce the doubled order parameters $\bbQ_{1,2\alpha\beta}$, defined to be purely spatial, i.e. $\bbbeta^\alpha \bbQ_{1,2\alpha\beta} = 0$. These fields are taken to be invariant under all the global spacetime symmetries, invariant under worldvolume gauge transformations, and covariant under worldvolume diffeomorphisms. The action of KMS symmetry on these fields is defined as
\begin{alignat}{2}
    \bbQ_{1\alpha\beta}(\sigma) &\KMSto  
    \Theta\bbQ_{1\alpha\beta}(\sigma), &\qquad
    \bbQ_{2\alpha\beta}(\sigma) &\KMSto 
    \Theta\bbQ_{2\alpha\beta}(\sigma + i\hbar\,\Theta\bbbeta)~,
\end{alignat}
with time-reversal eigenvalue $+1$. In the physical spacetime formulation, the associated average-difference basis can be used to define the spacetime order parameter and the associated noise field via $Q_{r,a\mu\nu} 
= \dow_\mu\sigma^\alpha\dow_\nu\sigma^\beta\bbQ_{r\alpha\beta}$, such that $\beta^\mu Q_{r,a\mu\nu} =0$. We can obtain the KMS transformation of these fields in the classical limit as
\begin{equation}
    Q_{r\mu\nu} \KMSto \Theta Q_{r\mu\nu} + \cO(\hbar)~, \qquad 
    Q_{a\mu\nu} \KMSto \Theta\hat Q_{a\mu\nu} \equiv
    \Theta\lb Q_{a\mu\nu} + i\lie_\beta Q_{r\mu\nu}  \rb + \cO(\hbar)~.
\end{equation}
We shall identify $Q_{r\mu\nu}$ with $Q_{\mu\nu}$ for the rest of this discussion. It is also useful to define a shifted noise field 
\begin{align}\label{eq:shiftedQ-defn}
    \cQ_{a\mu\nu} &= Q_{a\mu\nu} 
    - Q^\rho_{~(\mu} \cH_{a\nu)\rho}
    + n_{(\mu}Q_{\nu)}{}^\rho u^\sigma\cH_{a\sigma \rho}
    ~, \qquad 
    \beta^\mu\cQ_{a\mu\nu} = 0~,
\end{align}
using $\cH_{a\sigma \rho}$ was defined in \cref{eq:cB-def}. This has been used in \cref{sec:active_nematics}.

The nematic order parameter is generally taken to be traceless. However, it is a bit subtle to implement this as a constraint in the SK formalism because the traces $\bbh^{\alpha\beta}_{1,2} \bbQ_{1,2\alpha\beta}$ involve the dynamical spacetime fields $X^\mu$ through the pullback maps and yield highly non-trivial dynamical constraints to be implemented in the path integral. On the other hand, since $\bbbeta^\alpha$ is a fixed timelike vector, the constraints $\bbbeta^\alpha \bbQ_{1,2\alpha\beta} = 0$ are linear and can easily be implemented. As it turns out, we can actually use $\tr(\cQ_a) = h^{\mu\nu}\cQ_{a\mu\nu}$ as a Lagrange multiplier to set $\tr(Q) = h^{\mu\nu}Q_{\mu\nu}$ to zero onshell. To this end, consider the contribution to the action
\begin{align}\label{eq:nematic-trace-action}
    \cL_{\tr} &= \tr(Q)\tr(\cQ_a)
    + \half \tr(Q)^2
    \lb \half h^{\mu\nu}\cH_{a\mu\nu}
    + u^\mu N_{a\mu} \rb~.
\end{align}
The second term is required so that \cref{eq:nematic-trace-action} is KMS-invariant up to a total-derivative boundary term $\frac{i}{2} \nabla'_\mu(\beta^\mu \tr(Q)^2))$. The equations of motion for $\tr(Q)$ and $\tr(\cQ_a)$ mutually set each other to zero onshell, as desired, and any contribution from \cref{eq:nematic-trace-action} identically drops out. For the remainder of this appendix, we will consider these constraint to have been implemented and consider both $Q_{\mu\nu}$ and $\cQ_{a\mu\nu}$ to be traceless.

\paragraph{Nematic thermodynamics:} Let us take a quick detour to study nematic thermodynamics in the presence of background sources, which forms the backbone for the thermodynamic contributions to the SK effective Lagrangian we saw in \cref{eq:Qterms}. For simplicity, let us assume that the free energy density $\cF$ only depends on $Q_{\mu\nu}$, its first Galilean-covariant derivative $\nabla^{\text g}_\lambda Q_{\mu\nu}$ and the thermodynamic parameters $T$ and $\varpi$, mutually contracted using $h^{\mu\nu}$, and does not depend on any higher-derivatives. Specialising to flat space, an example of such a free energy is given in \cref{eq:F-nematic}. Generally, we may express the variations of $\cF$ as
\begin{align}
    \delta\cF 
    &= - s\,\delta T - \rho\,\delta\varpi
    + \frac{\dow\cF}{\dow Q_{\mu\nu}} \delta Q_{\mu\nu}
    + \frac{\dow\cF}{\dow\nabla^{\text g}_\lambda Q_{\mu\nu}} 
    \delta \nabla^{\text g}_\lambda Q_{\mu\nu}
    + \frac{\dow\cF}{\dow h^{\mu\nu}} \delta h^{\mu\nu} \nn\\
    &= - s\,\delta T - \rho\,\delta\varpi
    + \frac{\delta\cF}{\delta Q_{\mu\nu}} \delta Q_{\mu\nu}
    - 2\frac{\dow\cF}{\dow\nabla^{\text g}_\lambda Q_{\mu\nu}} 
    \delta \Gamma^{\text g}{}^\rho_{\lambda\mu} Q_{\rho\nu}
    + \frac{\dow\cF}{\dow h^{\mu\nu}}  \delta h^{\mu\nu}~,
    \label{eq:nem-thermo}
\end{align}
where we have utilised the Galilean connection introduced in \cref{eq:gal-connection}.
This is a generalisation of the thermodynamic relation in \cref{eq:thermoidentities}.
Since $\cF$ is a scalar, the variations are not all independent and satisfy
\begin{gather}
    \frac{\dow\cF}{\dow h^{\mu\nu}} 
    = \lb h_{(\mu}^\rho h_{\nu)\sigma} 
    + 2n_{(\mu}h_{\nu)\sigma} v^\rho \rb
    \lb \frac{\dow\cF}{\dow Q_{\tau\sigma}} Q_{\tau\rho} 
    + \frac{\dow\cF}{\dow\nabla^{\text g}_\lambda Q_{\tau\sigma}} 
    \nabla^{\text g}_\lambda Q_{\tau\rho}
    + \half \frac{\dow\cF}{\dow\nabla^{\text g}_\sigma Q_{\tau\lambda}}
    \nabla^{\text g}_\rho Q_{\tau\lambda} \rb
    , \nn\\
    \text{and}\qquad
    2 \frac{\dow\cF}{\dow Q_{\sigma[\nu}} Q_{\sigma}{}^{\mu]}
    + 2 \frac{\dow\cF}{\dow\nabla^{\text g}_\lambda Q_{\sigma[\nu}}  
    \nabla^{\text g}_\lambda Q_{\sigma}{}^{\mu]}
    + \nabla_{\rm g}^{[\mu} Q_{\sigma\lambda} \frac{\dow\cF}{\dow\nabla^{\text g}_{\nu]} Q_{\sigma\lambda}}
    = 0~.
    \label{eq:neamtic-conditions}
\end{gather}
One may check that these identities are satisfied for the variations arising from the free energy specified in \cref{eq:F-nematic}. Acting on \cref{eq:nem-thermo} with $\lie_\beta$, we can deduce
\begin{align}\label{eq:adiabaticequation}
    -\frac{1}{\sqrt{\gamma}}\lie_\beta \Big(\sqrt{\gamma}\,\cF \Big)
    &= - \varepsilon\, u^\mu \lie_\beta n_\mu
    + \lb \rho\,u^\mu u^\nu 
    - \cF\, h^{\mu\nu}
    \rb \lb 
    \half\lie_\beta h_{\mu\nu} 
    - \lie_\beta n_{(\mu}\vec u_{\nu)} \rb \nn\\
    &\qquad 
    + \frac{\dow\cF}{\dow\nabla^{\text g}_\mu Q_{\tau\lambda}} 
    \lb u^\rho\nabla^{\text g}_\rho Q_{\tau\lambda}
    \lie_\beta n_\mu 
    + \nabla_{\text g}^\nu Q_{\tau\lambda} 
    \lb \half\lie_\beta h_{\mu\nu} 
    - \lie_\beta n_{(\mu}\vec u_{\nu)} \rb\rb
    \nn\\
    &\qquad 
    + 2\frac{\dow\cF}{\dow\nabla^{\text g}_\mu Q_{\sigma[\rho}} Q_{\sigma}{}^{\nu]}
    \lb 
    \nabla'^{\text g}_\rho \lb 
    \lie_\beta h_{\mu\nu} - 2\lie_\beta n_{(\mu}\vec u_{\nu)} \rb 
    + h_{\lambda\nu} \nabla^{\text g}_{\rho} u^\lambda
    \lie_\beta n_{\mu}
    + 2h_{\lambda(\rho} \nabla^{\text g}_{\mu)} u^\lambda
    \lie_\beta n_{\nu}
    \rb
    \nn\\
    &\qquad 
    - \lb \frac{\dow\cF}{\dow Q_{\mu\nu}} 
    + \frac{\dow\cF}{\dow \Ng_\lambda Q_{\mu\nu}} \Ng_\lambda
    \rb
    \Big( \lie_\beta Q_{\mu\nu}   
    - \lb \lie_\beta h_{\mu\rho} 
    - 2\lie_\beta n_{(\mu}\vec u_{\rho)} \rb Q_{\nu}^{~\rho}
    \Big)~.
\end{align}
In deriving this, we have used the variations of the Galilean-invariant connection 
\begin{align}
    \delta\Gamma^{\text g}{}^\lambda_{\mu\nu} 
    &= u^\lambda \nabla^{\text g}_\mu \delta n_\nu
    + \half h^{\lambda\rho}\lb 
    2 \nabla^{\text g}_{(\mu}\!\lb\delta h_{\nu)\rho} 
    - \vec u_{\nu)}\delta n_\rho
    - \delta n_{\nu)} \vec u_{\rho}
    \rb
    - \nabla^{\text g}_\rho\!\lb\delta h_{\mu\nu} 
    - 2\vec u_{(\mu} \delta n_{\nu)}
    \rb \rb
    \nn\\
    &\qquad 
    + \delta n_{(\mu}\lb 
    \nabla^{\text g}_{\nu)} u^\lambda
    - h^{\lambda\rho}h_{\nu)\sigma}\nabla^{\text g}_{\rho}u^\sigma \rb 
    + h_{\rho(\mu}\nabla^{\text g}_{\nu)} u^\rho  h^{\lambda\sigma} \delta n_\sigma
    + \lb \ldots \rb^{\lambda}_{~\mu} n_{\nu}
    + \lb \ldots \rb^{\lambda}_{~\nu} n_{\mu}~,
    \label{eq:finalresult}
\end{align} 
evaluated at fixed $\beta^\mu$ and up to some temporal terms that do not affect our results. 

\paragraph{Effective action:} The SK effective Lagrangian for an active nematic liquid crystal can be written as 
\begin{align}\label{eq:SKaction-active-nematic}
    \cL
    &= 
    - \varepsilon\, u^\mu N_{a\mu}
    + \half\lb \rho\,u^\mu u^\nu - \cF\, h^{\mu\nu}\rb \cH_{a\mu\nu}
    - \lb \frac{\dow\cF}{\dow Q_{\mu\nu}} 
    + \frac{\dow\cF}{\dow \Ng_\lambda Q_{\mu\nu}} \Ng_\lambda
    \rb \cQ_{a\mu\nu}
    \nn\\
    &\qquad 
    + \frac{\dow\cF}{\dow\nabla^{\text g}_\mu Q_{\tau\lambda}} 
    \lb u^\rho\nabla^{\text g}_\rho Q_{\tau\lambda} N_{a\mu}
    + \half\nabla_{\text g}^\nu Q_{\tau\lambda} \cH_{a\mu\nu} \rb
    + 2\frac{\dow\cF}{\dow\nabla^{\text g}_\mu Q_{\sigma[\rho}} Q_{\sigma}{}^{\nu]}
    \lb \nabla^{\text g}_\rho \cH_{a\mu\nu}
    + h_{\lambda\nu} \nabla^{\text g}_{\rho} u^\lambda N_{a\mu}
    + 2h_{\lambda(\rho} \nabla^{\text g}_{\mu)} u^\lambda N_{a\nu}
    \rb \nn\\
    &\qquad 
    + i\kB T
    \begin{pmatrix}
        -N_{a\mu} \\
         \half\cH_{a\mu\nu} \\ 
         \cQ_{a\mu\nu} 
         -  \half \gamma_{\mu\nu}{}^{\tau\lambda} \cH_{a\tau\lambda} \\
         \blue{\beta_\env\Pi^\env_a}
    \end{pmatrix}^{\!\!\intercal}\!\!
    \begin{pmatrix}
        T\kappa^{\mu\rho} & 0 & 0 & \blue{0} \\
        0 & \eta^{\mu\nu\rho\sigma} & 0
        &\blue{ p_\env h^{\mu\nu} + \lambda_\env Q^{\mu\nu}} \\
        0 & 0 & \sigma_Q^{\mu\nu\rho\sigma} & 
        \blue{-a_\env Q^{\mu\nu}} \\
        \blue{0} & \blue{ -p_\env h^{\rho\sigma} - \lambda_\env Q^{\rho\sigma}}
        & \blue{a_\env Q^{\rho\sigma}}
        & \blue{\gamma_\env/\beta_\env}
    \end{pmatrix}\!\!
    \begin{pmatrix}
        -N_{a\rho} \\
        \half\hat\cH_{a\rho\sigma} \\
        \hat \cQ_{a\rho\sigma}
        -  \half \gamma_{\rho\sigma}{}^{\tau\lambda} \hat\cH_{a\tau\lambda} \\
        \blue{\beta_\env\hat\Pi^\env_a}
    \end{pmatrix} \nn\\
    &\qquad 
    + \blue{i\kB T \gamma_\fuel \Pi^\fuel_a\hat\Pi^\fuel_a}~.
\end{align}
The first two lines are KMS-invariant up to a boundary term $-\nabla'_\mu(\cF\beta^\mu + (\ldots)^\mu)$ due to \cref{eq:adiabaticequation}. The remaining terms are manifestly KMS-invariant.
Here we have introduced the thermal conductivity matrix $\kappa^{\mu\nu}$, viscosity tensor $\eta^{\mu\nu\rho\sigma}$, nematic conductivity tensor $\sigma_Q^{\mu\nu\rho\sigma}$, and the nematic shear coupling tensor $\gamma^{\mu\nu\rho\sigma}$. Note that these may have anisotropic components due to the nematic order. Assuming the nematic to be uniaxial~\footnote{A uniaxial nematic order parameter satisfies the identity $2Q^{\mu[\nu}Q^{\rho]\sigma} + \frac2d Q^{\mu[\nu}h^{\rho]\sigma} + \frac2d h^{\mu[\nu}Q^{\rho]\sigma} + \frac{2}{d^2} h^{\mu[\nu}h^{\rho]\sigma} = 0$, which can be used to eliminate higher-rank polynomial tensor structures.}, they take the general form
\begin{align}
    \kappa^{\mu\nu}
    &= \kappa\,h^{\mu\nu} + \kappa_1 Q^{\mu\nu}~, \nn\\
    \eta^{\mu\nu\rho\sigma}
    &= 2\eta\,h^{\rho\langle\mu}h^{\nu\rangle\sigma}
    + 2\eta_1 \lb h^{\rho(\mu}Q^{\nu)\sigma} 
    + Q^{\rho(\mu}h^{\nu)\sigma} 
    - \frac{2}{d} \lb Q^{\mu\nu}  h^{\rho\sigma}
    + h^{\mu\nu} Q^{\rho\sigma} \rb
    \rb
    + \eta_2 Q^{\mu\nu} Q^{\rho\sigma} \nn\\
    &\qquad 
    + \zeta\, h^{\mu\nu}h^{\rho\sigma}
    + \zeta_1 \lb Q^{\mu\nu}  h^{\rho\sigma}
    + h^{\mu\nu} Q^{\rho\sigma} \rb~, \nn\\
    \sigma_Q^{\mu\nu\rho\sigma}
    &= \sigma_Q\,h^{\rho\langle\mu}h^{\nu\rangle\sigma}
    + \sigma_{Q,1} \lb h^{\rho(\mu}Q^{\nu)\sigma} 
    + Q^{\rho(\mu}h^{\nu)\sigma} 
    - \frac{2}{d} \lb Q^{\mu\nu}  h^{\rho\sigma}
    + h^{\mu\nu} Q^{\rho\sigma} \rb
    \rb
    + \sigma_{Q,2} Q^{\mu\nu} Q^{\rho\sigma}~, \nn\\
    \gamma^{\mu\nu\rho\sigma}
    &= \gamma_1 h^{\rho\langle\mu} h^{\nu\rangle\sigma}
    + \gamma_2 \lb h^{\rho(\mu}Q^{\nu)\sigma} 
    + Q^{\rho(\mu}h^{\nu)\sigma} 
    - \frac{2}{d} \lb Q^{\mu\nu}  h^{\rho\sigma}
    + h^{\mu\nu} Q^{\rho\sigma} \rb
    \rb
    + \gamma_3 Q^{\mu\nu} Q^{\rho\sigma}
    + \gamma_4 Q^{\mu\nu} h^{\rho\sigma}~.
\end{align}
The coefficients $\kappa_1$, $\eta_1$, $\eta_2$, $\zeta_1$, $\sigma_{Q,1}$, and $\sigma_{Q,2}$ may be understood as anisotropic viscosities and conductivities in the nematic phase. Whereas, $\gamma_1$, $\gamma_2$, $\gamma_3$ are the nematic tumbling parameters that characterise the coupling of $Q_{\mu\nu}$ to the fluid shear tensor $\sigma^{\mu\nu}$, while $\gamma_4$ to the fluid expansion $\nabla'_\mu u^\mu$. A kinetic theory computation for 3d incompressible nematics predicts $3\gamma_1/2 = \gamma_2 =-\gamma_3/2 \equiv \xi$~\cite{Doostmohammadi2018,thampi2015intrinsic,beris1994thermodynamics}. In the active sector, we have introduced new coefficients $\lambda_\env$, $a_\env$, and $\gamma_\env$ that will play similar roles to their namesakes from the superfluid model.

\paragraph{Constitutive relations:} The equation of motion for the nematic order-parameter reads
\begin{equation}
    u^\lambda \Ng_\lambda Q^{\mu\nu}
    = \Gamma_Q^{\mu\nu\rho\sigma}\Big( \sfH_{\rho\sigma} \blue{\,-\, \aleph a_\env Q_{\rho\sigma}} \Big)
    + \sfS^{\mu\nu\rho\sigma} h_{\sigma\lambda} \nabla^{\text g}_\rho u^\lambda~,
\end{equation}
where we have defined
\begin{gather}
    \sfH^{\mu\nu} = - \frac{\delta\cF}{\delta Q_{\mu\nu}} 
    + \frac{1}{d} h^{\mu\nu} \lb h_{\rho\sigma} - 2n_\mu \vec u_\nu + n_\mu n_\nu \vec u^2 \rb \frac{\delta\cF}{\delta Q_{\mu\nu}}~, \qquad 
    \Gamma_Q^{\mu\nu\rho\sigma} \sigma^Q_{\lambda\tau\rho\sigma}
    = h_\lambda^{\langle\mu}h_\tau^{\nu\rangle}~, \nn\\
    \sfS^{\mu\nu\rho\sigma}
    = \gamma^{\mu\nu\rho\sigma}
    + Q^{\rho(\mu} h^{\nu)\sigma} - Q^{\sigma(\mu} h^{\nu)\rho}~.
\end{gather}
Similarly, we can obtain the constitutive relations for an active nematic liquid crystal
\begin{align}\label{eq:fullstress}
    \epsilon^\mu 
    &= \lb \varepsilon + \half\rho \vec u^2 \rb u^\mu 
    - \lb \cF \blue{\,-\,\aleph\,p_\env}\rb \vec u^\mu 
    \blue{\,+\,\aleph\lambda_\env Q^{\mu\nu} \vec u_\nu}
    - T\kappa\, \sfV^\mu_{\epsilon}
    - \eta^{\mu\nu\rho\sigma}\vec u_\nu h_{\sigma\lambda}\Ng_\rho u^\lambda 
    + \sfS^{\rho\sigma\mu\nu}\sfH_{\rho\sigma}\vec u_\nu
    \nn\\
    &\qquad
    - \frac{\dow\cF}{\dow\nabla^{\text g}_\mu Q_{\tau\lambda}} v^\nu\Ng_\nu Q_{\tau\lambda}
    - \nabla'^{\text g}_\rho\!\lb\cX^{[\rho\mu]\nu}u^\sigma\rb h_{\nu\sigma}~, \nn\\
    \tau^{\mu\nu} 
    &= \rho\,\vec u^\mu \vec u^\nu 
    - \lb \cF\,h^{\mu\nu}
    \blue{\,-\,\aleph\, p_\env h^{\mu\nu} - \aleph\lambda_\env Q^{\mu\nu}}
    \rb 
    - \eta^{\mu\nu\rho\sigma}h_{\sigma\lambda}\Ng_\rho u^\lambda
    + \sfS^{\rho\sigma\mu\nu}\sfH_{\rho\sigma}
    + \frac{\dow\cF}{\dow\nabla^{\text g}_{\mu} Q_{\rho\sigma}} \nabla_{\text g}^{\nu} Q_{\rho\sigma}
    - \nabla'^{\text g}_\rho\cX^{[\rho\mu]\nu}~, \nn\\
    j^\mu 
    &= \rho\,u^\mu~, \qquad 
    \pi^\mu = \rho\vec u^\mu~, \nn\\
    r_\env 
    &= \blue{\ell\gamma_\env \Delta E_\env
    + \beta_\env p_\env  \nabla^{\text g}_\mu u^\mu
    + \beta_\env\lambda_\env Q^{\mu\nu}  h_{\mu\lambda} \nabla^{\text g}_\nu u^\lambda
    - \beta_\env a_\env \Gamma_Q^{\mu\nu\rho\sigma}Q_{\mu\nu}
    \Big( \sfH_{\rho\sigma} - \aleph a_\env Q_{\rho\sigma} \Big)}~, \nn\\
    r_\fuel 
    &= \ell\gamma_\fuel \Delta E~,
\end{align}
where we have used the identities in \cref{eq:neamtic-conditions} and defined
\begin{align}
    \cX^{[\rho\mu]\nu}
    &= 2\frac{\dow\cF}{\dow\nabla^{\text g}_{[\mu} Q_{\rho]\sigma}} Q_{\sigma}{}^{\nu}
    + 2Q_{\sigma}{}^{[\mu}\frac{\dow\cF}{\dow\nabla^{\text g}_{\rho]} Q_{\sigma\nu}} 
    + 2\frac{\dow\cF}{\dow\nabla^{\text g}_\nu Q_{\sigma[\rho}} Q_{\sigma}{}^{\mu]}~.
\end{align}
One may verify that the stress tensor $\tau^{\mu\nu}$ in \cref{eq:fullstress} is symmetric, but we have cast it in the canonical form by isolating a total-derivative improvement term $\nabla'^{\text g}_\rho\cX^{[\rho\mu]\nu}$ that identically drops out from the conservation equations in flat spacetime. The remaining ``canonical'' stress tensor is asymmetric~\cite{Doostmohammadi2018,starklubensky}. Similarly, the last term in the energy current $\epsilon^\mu$ drops out from the conservation equations in flat spacetime.

Around a uniaxial nematic state, we can parametrise the fluctuations of the order parameter as 
\begin{align}
    Q^{\mu\nu} = Q_\sfS\lb p^\mu p^\nu - \frac1d h^{\mu\nu} \rb + Q^{\mu\nu}_\sfT~,
\end{align}
where $\vec p^2 \equiv p_\mu p^\mu = 1$, $p_\mu Q^{\mu\nu}_\perp = h_{\mu\nu}Q^{\mu\nu}_\perp = 0$.
The $Q_\sfS$ and $Q_\sfT^{\mu\nu}$ components of the order parameter are gapped and can be integrated out setting them to $Q_0$ and $0$ respectively. This leaves us with an equation of motion for the director $p_\mu$ that takes the form
\begin{equation}
    u^\lambda \Ng_\lambda p^{\mu}
    = - \frac{1}{\sigma_p}\frac{\delta\cF}{\delta p_\mu}
    + \lb \bar p^{\mu[\sigma}p^{\rho]} + \gamma \bar p^{\mu(\sigma}p^{\rho)} \rb
    h_{\sigma\lambda} \nabla^{\text g}_\rho u^\lambda~,
\end{equation}
where $\bar p^{\mu\nu} = h^{\mu\nu}-p^\mu p^\nu$, and we have identified
\begin{align}
    \sigma_p = \frac{4Q_0^2}{d-1}p_\mu p_\rho \bar p_{\nu\sigma}\sigma_Q^{\mu\nu\rho\sigma}
    = 2Q_0^2\sigma_Q + 2Q_0^3\frac{d-2}{d}\sigma_{Q,1} ~, \qquad
    \gamma = \frac{2/Q_0}{d-1}p_\mu p_\rho \bar p_{\nu\sigma}\gamma^{\mu\nu\rho\sigma}
    = \frac{1}{Q_0}\gamma_1 
    + \frac{d-2}{d} \gamma_2~.
\end{align}

\paragraph{Director theory:} Following cue from our superfluid discussion, we can also write down a SK description for the nematic director directly. To this end, we introduce the doubled spatial director fields ${\mathbb p}_{1,2\alpha}$, satisfying $\bbbeta^\alpha{\mathbb p}_{1,2\alpha} = 0$, transforming under the KMS symmetry as
\begin{alignat}{2}
    {\mathbb p}_{1\alpha}(\sigma) &\KMSto  
    \Theta{\mathbb p}_{1\alpha}(\sigma), &\qquad
    {\mathbb p}_{2\alpha}(\sigma) &\KMSto 
    \Theta{\mathbb p}_{2\alpha}(\sigma + i\hbar\,\Theta\bbbeta)~,
\end{alignat}
with time-reversal eigenvalue $+1$. The director and its noise partner on the physical spacetime are defined as $p_{r,a\mu} = \frac{\dow\sigma^\alpha}{\dow x^\mu} {\mathbb p}_{r,a\alpha}$, such that $\beta^\mu p_{r,a\mu} = 0$, with the KMS transformation in the classical limit
\begin{equation}
    p_{r\mu} \KMSto \Theta p_{r\mu} + \cO(\hbar), \qquad 
    p_{a\mu} \KMSto \Theta\hat p_{a\mu} \equiv
    \Theta\lb p_{a\mu} + i\lie_\beta p_{r\mu} \rb+ \cO(\hbar)~.
\end{equation}
Furthermore, it is useful to define the shifted noise director given by
\begin{align}\label{eq:shiftedp-defn}
    \calp_{a\mu} &= p_{a\mu} 
    - \half \cH_{a\mu\nu}p^\nu
    + \half n_{\mu}u^\sigma\cH_{a\sigma\nu} p^\nu
    ~, \qquad 
    \beta^\mu\calp_{a\mu} = 0~,
\end{align}

To impose the normalisation of $p_\mu$, we can introduce the Lagrange multiplier terms similar to \cref{eq:nematic-trace-action}, i.e.
\begin{align}
    \cL_{\text{norm}}
    = (p^\mu p_\mu-1)(p^\nu\calp_{a\nu})
    + \frac14 (p^\mu p_\mu-1)^2
    \lb \half h^{\mu\nu}\cH_{a\mu\nu}
    + u^\mu N_{a\mu} \rb~,
\end{align}
which are KMS-invariant, and set $p^\mu p_\mu = 1$ and $p^\mu\calp_{a\mu}=0$ onshell.

An argument similar to that employed for \cref{eq:adiabaticequation} can be used to show
\begin{align}
    -\frac{1}{\sqrt{\gamma}}\lie_\beta \Big(\sqrt{\gamma}\,\cF \Big)
    &= - \varepsilon\, u^\mu \lie_\beta n_\mu
    + \lb \rho\,u^\mu u^\nu 
    - \cF\, h^{\mu\nu}
    \rb \lb 
    \half\lie_\beta h_{\mu\nu} 
    - \lie_\beta n_{(\mu}\vec u_{\nu)} \rb \nn\\
    &\qquad 
    + \frac{\dow\cF}{\dow\nabla^{\text g}_\mu p_{\lambda}} 
    \lb u^\rho\nabla^{\text g}_\rho p_{\lambda}
    \lie_\beta n_\mu 
    + \nabla_{\text g}^\nu p_\lambda
    \lb \half\lie_\beta h_{\mu\nu} 
    - \lie_\beta n_{(\mu}\vec u_{\nu)} \rb\rb
    \nn\\
    &\qquad 
    + \frac{\dow\cF}{\dow\nabla^{\text g}_\mu p_{[\rho}} p^{\nu]}
    \lb 
    \nabla'^{\text g}_\rho \lb 
    \lie_\beta h_{\mu\nu} - 2\lie_\beta n_{(\mu}\vec u_{\nu)} \rb 
    + h_{\lambda\nu} \nabla^{\text g}_{\rho} u^\lambda
    \lie_\beta n_{\mu}
    + 2h_{\lambda(\rho} \nabla^{\text g}_{\mu)} u^\lambda
    \lie_\beta n_{\nu}
    \rb
    \nn\\
    &\qquad 
    - \lb \frac{\dow\cF}{\dow p_{\mu}} 
    + \frac{\dow\cF}{\dow \Ng_\lambda p_{\mu}} \Ng_\lambda \rb
    \lb 
    \lie_\beta p_\mu
    -\half\lie_\beta h_{\mu\rho} p^{\rho}
    + \lie_\beta n_{(\mu}\vec u_{\rho)} p^{\rho} \rb ~,
\end{align}
for a free energy density $\cF$ that only depends on $T$, $\varpi$, $p_\mu$, $\Ng_\lambda p_\mu$, and $h^{\mu\nu}$. This allows one to immediately write down the effective action
\begin{align}\label{eq:SKaction-active-nematic-vector}
    \cL
    &= 
    - \varepsilon\, u^\mu N_{a\mu}
    + \half\lb \rho\,u^\mu u^\nu - \cF\, h^{\mu\nu}\rb \cH_{a\mu\nu}
    - \lb \frac{\dow\cF}{\dow p_{\mu}} 
    + \frac{\dow\cF}{\dow \Ng_\lambda p_{\mu}} \Ng_\lambda \rb \calp_{a\mu}
    \nn\\
    &\qquad 
    + \frac{\dow\cF}{\dow\nabla^{\text g}_\mu p_{\lambda}} 
    \lb u^\rho\nabla^{\text g}_\rho p_{\lambda} N_{a\mu}
    + \half\nabla_{\text g}^\nu p_{\lambda} \cH_{a\mu\nu} \rb
    + \frac{\dow\cF}{\dow\nabla^{\text g}_\mu p_{[\rho}} p^{\nu]}
    \lb \nabla^{\text g}_\rho \cH_{a\mu\nu}
    + h_{\lambda\nu} \nabla^{\text g}_{\rho} u^\lambda N_{a\mu}
    + 2h_{\lambda(\rho} \nabla^{\text g}_{\mu)} u^\lambda N_{a\nu}
    \rb \nn\\
    &\qquad 
    + i\kB T
    \begin{pmatrix}
        -N_{a\mu} \\
         \half\cH_{a\mu\nu} \\ 
         \calp_{a\mu} - \half\gamma\bar p_{\mu}^\lambda p^\tau \cH_{a\tau\lambda} \\
         \blue{\beta_\env\Pi^\env_a}
    \end{pmatrix}^{\!\!\intercal}\!\!
    \begin{pmatrix}
        T\kappa^{\mu\rho} & 0 & 0 & \blue{0} \\
        0 & \eta^{\mu\nu\rho\sigma} & 0
        &\blue{ p_\env h^{\mu\nu} + \lambda_\env p^{\langle\mu}p^{\nu\rangle}} \\
        0 & 0 & \sigma_p \bar p^{\mu\rho} & \blue{0} \\
        \blue{0} & \blue{ -p_\env h^{\rho\sigma} - \lambda_\env p^{\langle\rho}p^{\sigma\rangle}}
        & \blue{0} & \blue{\gamma_\env/\beta_\env}
    \end{pmatrix}\!\!
    \begin{pmatrix}
        -N_{a\rho} \\
        \half\hat\cH_{a\rho\sigma} \\
        \!\!\!\!\hat{~~\calp_{a\rho}} - \half\gamma\bar p_{\rho}^\lambda p^\tau \hat\cH_{a\tau\lambda} \\
        \blue{\beta_\env\hat\Pi^\env_a}
    \end{pmatrix} \nn\\
    &\qquad 
    + \blue{i\kB T \gamma_\fuel \Pi^\fuel_a\hat\Pi^\fuel_a}~,
\end{align}
yielding \cref{eq:nematic-action-q} in the main text.

\newpage

\section{Glossary of notation}
\label{eq:glossary}

For accessibility, in this appendix we provide a comprehensive glossary of symbols and notation employed in the main text, as well as the appendices.

{\centering
\vspace{1em}

\begin{tabular}{c|l}
    \multicolumn{2}{c}{\textbf{Glossary I: General notation}} \\
    \hline\hline 
    $(x^\mu)=(t,x^i)$ & Time and space coordinates \\
    $(\dow_\mu)=(\dow_t,\dow_i)$ & Time and space derivatives \\
    $\sigma^\alpha$ & Worldvolume coordinates in SK-EFT \\
    $\omega,k^i$ & Frequency and wavevector of linearised fluctuations \\
    $T_0$, $\beta_0=1/(\kB T_0)$ & Equilibrium/steady-state temperature and inverse temperature ($\kB$: Boltzmann constant) \\
    \hline
    $\cO$ & Stochastic operators \\
    $G^\rmS_{\cO\cO'}$ & Symmetric correlation functions (stochastic variances) \\
    $G^\rmR_{\cO\cO'}$ & Retarded correlation functions (linear response functions) \\
    \hline
    $L$, $\cL$, $\cZ$ & SK-EFT Lagrangian, Lagrangian density, and generating functional \\
    $\psi,\psi_a$ & Dynamical fields in SK-EFT \\
    $\sfs_{r,a}$ & Doubled background fields coupled to $\cO$ \\
    $\underline\sfs_{r,a}$ & Collective notation including fuel and environment background fields $(\sfs_{r,a},\ell\Phi^\fuel_{r,a},\ell\Phi^\env_{r,a})$ \\
    \hline 
    $f_r$ & Average of SK doubled fields $(f_1+f_2)/2$ \\
    $f_a$ & Difference of SK doubled fields $(f_1-f_2)/\hbar$ ($\hbar$: reduced Planck's constant) \\
    $\hat f_a$ & KMS transform of $f_a$ \\
    $\KMSto$ & KMS transformation operation ($f_r\KMSto \eta_\Theta f_r$, $f_a\KMSto \eta_\Theta \hat f_a$) \\
    $\eta_\Theta$ & Time-reversal eigenvalue \\
    \hline 
    $(\ldots)^*$ & Complex-conjugation \\
    $(\ldots)^\dagger\equiv (\ldots)^*|_{f_a\to-f_a}$ & SK-unitarity-conjugation \\
    \hline 
    $\uppsi$ & The state of field $\psi$ at some fixed time \\
    $\bbP(\psi|\uppsi_\i,t_\i)$ & Probability of traversing a path in phase space (\cref{eq:path-distribution}) \\
    $\bbP(\uppsi_\f,t_\f|\uppsi_\i,t_\i)$ & Total transition probability (\cref{eq:transition-probability}) \\
    $W_\psi$, $W$ & Work done along a path (\cref{eq:work-done-along-path}) and average external work (\cref{eq:work-done}) \\
    $\Omega$, $\cF$ & Grand canonical free energy and free energy density \\
    $\Delta\Omega$ & Free energy differential during a transition \\
    \hline\hline
\end{tabular}

\vspace{1em}

\begin{tabular}{c|c|l}
    \multicolumn{3}{c}{\textbf{Glossary II: Conserved currents and background fields}} \\
    \hline\hline
    $\epsilon$, $\epsilon^i$ & $\epsilon^\mu$ & Energy density and flux \\
    $\pi_i$, $\tau^{ij}$ & $\pi_\mu$, $\tau^{\mu\nu}$ & Momentum density and stress tensor
    ($\pi_\mu v^\mu=\tau^{\mu\nu}n_\nu=0$) \\
    $n$, $j^i$ & $j^\mu$ & Particle number/charge density and flux \\
    $\rho=n$, $\pi^i=j^i$ & $\rho^\mu=j^\mu$ & Mass density and flux for Galilean fluids ($\rho^\mu h_{\mu\nu} = \pi_\nu$) \\
    $s$, $s^i$ & $s^\mu$ & Entropy density and flux \\
    \hline 
    $n_t,n_i \equiv n_{rt},n_{ri}$ & $n_\mu\equiv n_{r\mu}$ & Background clock-field coupled to $\epsilon,\epsilon^i$ \\
    $v^i,h_{ij}\equiv v^i_{r},h_{rij}$ & $h_{\mu\nu}\equiv h_{r\mu\nu}$ & Background velocity and spatial-metric coupled to $\pi_i,\tau^{ij}$ \\
    & $v^\mu\equiv v^\mu_r$ & Newton-Cartan frame velocity ($h^{\mu\nu} n_\mu = 0$, $h^{\mu\nu} h_{\nu\rho}=\delta^\mu_\rho - v^\mu n_\rho $) \\
    & $h^{\mu\nu}\equiv h^{\mu\nu}_r$ & Newton-Cartan inverse spatial-metric ($v^\mu n_\mu = 1$, $v^\mu h_{\mu\nu}=0 $) \\
    $A_{t},A_{i}\equiv A_{rt}, A_{ri}$ & $A_\mu\equiv A_{r\mu}$ & Background gauge field coupled particle $n$, $j^i$ \\
    $n_{at},A_{at},\ldots$ & $n_{a\mu},A_{a\mu},\ldots$ & Noise partners of the background fields in SK-EFT \\
    \hline 
    & $\Gamma^\lambda_{\mu\nu}$, $\nabla_\mu$ & Newton-Cartan connection and covariant derivative (\cref{eq:connection}) \\
    & $\Gamma^{\rm g}{}^\lambda_{\mu\nu}$, $\nabla^{\rm g}_\mu$ & Galilean-invariant connection and covariant derivative \\
    & $\nabla'_\mu$, $\nabla'^{\rm g}_\mu$
    & Covariant derivatives shifted with $+F^n_{\mu\nu}v^\nu$ \\
    $\Df_t$, $\Df_i$ & $\Df_\mu$ & Gauge-covariant derivatives \\
    & $F^n_{\mu\nu}\equiv\dow_\mu n_\nu -\dow_\nu n_\mu$ & Background clock field strength \\
    & $F_{\mu\nu}\equiv\dow_\mu A_\nu -\dow_\nu A_\mu$ & Background gauge field strength \\
    $E_i\equiv \dow_iA_t - \dow_t A_i$ & $E_\mu\equiv F_{\mu\nu}v^\nu$ & Background electric field ($E_\mu v^\mu = 0$)\\
    \hline\hline
    $\chi^t,\chi^i,b^i,\Lambda\equiv\chi^t_r,\chi^i_r,b_r^i,\Lambda_r$ & $\chi^\mu,b^\mu,\Lambda\equiv\chi^\mu_r,b_r^\mu,\Lambda_r$ & Time-translation, space-translation, boost, and U(1) parameters \\
    $\chi^t_a$, $\chi^i_a$, $b^i_a$, $\Lambda_a$ & $\chi^\mu_a$, $b^\mu_a$, $\Lambda_a$ & Stochastic symmetry transformations in SK-EFT \\
    \hline\hline
\end{tabular}

\vspace{1em}

\begin{tabular}{c|c|l}
    \multicolumn{3}{c}{\textbf{Glossary III: Ordinary fluids}} \\
    \hline\hline 
    $T$, $u^i$, $\mu$ & $T$, $u^\mu$, $\mu$ & Dynamical temperature, fluid velocity ($u^\mu n_\mu = 1$), chemical potential \\
    \cline{1-2}
    \multicolumn{2}{c|}{$\varpi=\mu+\half\vec u^2$} & Galilean-invariant mass chemical potential \\
    \cline{1-2}
    $T_0$, $u^i_0$, $\mu_0$, $\varpi_0$
    & $T_0$, $u^\mu_0$, $\mu_0$, $\varpi_0$
    & Equilibrium/steady state \\
    $X^t_a$, $X^i_a$, $\varphi_a$ & $X^\mu_a$, $\varphi_a$ & Dynamical noise partners for $T$, $u^i$, $\mu$ in SK-EFT \\
    \hline
    
    $N_{at},N_{ai},B_{at},B_{ai},H_{ati},H_{aij}$ & $N_{a\mu},H_{a\mu\nu},B_{a\mu}$ & Noise invariants in SK-EFT defined in \cref{eq:invariant-defs-noncov} \\
    $\cB_{at}$, $\cB_{ai}$, $\cH_{aij}$ & $\cB_{a\mu}$, $\cH_{a\mu\nu}$ & Galilean-covariant noise invariants in SK-EFT defined in \cref{eq:cB-def} \\
    \hline\hline
    
    \multicolumn{2}{c|}{$p$} & Thermodynamic pressure \\
    \multicolumn{2}{c|}{$\varepsilon=\epsilon-\half\rho\vec u^2$} & Internal energy density (for Galilean fluids)\\
    \multicolumn{2}{c|}{$\eta,\zeta$} & Shear and bulk viscosities \\
    \multicolumn{2}{c|}{$\kappa$} & Thermal conductivity \\
    \multicolumn{2}{c|}{$\sigma$} & Particle/charge conductivity \\
    \multicolumn{2}{c|}{$\sigma_\times$} & Cross-conductivity between energy and particle/charge flux \\
    \hline
    \multicolumn{2}{c|}{$c_v=\dow\epsilon/\dow T$} & Heat capacity \\
    \multicolumn{2}{c|}{$\chi=\dow n/\dow\mu$} & Particle/charge susceptibility \\
    \multicolumn{2}{c|}{$D_\epsilon$} & Energy diffusivity (\cref{eq:D-defn}) \\
    \multicolumn{2}{c|}{$D_n$} & Particle/charge diffusivity (\cref{eq:D-defn}) \\
    \hline\hline
\end{tabular}

\vspace{1em}

\begin{tabular}{c|l}
    \multicolumn{2}{c}{\textbf{Glossary IV: Ordinary active fluids and energy relaxation}} \\
    \hline\hline
    $r_{\fuel,\env}$ & Rates of fuel consumption and entropy loss (divided by $\kB$) \\
    \hline
    $\Delta E$ & Energy differential driving activity \\
    $T_\env$ & Environment temperature \\
    $\Delta T=T-T_\env$ & Temperature differential induced by activity \\
    \hline 
    $\Phi^{\env,\fuel}\equiv\Phi^{\env,\fuel}_{r}$ & Environment and fuel background fields coupled to $\ell r_{\env,\fuel}$ \\
    $\Phi^{\env,\fuel}_{a}$ & Noise partners of the background fields $\Phi^{\env,\fuel}_{r}$ in SK-EFT \\
    \hline 
    $\Pi^{\env,\fuel}_{a}$ & SK symmetry-invariants defined in \cref{eq:cF-def,eq:cF-mod,eq:cF-def-general} \\
    \hline\hline
    $\gamma_{\fuel}$ & Fuel consumption response to energy differential \\
    $\gamma_{\env}$ & Entropy loss response to temperature differential \\
    $p_\env$ & Active parameter affecting mechanical pressure in the stress tensor \\
    \hline
    $\Gamma_\epsilon$ & Energy relaxation rate (\cref{eq:energy-relaxation-rate}) \\
    \hline 
    $\ell$ & Bookkeeping parameter for energy non-conservation \\
    $\aleph=\ell\Delta T/T_\env$ & Dynamical measure of activity \\
    $\hat\aleph=\ell\Delta E/(\kB T_\env)$ & External measure of activity \\
    \hline\hline
\end{tabular}

\vspace{1em}

\begin{tabular}{c|c|l}
    \multicolumn{3}{c}{\textbf{Glossary V: (Active) superfluids}} \\
    \hline
    \hline 
    \multicolumn{2}{c|}{$\Psi\equiv\Psi_r$} & Complex scalar order parameter for superfluidity \\
    \multicolumn{2}{c|}{$\Psi_0$} & Equilibrium magnitude of the order parameter \\
    \multicolumn{2}{c|}{$\phi\equiv\phi_r$} & Goldstone phase of the order parameter \\
    \multicolumn{2}{c|}{$\Psi_a$, $\phi_a$} & Dynamical noise partners for $\Psi_r$, $\phi_r$ in SK-EFT \\
    \hline 
    $\xi_t\equiv \dow_t\phi+A_t$ & \multirow{2}{2em}{\centering$\xi_\mu$} & Covariant time-derivative of $\phi$ \\
    $\xi_i\equiv \dow_i\phi+A_i$ & & Superfluid velocity -- covariant space-derivative of $\phi$ \\
    \hline 
    \multicolumn{2}{c|}{$V$} & Superfluid potential in \cref{eq:potential} \\
    \multicolumn{2}{c|}{$a$, $a_4$} & Parameters in $V$ that determine $\Psi_0$ in \cref{eq:Psi-ground} \\
    \multicolumn{2}{c|}{$f_s \equiv 2\Psi_0^2f_\Psi$} & Superfluid density parameter \\
    \multicolumn{2}{c|}{$\sigma_\phi \equiv 2\Psi_0^2\sigma_\Psi$} & Superfluid diffusion parameter \\
    \multicolumn{2}{c|}{$\lambda_{n\phi}$} 
    & Cross-conductivity between the charge flux and Josephson equation \\
    \multicolumn{2}{c|}{$\lambda_{\epsilon\phi}$} 
    & Cross-conductivity between energy flux and Josephson equation \\
    \hline 
    \multicolumn{2}{c|}{$v_s$} & Speed of superfluid sound (\cref{eq:D-defn}) \\
    \multicolumn{2}{c|}{$D_\phi$} & Superfluid diffusivity (\cref{eq:D-defn}) \\
    \hline\hline
    \multicolumn{2}{c|}{$a+\aleph a_\env$} & Active parameter affecting the equilibrium state $\Psi_0$ in \cref{eq:Psi-ground} \\
    \multicolumn{2}{c|}{$\lambda_\phi \equiv 1+\aleph\lambda_{\phi\env}$} & Active parameter screening the effective chemical potential in the Josephson equation \\
    \multicolumn{2}{c|}{$\lambda_n \equiv 1+\aleph\lambda_{n\env}$} & Active parameter screening the superfluid density in the charge flux \\
    \multicolumn{2}{c|}{$\lambda_\epsilon \equiv 1+\aleph\lambda_{\epsilon\env}$}
    & Active parameter screening the superfluid density in the energy flux \\
    \hline\hline
\end{tabular}

\vspace{1em}

\begin{tabular}{c|c|l}
    \multicolumn{3}{c}{\textbf{Glossary VI: (Active) nematics}} \\
    \hline\hline 
    $Q_{ij}\equiv Q_{rij}$ & $Q_{\mu\nu}\equiv Q_{r\mu\nu}$ & Nematic order parameter ($u^\mu Q_{\mu\nu}=0$) \\
    \cline{1-2}
    \multicolumn{2}{c|}{$Q_0$} & Strength of nematic alignment \\
    \cline{1-2}
    $p_i\equiv p_{ri}$ & $p_\mu\equiv p_{r\mu}$ & Goldstone director for nematic order ($u^\mu p_{\mu}=0$) \\
    $Q_{aij}$, $p_{ai}$ & 
    $Q_{a\mu\nu}$, $p_{a\mu}$ &
    Dynamical noise partners for $Q_{rij}$, $p_{ri}$ in SK-EFT ($u^\mu Q_{a\mu\nu}=u^\mu p_{a\mu}=0$) \\
    $\cQ_{aij}$, $\calp_{ai}$ & 
    $\cQ_{a\mu\nu}$, $\calp_{a\mu}$ & 
    Shifted noise partners defined in \cref{eq:shiftedQ-defn,eq:shiftedp-defn}
    ($u^\mu \cQ_{a\mu\nu}=u^\mu \calp_{a\mu}=0$) \\
    \hline 
    \multicolumn{2}{c|}{$V$} & Elastic potential in \cref{eq:potential-nem} \\
    \multicolumn{2}{c|}{$a$, $a_3$, $a_4$} & Parameters in $V$ that determine $Q_0$ in \cref{eq:def-Q0} \\
    \multicolumn{2}{c|}{$K\equiv 2Q_0^2K_Q$} & Frank elastic constant \\
    \multicolumn{2}{c|}{$\sigma_p\equiv 2Q_0^2\sigma_Q$} & Nematic conductivity \\
    \multicolumn{2}{c|}{$\gamma$} & Nematic shear coupling \\
    \hline\hline
    \multicolumn{2}{c|}{$a+\aleph a_\env$} & Active parameter affecting the equilibrium state $Q_0$ in \cref{eq:def-Q0} \\
    \multicolumn{2}{c|}{$\lambda_\env$} & Active parameter controlling the nematic stress $\tau^{ij}\sim Q^{ij}$
    (often called $\zeta$ in literature)
    \\
    \hline\hline
\end{tabular}

}
\end{document}